\documentclass[aps,prl,twocolumn,floatfix,superscriptaddress,balancelastpage,footinbib,amsmath,preprintnumbers]{revtex4-1}[11pt]
\pdfoutput=1
\usepackage{graphicx}
\usepackage{amsmath,amssymb,mathrsfs}
\usepackage{bbm}
\usepackage{color}
\usepackage{xcolor}
\usepackage{dsfont}
\usepackage{cancel}
\usepackage{dsfont}
\usepackage{epstopdf}
\usepackage{epsfig}
\usepackage{bm}
\usepackage{dcolumn}
\usepackage{hyperref}
\usepackage{enumitem}

\newcommand{\ben}{\begin{enumerate}}
\newcommand{\een}{\end{enumerate}}
\newcommand{\bit}{\begin{itemize}}
\newcommand{\eit}{\end{itemize}}

\newcommand{\beqa}{\begin{eqnarray}}
\newcommand{\eeqa}{\end{eqnarray}}
\newcommand{\beq}{\begin{equation}}
\newcommand{\eeq}{\end{equation}}
\newcommand{\bay}{\begin{array}}
\newcommand{\eay}{\end{array}}

\def\ifmath#1{\relax\ifmmode #1\else $#1$\fi}

\def\gsim{\ \rlap{\raise 3pt \hbox{$>$}}{\lower 3pt \hbox{$\sim$}}\ }
\def\lsim{\ \rlap{\raise 3pt \hbox{$<$}}{\lower 3pt \hbox{$\sim$}}\ }

\def\ls#1{\ifmath{_{\lower1.5pt\hbox{$\scriptstyle #1$}}}}
\def\lsup#1{^{\lower 6pt\hbox{$\scriptstyle#1$}}}

\def\bracket#1#2 {\mathinner{\langle{#1}|{#2}\rangle}}

\def\bracket#1#2 {\mathinner{\langle{#1}|{#2}\rangle}}

\newcommand{\bea}{\begin{eqnarray}}
\newcommand{\eea}{\end{eqnarray}}

\renewcommand{\vec}[1]{\mathbf{#1}}
\renewcommand{\vec}[1]{\mbox{\boldmath${#1}$}}

\graphicspath{{figs/}}
\begin{document}


\title{Direct Detection of sub-GeV Dark Matter with Scintillating Targets}

\author{Stephen Derenzo}
\email{sederenzo@lbl.gov}
\affiliation{Lawrence Berkeley National Laboratory, Mail Stop 55-121, Berkeley, CA 94720}

\author{Rouven Essig}
\email{rouven.essig@stonybrook.edu}
\affiliation{C.N. Yang Institute for Theoretical Physics, Stony Brook University, Stony Brook, NY 11794}

\author{Andrea Massari}
\email{andrea.massari@stonybrook.edu}
\affiliation{C.N. Yang Institute for Theoretical Physics, Stony Brook University, Stony Brook, NY 11794}

\author{Adri\'an Soto}
\email{adrian.soto-cambres@stonybrook.edu}
\affiliation{Department of Physics and Astronomy, Stony Brook University, Stony Brook, NY 11794-3800}
\affiliation{Institute for Advanced Computational Science, Stony Brook University, Stony Brook, NY 11794}

\author{Tien-Tien Yu}
\email{chiu-tien.yu@stonybrook.edu}
\affiliation{C.N. Yang Institute for Theoretical Physics, Stony Brook University, Stony Brook, NY 11794}

\preprint{YITP-SB-16-25}

\begin{abstract}
We describe a novel search for MeV-to-GeV-mass dark matter, in which the dark matter scatters off electrons in a scintillating target.  
The excitation and subsequent de-excitation of the electron produces one or more photons, which could be detected with 
an array of cryogenic low-noise photodetectors, such as transition edge sensors (TES) or microwave kinetic inductance devices (MKID). 
Scintillators may have distinct advantages over other experiments searching for a low ionization signal from sub-GeV DM. 
First, the detection of one or a few photons may be technologically easier.  
Second, since no electric field is required to detect the photons, there may be far fewer dark counts 
mimicking a DM signal. 
We discuss various target choices, but focus on calculating the expected dark matter-electron scattering rates in three scintillating crystals,  
sodium iodide (NaI), cesium iodide (CsI), and gallium arsenide (GaAs). 
Among these, GaAs has the lowest band gap (1.52~eV) compared to NaI (5.9~eV) or CsI (6.4~eV), 
allowing it to probe dark matter masses possibly as low as $\sim 0.5$~MeV, compared to $\sim 1.5$~MeV with NaI or CsI.  
We compare these scattering rates with those expected in silicon (Si) and germanium (Ge). 
The proposed experimental concept presents an important complementary path to existing efforts, 
and its potential advantages may make it the most sensitive direct-detection probe of DM down to MeV masses. 
\end{abstract}

\maketitle

\section{INTRODUCTION} \vskip -3mm
Dark matter (DM) with a mass in the MeV--GeV range is phenomenologically viable and has received increasing attention in recent years~\cite{Essig:2011nj,Essig:2015cda, Graham:2012su, Lee:2015qva, Essig:2012yx, Hochberg:2015pha,*Hochberg:2015fth,*Schutz:2016tid,*Hochberg:2016ntt, 
Essig:2013lka,*Bird:2004ts,*Borodatchenkova:2005ct,*McElrath:2005bp, *Fayet:2006sp, *Bird:2006jd, *Kahn:2007ru,*Fayet:2007ua,*Essig:2009nc,*Bjorken:2009mm,*Reece:2009un,*Fayet:2009tv,*Yeghiyan:2009xc,*Badin:2010uh,*Echenard:2012iq,*MarchRussell:2012hi,*Essig:2013vha, *Essig:2013goa,*Boehm:2013jpa,*Nollett:2013pwa,*Andreas:2013lya,*Izaguirre:2013uxa,*Battaglieri:2014qoa, *Izaguirre:2014bca,*Batell:2014mga,*Kahn:2014sra,*Krnjaic:2015mbs,*Batell:2009di,*Izaguirre:2015yja,      
Boehm:2003hm,*Strassler:2006im,*ArkaniHamed:2008qn,*Pospelov:2008jd,*Hooper:2008im,*Feng:2008ya,*Morrissey:2009ur,*Essig:2010ye,*Cohen:2010kn,*Lin:2011gj,*Chu:2011be,*Hochberg:2014dra,*Hochberg:2014kqa}. 
An important probe for DM is with {\it direct detection} experiments, in which a DM particle in 
the Milky-Way halo interacts with some target material in a detector, producing an observable signal in the form of heat, 
phonons, electrons, or photons~\cite{Cushman:2013zza}. 
The traditional technique of searching for nuclear recoils loses sensitivity rapidly 
for DM masses below a few GeV, since the DM is unable to transfer enough of its energy to the nucleus, resulting in no observable signal 
above detector thresholds. 
However, DM scattering off electrons, whose mass is much less than a nucleus, can lead to observable signals for masses 
well below 1~GeV~\cite{Essig:2011nj}, 
opening up vast new regions of parameter space for experimental exploration.  

DM-electron scattering in direct detection experiments has been investigated for noble liquid 
targets~\cite{Essig:2011nj,Essig:2012yx} and was demonstrated explicitly to have sensitivity down to 
DM masses of a few MeV and cross-sections of 
$\sim 10^{-37}~\rm{cm}^2$~\cite{Essig:2012yx} using published XENON10 data~\cite{Angle:2011th}.  
Semiconductor targets like silicon (Si) and germanium (Ge) 
could probe potentially several orders of magnitude of unexplored DM parameter space for masses as low as a few 
hundred keV~\cite{Essig:2011nj,Graham:2012su,Lee:2015qva,Essig:2015cda}.  
The feasibility of the required detector technology to detect small ionization signals  
still needs to be demonstrated and may become available in the next few years, 
e.g.~with SuperCDMS~\cite{Agnese:2015nto} and DAMIC~\cite{Moroni:2011xs}.  
In the future, even lower masses could be probed using superconductors or 
superfluids~\cite{Hochberg:2015pha,Hochberg:2015fth,Schutz:2016tid}.  

In this letter, we explore using a \emph{scintillator}
as the target material to search for dark matter with masses as low as a few hundred keV. 
One or more scintillation photons are emitted when an electron excited by a DM-electron scattering interaction relaxes to 
the ground state~\cite{Essig:2011nj}\footnote{Note that~\cite{Starkman:1994gf} proposed the search of one or more photons from 
Weak-scale dark matter through atomic excitations.}. Scintillation photons with an energy of $\mathcal{O}$(few~eV) could be detected by an array of transition edge sensors (TES) 
or microwave kinetic inductance detectors (MKIDs) operated at cryogenic temperatures, which surround a scintillating target of 
volume $\sim \mathcal{O}(({\rm few~ cm})^3)$.  
The development of such a large array of photodetectors sensitive to single photons is an active area of research~\cite{Pyle}.
The target itself should be cooled to cryogenic temperatures to avoid excitations induced by thermal fluctuations and 
large thermal gradients between it and the detector array. 

Several good scintillating materials exist.  
In this letter, we focus on three crystals,  
sodium iodide (NaI), cesium iodide (CsI), and gallium arsenide (GaAs).  
Other materials will be mentioned briefly. 

\vspace{-4mm}
\section{SCINTILLATORS: ADVANTAGES \& CHALLENGES} \vskip -3mm

Several signals are possible when sub-GeV DM scatters off a bound electron in an atom or a crystal, 
exciting the electron to a higher energy level or an unbound state~\cite{Essig:2011nj}. Depending on the target material, an experiment can either measure an ionization signal, which is obtained by manipulating the electron with an electric field, or one or more scintillation photons, which are emitted as the electron relaxes back to its ground state. Until now, the latter approach has not been considered in detail. 

Measuring the ionization signal has already 
constrained DM as light as a few MeV~\cite{Essig:2012yx}, using XENON10's two-phase xenon time projection chamber (TPC).  
Unfortunately, several possible detector-specific backgrounds exist, so one cannot currently claim 
that the observed one- and few-electron events are from DM~\cite{Essig:2012yx,Angle:2011th,Aprile:2016wwo}. 
Using semiconductors, CDMSlite~\cite{Agnese:2015nto} applied a bias voltage, forcing  
a conduction-band electron to traverse the material and generate 
enough Neganov-Luke phonons \cite{Luke:1990ir,Neganov:1985} to be measured by phonon detectors.  
The CDMSlite setup with improved phonon detectors may in the future surpass xenon-based TPCs in their sensitivity to 
sub-GeV DM. 
However, while there may be fewer dark counts than for two-phase xenon TPCs, 
the presence of an electric field may create spontaneous electron-hole pairs that could mimic a DM signal. 
Therefore, more work is needed to establish the potential of the CDMSlite setup. 

Sub-GeV DM searches are unlikely to be limited by traditional backgrounds like  
Compton scattering, cosmogenics, or neutrons. 
These backgrounds typically produce electron recoils at higher energies, and $\le 1$ event/kg/year is expected in SuperCDMS in the $\sim 1-50$~eV range~\cite{GolwalaTalk}. Great care must be taken to limit the material's surface exposure and  
radioactive contaminants. 
Coherent nuclear scattering of solar neutrinos is similarly insignificant. 
Instead, the discussion above highlights that understanding and controlling detector dark counts 
will likely determine the sensitivity.  

Instead of searching for an ionization signal, one could search for  
one or more scintillation photons. 
Scintillators possibly have two distinct advantages.  
First, the detection of such a low number of photons may turn out to be technologically easier than detecting a low number of electrons 
with the CDMSlite setup (or with the DAMIC setup~\cite{Moroni:2011xs}).  
Second, no electric field is required to detect the photons, which may lead to fewer dark counts.

A potential background for scintillators  
is phosphorescence induced from a previous interaction (afterglow). 
Our candidate targets scintillate on nano-to-millisecond timescales, but some 
photons could arise from excited states whose lifetimes are much longer (phosphorescent) due to a ``forbidden'' radiative transition.  
The phosphorescent photons typically have a lower energy, so if the photodetector's energy resolution is too low, 
a narrow-band optical filter could be placed between scintillator and photodetector 
to remove phosphorescent photons. 

\vspace{-4mm}
\section{SCHEMATIC EXPERIMENTAL CONCEPT} \vskip -3mm
\begin{figure}[t]
\vspace{-0.5cm}
\includegraphics[width=0.4\textwidth]{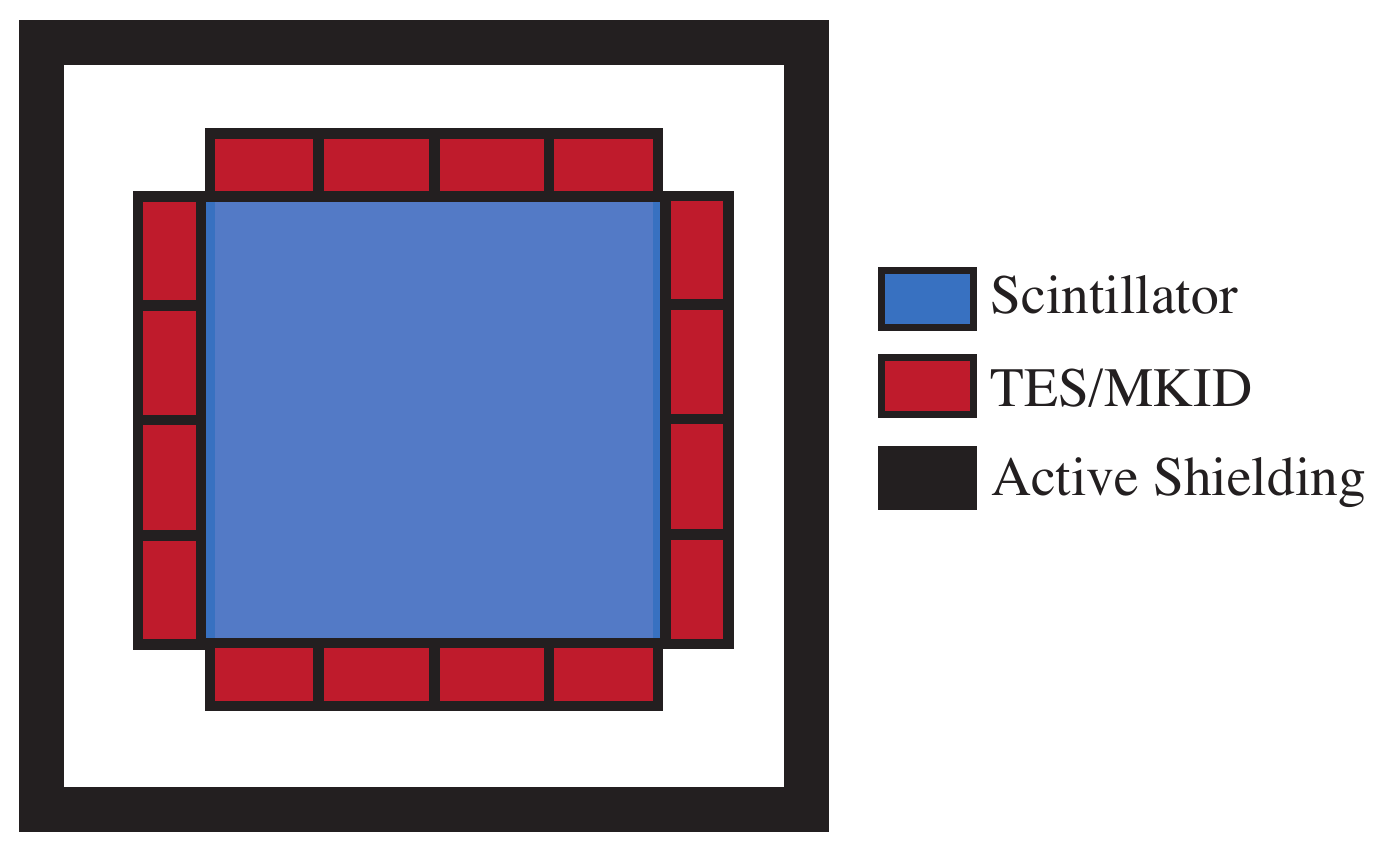}
\caption{Schematic experimental concept: a DM particle scatters off an electron in a scintillating target, exciting it to a higher-energy level; 
one or more scintillation photons from the relaxation of the electron to the ground state are observed by a surrounding 
photodetector array. 
The detector is encased in an active shield to eliminate environmental backgrounds.
No electric field is needed, reducing or eliminating many potential detector-specific backgrounds. }
\vspace{-0.5cm}
\label{fig:setup}
\end{figure}

Fig.~\ref{fig:setup} shows the experimental concept: a scintillating target is surrounded by a detector-array sensitive to single photons.  
An active shield surrounds the detector to veto radioactive backgrounds, including gamma rays that Compton-scatter in the target material. 
An optical filter between the scintillator and the photodetector could ensure passage of only the expected photon wavelengths. 

Detectors with single-photon sensitivity and no dark counts exist, e.g.~MKIDs~\cite{Mazin:12} and TESs~\cite{TES}, which operate at $\mathcal{O}$(100~mK) temperatures.  
These detectors can have few-percent energy resolution and microsecond time resolution~\cite{Mazin:12}. 
MKIDs (TESs) have demonstrated single-photon sensitivity at photon energies of 
$\sim 0.25-12.4$~eV~\cite{Mazin:12} ($\sim 0.04-3.1$~eV~\cite{TES3}), with the potential to be sensitive to meV phonon energies~\cite{TES4,Hochberg:2015fth}. 
Currently the most sensitive single-photon devices~\cite{TES1,Goldie,Miller,Hochberg:2015fth} 
are small in size, $\sim\!((5-125)~\mu\rm{m})^3$, but efforts exist to enlarge them~\cite{Pyle}. 
CRESST-II currently has the best detector of few-cm size, sensitive to $\mathcal{O}(10)$ photons, 
which uses a TES read out by SQUIDs~\cite{Angloher:2015ewa,Angloher:2016hbv}. 
Silicon photomultipliers (SiPM) are possible photodetectors and operate well at cryogenic temperatures, but the dark-count rate 
may be too large~\cite{Biroth,Achenbach}.  

\vspace{-4mm}
\section{DISTINGUISHING SIGNAL FROM BACKGROUNDS} \vskip -3mm
A few handles exist to distinguish a DM signal from a background. 
First, the signal rate modulates annually and daily due to the motion of the Earth~\cite{Drukier:1986tm}. 
The modulation is larger than for elastic WIMP-nucleus recoils, since the scattering is inelastic~\cite{TuckerSmith:2001hy}, 
and increases with threshold.  
Backgrounds are not expected to have the same phase, amplitude, period, and energy dependence.  
Second, the DM-induced electron-recoil spectrum is distinctive and unlikely to be mimicked by a background.  
Third, the DM signal scales with the target volume, in contrast to many potential backgrounds arising
from the surrounding detector package.  
This can be confirmed by using the same detector but with 
a hollow crystal~\footnote{We acknowledge Matthew Pyle for insightful discussions.}.  
Backgrounds that scale with the target volume, such as external gammas and phosphorescence, 
can be determined by measuring the change in signal when a gamma ray source is placed outside the detector. 

\vspace{-4mm}
\section{DARK MATTER-ELECTRON SCATTERING} \vskip -3mm

To inform our choice of scintillating materials, we review here the scattering of sub-GeV DM off a bound electron in a crystal.  The salient features emphasized below also apply to atoms. See~\cite{Essig:2015cda} for details. 

The rate for DM-electron scattering to excite an electron from level $i$ to $f$ is 
\begin{align}
&\frac{d R_{\rm crystal}}{d \ln E_e} =
\frac{\rho_\chi}{m_\chi}\ N_{\rm cell} \ \overline{\sigma}_e \ \alpha\
  \frac{m_e^2}{\mu_{\chi e}^2}\ \times \\
&\int\! d \ln q \, \bigg(\frac{E_e}{q} \eta \big( v_{\rm min}(q, E_e) \big)\bigg)
F_{\rm DM}(q)^2 \big| f_{\rm crystal}(q, E_e) \big|^2 \, ,\nonumber
\end{align}
where $\alpha\simeq 1/137$ is the fine-structure constant, $m_\chi$ ($m_e$) denotes the DM (electron) mass, $\rho_\chi\simeq 0.4$~GeV/cm$^3$ is the local DM density, $E_e$ is the total energy deposited, $q$ is the DM-to-electron momentum transfer,
$N_{\rm cell} = M_{\rm target}/M_{\rm cell}$ is the number of unit cells in the target crystal of total (cell) mass $M_{\rm target}$ 
($M_{\rm cell}$), and $\mu_{\chi e}$ is the DM-electron reduced mass. 
The \emph{crystal form-factor} is 
\begin{align}
\big| f_{\rm crystal}(q, E_e) \big|^2 &=
\frac{2\pi^2 V_{\rm cell}}{\alpha m_e^2}  \sum_{i\,  f} \! \int_{\rm BZ} \frac{d^3 k\  d^3 k'}{(2 \pi)^6} \delta(E_e - \Delta E) \\
&  \times  \sum_{\vec G'} q \delta(q - |\vec k' - \vec k + \vec G'|)
\big| f_{[i \vec k , f \vec k', \vec G']} \big|^2 \, ,\nonumber
\end{align}
where $\Delta E=E_{f \vec k'} - E_{i \vec k}$, $V_{\rm cell}$ is the volume of the unit cell, $\vec k, \vec k'$ are wavevectors in the first Brillouin Zone (BZ), and $\vec G'$ is the reciprocal lattice vector.
The reference cross-section $\overline\sigma_e$ and DM form factor $ |F_{\rm{DM}}(q)|^2$ are parameterizations of the DM-electron interaction defined as 
\bea
\overline{|{\cal{M}}_{\rm{free}}(\vec q)|^2}&\equiv&\overline{|{\cal{M}}_{\rm{free}}(\alpha m_e)|^2}\times  |F_{\rm{DM}}(q)|^2 
\label{eq:sigmaebar} \\
\overline\sigma_e&\equiv&\frac{\mu_{\chi e}^2\overline{|{\cal{M}}_{\rm{free}}(\alpha m_e)|^2}}{16\pi m_\chi^2 m_e^2}, \label{eq:FDM} 
\eea
where $\overline{|{\cal{M}}_{\rm{free}}|^2}$ is the absolute value squared of the elastic DM-free-electron scattering matrix element, averaged over initial-, and summed over final-state particle spins. The DM-halo profile is 
\begin{eqnarray}
\eta(v_{min}) & =&\int d^3 v_\chi \,g_\chi(\vec v_\chi) \frac{1}{v_\chi} \Theta(v_\chi - v_{\rm min})
\label{eq:etavMin1}\\
&=&\frac{1}{K}\int d\Omega \, dv_\chi \, v_\chi ~e^{-|\vec v_\chi-\vec v_{\rm E}|^2/v_0^2}\\
& & \times\ \Theta(v_\chi - v_{\rm min})  \Theta(v_{\rm esc} - v_\chi)\,,\nonumber
\label{eq:etavmin}
\end{eqnarray}
where in Eq.~(\ref{eq:etavMin1}) we chose for $g_\chi(\vec v_\chi)$ the standard Maxwell-Boltzmann distribution with a sharp cutoff. 
We take $v_0=230$~km/s, the Earth velocity about the galactic center $\vec v_{\rm E} = 240$~km/s, and the DM escape velocity from the galaxy as $v_{\rm esc} = 600$~km/s. $K=6.75\times10^{22}(\rm{cm/s})^3$ is the normalization factor. 
The minimum velocity required for scattering is
\begin{equation}
v_{\rm min}(q, E_e) = \frac{E_e}{q} + \frac{q}{2 m_\chi} \, .
\label{eq:vmin}
\end{equation}

There are four salient features worth emphasizing for sub-GeV DM scattering off electrons: 
\begin{itemize}
\item First, since the bound electron's momentum can be arbitrarily high (albeit with suppressed probability), 
$q$ can be arbitrarily high, so that in principle all of the DM's kinetic energy can be transferred to the electron 
(in sub-GeV DM scattering off nuclei only a fraction is transferred to a much heavier nucleus).  
Thus, $E_\chi = \frac{1}{2} m_\chi v_\chi^2 \ge E_e$ implies $m_\chi \gtrsim 250~{\rm keV} \times (E_e/1~{\rm eV})$ for $v_\chi\lesssim v_{\rm esc}+v_{\rm E}$.  
Therefore, smaller ionization energies or band gaps can probe lower DM masses, with crystal targets being 
sensitive down to a few hundred keV.  
\item Second, since the electron moves at a speed of $\sim \alpha$, much faster than the DM ($\sim 10^{-3}$), the electron  
determines the typical $q$, $q_{\rm typ}$.  
A rough estimate for $q_{\rm typ}$ is the crystal momentum, $2\pi/a\sim 2.3$~keV, where $a\sim 10 \alpha m_e$ is 
the lattice constant for our target choices (see below). 
Since $E_e \sim \vec{q} \cdot \vec{v}_\chi$, the minimum $q$ to obtain a particular $E_e$ 
is given by $q \gtrsim q_{\rm typ} \times E_e/(2.3~{\rm eV})$.  
A similar estimate holds for atoms~\cite{Essig:2015cda}.  
The signal rate is thus larger in semiconductors with low band gaps ($\Delta E \sim 1-2$~eV) 
than insulators ($\Delta E \gtrsim 5$~eV) or noble liquids ($\Delta E \sim 12$, 16, 25~eV for xenon, argon, helium, respectively).  
\item Third, while the value of $q$ is naturally $q_{\rm typ}$, $q$ can in fact be much larger as mentioned above. 
This allows for much larger momentum transfers and recoil energies, although these 
are strongly suppressed.  
\item Fourth, since the scattering is inelastic, the annual modulation of the signal rate is larger than for 
typical WIMP elastic scattering~\cite{TuckerSmith:2001hy}.  
\end{itemize}

\vspace{-4mm}
\section{SCINTILLATING TARGETS} \vskip -3mm

The previous discussion suggests using scintillating crystals with low band gaps.  
However, the crystals must also have high purity, high radiative efficiency (i.e.~little non-radiative recombination of excited electron-hole pairs), 
and few native defects, all while being grown to large sizes ($\gtrsim 1$~kg).  
We thus focus on NaI and CsI, but include GaAs, which may also satisfy these criteria. 
Table~\ref{tab:materials}  ({\it top}) lists salient features.  

NaI and CsI are insulators that scintillate efficiently through the decay of 
self-trapped excitons.
They are used extensively due to their high light output and ease of production~\cite{Moszynski2005357,tagkey1970iv,PhysRev.75.796,PhysRev.75.1611.2,SaintGobain,4545171,2012EPJC72.2061S,PSSB:PSSB201451464}.  
Pure CsI is being considered for a DM-nucleus-recoil search~\cite{Angloher:2016hbv}.
Early measurements of GaAs, a direct-gap semiconductor, 
showed a radiative efficiency (internal) of $\sim 0.6$ at 77~K when doped with donors or 
acceptors~\cite{Cusano}. 
Conventional coupling to photodetectors is inefficient due to the high refractive index ($\sim 3.8$) 
but one could apply photonic coatings or deposit the photodetectors directly onto the surfaces of the 
GaAs crystal to reduce internal reflection~\cite{Knapitsch}. 

\begin{table}[t]
\begin{center}
\begin{tabular}{|c||c|c|c|c|c|}
\hline
Material & $E_g$ [eV] & Rad.~Eff. & $E_{\rm em}^{\rm max}$  [eV] & $\tau$ [ns] & Mechanism\\
\hline \hline
NaI~\cite{Moszynski2003}& 5.9 & 0.95 & 4.1 & 300 & SX\\ \hline 
CsI~\cite{Moszynski2005357,Amsler2002494}& 6.4 & $\sim 1$ & 4.0 & $10^3$ & SX \\ \hline
GaAs~\cite{Cusano} & 1.52 & $\sim 0.6$ & $\sim 1.5$ & $10^3$$^{(a)}$ & BE  \\
\hline
\end{tabular}
\vskip 4mm
\begin{tabular}{|c||c|c|c|c|c|}
\hline
Material & $E_g$ [eV] & Rad.~Eff. & $E_{\rm em}^{\rm max}$  [eV] & $\tau$ [ns] & Mechanism\\
\hline \hline
PVT~\cite{Trilling1970}& 4.8 & 0.10 & 3.0 & 2 & organic \\ \hline 
CaWO$_4$~\cite{Zdesenko2005657}& 4.2 & 0.21 & 2.9 & 8000 & CX \\ \hline 
Xe~\cite{Aprile:2008rc,Chepel:2012sj} & $12.1^{(b)}$ & 0.30 & 7.1  & 30$^{(c)}$ & excimers \\ \hline
Ar~\cite{Ar,Chepel:2012sj} & $15.8^{(b)}$ & 0.40 & 9.9 &  $10^3$$^{(c)}$  & excimers  \\ \hline
He~\cite{Guo:2013dt} & $24.6^{(b)}$ & 0.29 & 15.5  & 10$^{(d)}$  & excimers  \\ \hline
NaI:Tl~\cite{deHaas2008}$^{(e)}$& 5.9 & 0.50 & 3.0 & 115 & Tl$^+$\\ \hline
CsI:Tl~\cite{deHaas2008}$^{(e)}$ & 6.4 & $\sim1$ & 2.2 & 980 & Tl$^+$ \\ 
\hline 
\end{tabular}
\vskip 4mm
\begin{tabular}{|c||c|}
\hline
Material & $E_g$ [eV] \\
\hline \hline
Si& 0.67 \\ \hline 
Ge & 1.1 \\ 
\hline 
\end{tabular}
\end{center}
\caption{\label{tab:materials} 
Band gap ($E_g$), radiative efficiency, 
photon emission energy peak ($E_{\rm em}^{\rm max}$), radiative recombination time ($\tau$), 
and scintillation mechanism (SX = self-trapped excitons, Tl$^+$ = thallium ion luminescent center, 
CX = charge-transfer emissions, BE = bound excitons, excimers = short-lived, excited dimeric molecule) 
for candidate scintillators.  
We focus on ({\it top table}): pure NaI, pure CsI, 
and GaAs (doped with acceptors or donors). 
Other scintillators may also be suitable targets ({\it middle table}): 
polyvinyltoluene (PVT, i.e.~C$_{27}$H$_{30}$), calcium tungstate (CaWO$_4$), xenon (Xe), argon (Ar), and 
helium (He).  NaI and CsI, doped with thallium (NaI:Tl, CsI:Tl), scintillate at room temperature. 
Si and Ge {(\it bottom table}) are used for comparison, and 
suitable dopants could allow them to scintillate. 
$^{(a)}$Expected (no measurement).
$^{(b)}$Ionization energy of outer-shell electron~\cite{NIST}. 
$^{(c)}$Triplet lifetime. 
$^{(d)}$Singlet lifetime. 
$^{(e)}$Room temperature values. 
\vspace{-5mm}
}
\end{table}

Other scintillator targets are possible, but not considered further (Table~\ref{tab:materials}, {\it middle}). 
Plastic scintillators, e.g.~PVT, have a low radiative efficiency, but this may be 
offset by their low production cost. 
CaWO$_4$ also has a low radiative efficiency~\cite{Moszynski}. 
Noble liquids can be scaled up relatively easily to large masses. 
At room temperature, phonons reduce radiative relaxation (i.e. quenching) in NaI and CsI, and Tl$^+$ 
doping is commonly used to provide efficient radiative centers.
We include them to compare with the undoped cases.  
All listed materials (except PVT) are used for DM-nuclear recoil searches~\cite{Angloher:2008jj,Angloher:2015ewa,Aprile:2016wwo,Akerib:2016lao,Abe:2013tc,Aprile:2015uzo,Akerib:2015cja,Aalseth:2015mba,Guo:2013dt,Bernabei:2010mq,Froborg:2016ova,deSouza:2016fxg,Lee.:2007qn}, but the photodetectors are not sensitive to single 
photons~\footnote{DM-electron scattering in e.g.~xenon TPCs could produce two photons in a multi-step de-excitation process.  
However the efficiency to detect a photon is low (e.g. $\sim 10\%$ in LUX). 
Moreover, the PMTs are not sensitive to the second photon, which is in the infrared.}.  

Other suitable low band gap materials may exist.  
Crystals with band gaps $\lesssim {\rm few~eV}$ are likely semiconductors~\cite{Derenzo}. 
Among these, direct-gap semiconductors have a high radiative efficiency, but no obvious  
candidates exist besides GaAs. 
Indirect-gap semiconductors are more common, but their scintillation is slow and inefficient without doping. 
However, luminescence has been reported from Si~\cite{Steger,Davies1994} and Ge~\cite{Davies1992} at 
cryogenic temperatures (Table~\ref{tab:materials}, {\it bottom}). 
More research could reveal suitable dopants to achieve high radiative efficiency. 
We show results for Ge and Si below since they are potential scintillators and are also used in 
current experiments sensitive to an ionization signal, like SuperCDMS and DAMIC. 

The supplementary materials will review the scintillation mechanisms of the substances in Table~\ref{tab:materials}. 

\begin{table}[t]
\begin{center}
\begin{tabular}{|l|c|c|c|c|c|}
\hline
     & $a$ (bohr) & $V_{\rm cell}$ (bohr$^3$) & $N_{\rm bands}$ & $N_e$ & $N_k$  \\ \hline
CsI  & 8.6868     & 655.51                & 80          & $8_v+8_{c, \rm Cs}$   & $30\times 125$ \\ \hline
NaI  & 12.927     & 464.88                & 50          & $8_v$     & $30\times 216$ \\ \hline
GaAs & 10.8690    & 321.00                & 60          & $8_v+10_{c, {\rm Ga}}$ 
											    & $30\times 216$ \\ \hline 
Ge &  10.8171 &  316.4269 & 66 & $8_v+20_c$  &  1$\times$ 243 \\ \hline
Si &10.3305 & 275.6154 & 56 & $8_v$ & 1$\times$ 243 \\ \hline
\end{tabular}
\caption{\label{tab:qedark-parameters}
Computational parameters for various materials. 
Lattice constant ($a$), cell volume ($V_{\rm cell}$), number of valence+conduction bands ($N_{\rm bands}$), 
number of valence $v$ and core $c$ electrons ($N_e$), and number of runs with independent random $k$-point meshes 
times number of $k$-points in each mesh ($N_k$). Note that there are two atoms per unit cell. 
\vspace{-5mm}
}
\end{center}
\end{table}

\begin{figure*}[t!]
\includegraphics[width=0.48\textwidth]{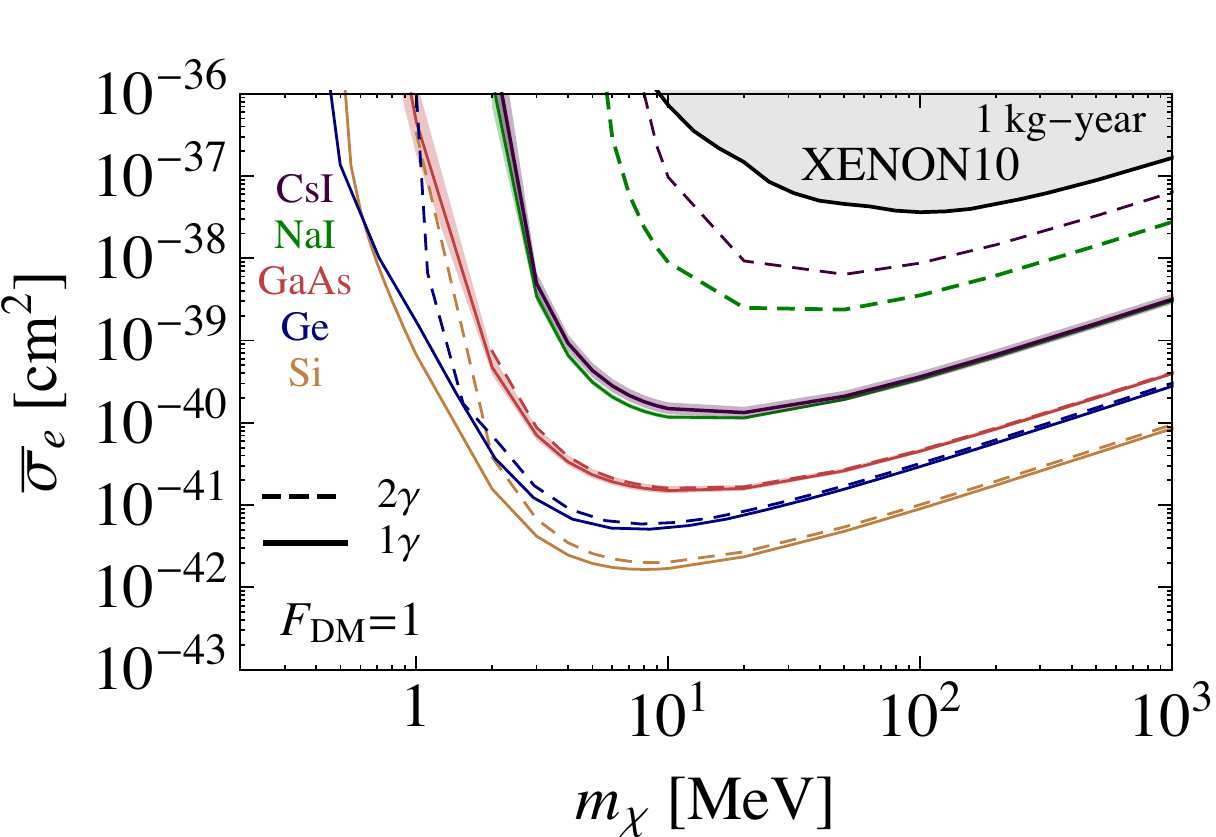} ~~
\includegraphics[width=0.48\textwidth]{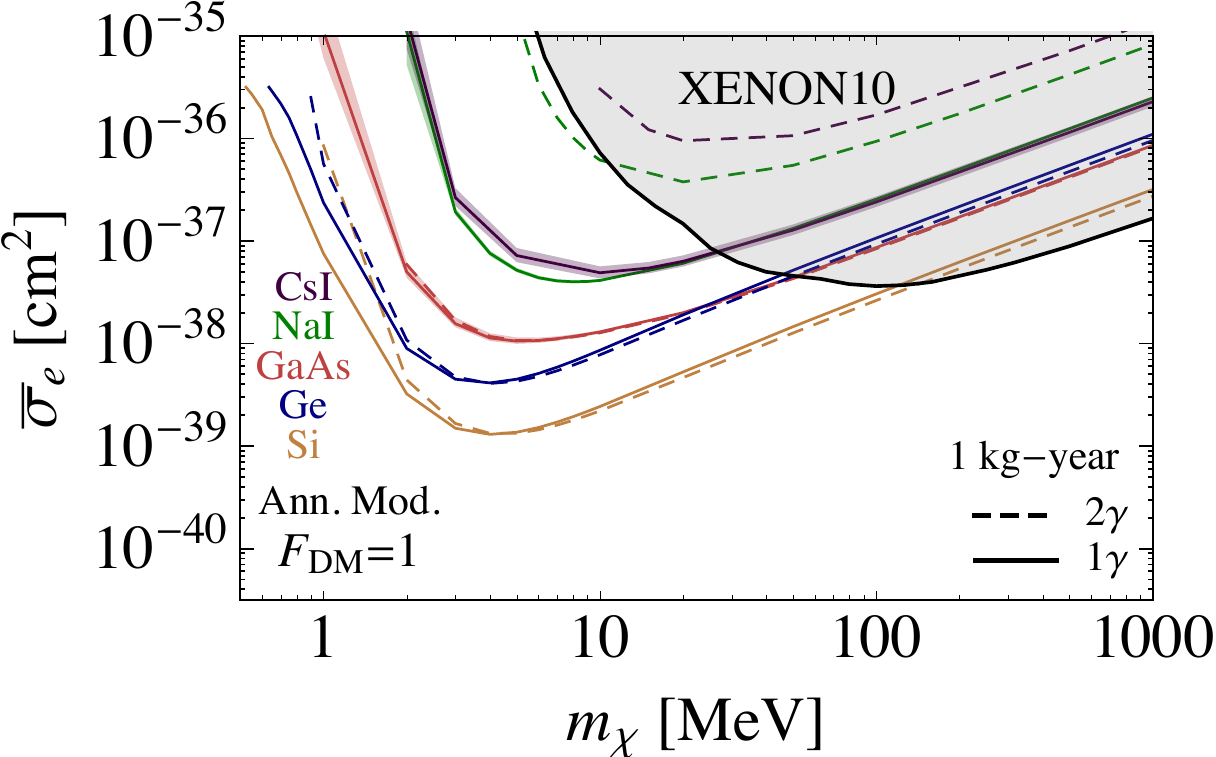}\\
\vskip -4mm
\includegraphics[width=0.48\textwidth]{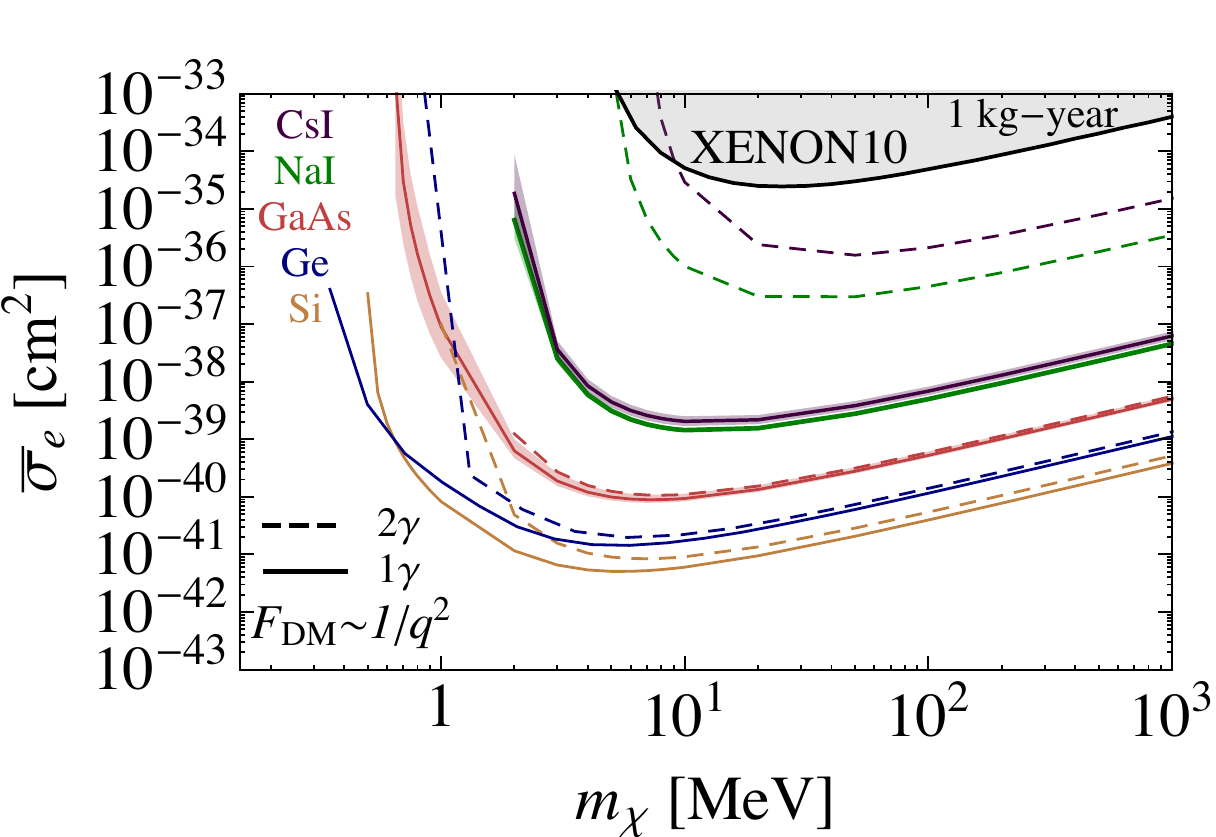} ~~
\includegraphics[width=0.48\textwidth]{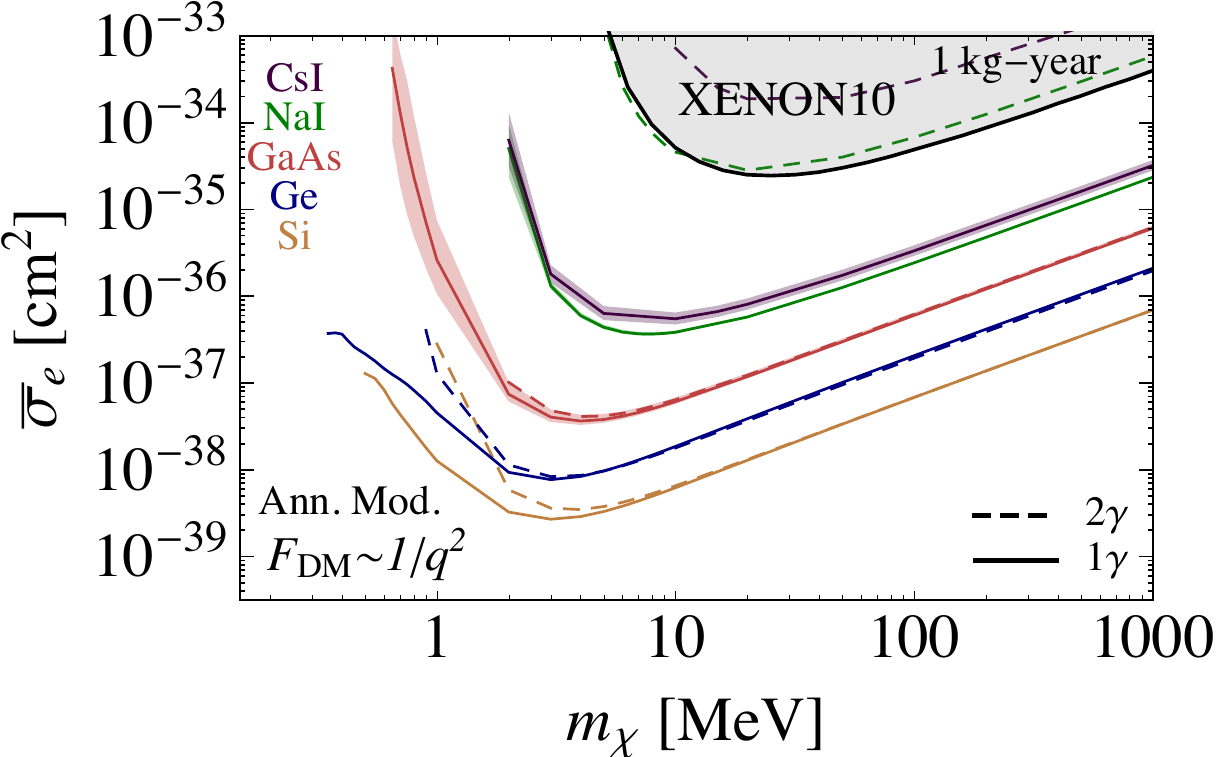}
\caption{DM-electron-scattering-cross-section ($\overline\sigma_e$) reach vs.~DM mass ($m_\chi$) for 
$F_{DM}(q)=1$ ({\it top}) and $F_{DM}(q)=1/q^2$ ({\it bottom}), assuming an exposure of 1~kg for 1~year and a radiative efficiency of 1.  
{\it Left:} Solid (dashed) lines show 3.6 events for a threshold of one (two) photons, corresponding to the 95\% c.l.~reach 
for zero background events in CsI (purple), NaI (green), and GaAs (red). 
Bands around solid lines show the numerical uncertainty.  
Solid (dashed) lines for Ge (blue) and Si (gold) are the one(two)-electron threshold lines 
from~\cite{Essig:2015cda}.  
{\it Right:} Solid (dashed) lines show $5\sigma$-discovery reach using annual modulation for a threshold of one (two) photons, assuming zero backgrounds.  
The gray region is excluded by XENON10~\cite{Essig:2012yx}. 
}
\label{fig:reach}
\end{figure*}

\vspace{-4mm}
\section{CALCULATIONS} \vskip -3mm

We calculate the DM-electron scattering rates in NaI, CsI, and GaAs using the {\tt{QEdark}} module
developed in~\cite{Essig:2015cda}. We use PBE functionals ~\cite{PhysRevLett.77.3865}, 
norm-conserving pseudopotentials~\cite{PhysRevLett.43.1494}, and adjust the band gaps to the values in Table~\ref{tab:materials} using a scissor correction ~\cite{PhysRevLett.63.1719,PhysRevB.43.4187}. Table~\ref{tab:qedark-parameters} lists the required calculation parameters.  
We include in the density functional theory (DFT) calculation all electrons with binding energies $E_B$ as low as the $3d$-shell of Ga (binding energy $E_B \sim 32$~eV), the $5p$- and $5s$-shell electrons ($E_B \sim 13$~eV and $\sim 23$~eV, respectively) of Cs, 
and the $3d$-shell electrons of Ge
as in~\cite{Essig:2015cda,Lee:2015qva} 
(deeper electrons are irrelevant).
The numerical uncertainty is estimated by choosing 30 random $k$-point meshes. 
The sensitivity lines for Ge and Si are from~\cite{Essig:2015cda} 
(only one mesh is shown, but the uncertainty is small~\cite{Essig:2015cda}).  

Our calculations do not include exciton effects. In the supplementary materials, we argue that exciton effects are negligible for the low-gap 
materials GaAs, Ge, or Si, and may have an $\mathcal{O}(1)$ effect for NaI and CsI. 

\vspace{-4mm}
\section{RESULTS} \vskip -3mm

Fig.~\ref{fig:reach} ({\it left}) shows the potential sensitivity to $\overline\sigma_e$ (Eq.~(\ref{eq:sigmaebar})) for 
two different $F_{\rm DM}$ (Eq.~(\ref{eq:FDM})), various materials, two thresholds, 
and data taken over one year with 1~kg of material.  We assume a radiative efficiency of 1. 
The low-gap materials GaAs, Si, and Ge can reach potentially DM masses as low as a few hundred keV, 
whereas the reach of NaI and CsI is 1--2 MeV.  
This could probe lower masses than XENON10~\cite{Essig:2012yx}, 
and extend the high-mass reach by one to several orders of magnitude. 

The signal in GaAs, NaI, and CsI consists of one or more photons, while in Ge and Si it consists of either one or more electrons, or (if suitable dopants can provide a high radiative efficiency) one or more photons. 
We show two thresholds: ``$1\gamma$'' requires $E_e \ge E_g$, while ``$2\gamma$'' 
requires $E_e \ge E_g+\langle E \rangle$, where $\langle E\rangle$ is the mean energy needed for the recoiling electron to form another electron-hole pair.  
A phenomenological approach gives $\langle E\rangle \sim 2.9$~eV (3.6~eV, 4.2~eV) for 
Ge (Si, GaAs)~\cite{Klein:1968,Knoll:1300754,Essig:2015cda}.  
Precise values for CsI and NaI are unavailable, so we show $\langle E\rangle = 3 E_g$ ~\cite{Knoll:1300754}. 
More theoretical work and an experimental calibration can better quantify the number of photons produced by low-energy 
electron recoils. 
The mass threshold is different for the $1\gamma$ and $2\gamma$ lines.  However, the low-gap materials have a 
similar high-mass reach for either threshold, since $E_e$ is typically several eV and more likely to produce two rather than one photon. 
Resolving two photons in coincidence can help reduce backgrounds. 

The annual modulation of the signal rate can be used as a discriminant from background~\cite{Drukier:1986tm}. 
Fig.~\ref{fig:reach} ({\it right}) shows $5\sigma$ discovery lines for which $\Delta S/\sqrt{S_{tot}+B}=5$ with $B=0$.  
Here $\Delta S$ is the modulation amplitude and $S_{tot}$ ($B$) is the total number of signal (background) events.  
The sensitivity weakens $\propto\sqrt{B}$, assuming $B$ is constant in time.  

\vskip 3mm
To summarize, we described a novel search for sub-GeV DM, using scintillators.  
Scintillators provide a complementary path with potential advantages over other approaches
searching for a low ionization signal: the detection of photons may be technologically easier with fewer dark counts. 

\vspace{-4mm}
\section{Acknowledgments} \vskip -3mm
We are grateful to Matthew Pyle for numerous insightful discussions, including 
discussing potential backgrounds (like afterglow), and for suggesting how these could be characterized. 
We are also grateful to Jeremy Mardon and Tomer Volansky for numerous stimulating conversations.  
We thank Philip Allen, Jeremy Mardon, and Matthew Pyle for comments on a draft of this manuscript. 
We have benefited from many useful conversations with 
Thomas Allison, Mariv\'i Fern\'andez-Serra,  Eden Figueroa, Enectali Figueroa-Feliciano, Lauren Hsu, 
Mark Hybertsen, Serge Luryi, Aaron Manalaysay, Ben Mazin, Daniel McKinsey, Laszlo Mihaly, Florian Reindl, 
Karoline Sch\"affner, and Craig Woody.  
R.E.~is supported by the DoE Early Career research program DESC0008061 and through a Sloan Foundation Research Fellowship. T.-T.Y.~is also supported by grant DESC0008061.  A.M.~acknowledges support from NSF grant PHY1316617.  
A.S.~acknowledges support from DoE grant DE-FG02-09ER16052. S.D.~acknowledges support from the U.S.~Department of Homeland Security, Domestic Nuclear Detection Office. This research used resources of the National Energy Research Scientific Computing Center, a DoE Office of Science User Facility supported by the Office of Science of the U.S.~Department of Energy under Contract No.~DE-AC02-05CH11231 and the Handy and LIred computer clusters at the Stony Brook University Institute for Advanced Computational Science.

\bibliography{SinglePhoton}

\begin{thebibliography}{125}%
\makeatletter
\providecommand \@ifxundefined [1]{%
 \@ifx{#1\undefined}
}%
\providecommand \@ifnum [1]{%
 \ifnum #1\expandafter \@firstoftwo
 \else \expandafter \@secondoftwo
 \fi
}%
\providecommand \@ifx [1]{%
 \ifx #1\expandafter \@firstoftwo
 \else \expandafter \@secondoftwo
 \fi
}%
\providecommand \natexlab [1]{#1}%
\providecommand \enquote  [1]{``#1''}%
\providecommand \bibnamefont  [1]{#1}%
\providecommand \bibfnamefont [1]{#1}%
\providecommand \citenamefont [1]{#1}%
\providecommand \href@noop [0]{\@secondoftwo}%
\providecommand \href [0]{\begingroup \@sanitize@url \@href}%
\providecommand \@href[1]{\@@startlink{#1}\@@href}%
\providecommand \@@href[1]{\endgroup#1\@@endlink}%
\providecommand \@sanitize@url [0]{\catcode `\\12\catcode `\$12\catcode
  `\&12\catcode `\#12\catcode `\^12\catcode `\_12\catcode `\%12\relax}%
\providecommand \@@startlink[1]{}%
\providecommand \@@endlink[0]{}%
\providecommand \url  [0]{\begingroup\@sanitize@url \@url }%
\providecommand \@url [1]{\endgroup\@href {#1}{\urlprefix }}%
\providecommand \urlprefix  [0]{URL }%
\providecommand \Eprint [0]{\href }%
\providecommand \doibase [0]{http://dx.doi.org/}%
\providecommand \selectlanguage [0]{\@gobble}%
\providecommand \bibinfo  [0]{\@secondoftwo}%
\providecommand \bibfield  [0]{\@secondoftwo}%
\providecommand \translation [1]{[#1]}%
\providecommand \BibitemOpen [0]{}%
\providecommand \bibitemStop [0]{}%
\providecommand \bibitemNoStop [0]{.\EOS\space}%
\providecommand \EOS [0]{\spacefactor3000\relax}%
\providecommand \BibitemShut  [1]{\csname bibitem#1\endcsname}%
\let\auto@bib@innerbib\@empty
\bibitem [{\citenamefont {Essig}\ \emph
  {et~al.}(2012{\natexlab{a}})\citenamefont {Essig}, \citenamefont {Mardon},\
  and\ \citenamefont {Volansky}}]{Essig:2011nj}%
  \BibitemOpen
  \bibfield  {author} {\bibinfo {author} {\bibfnamefont {R.}~\bibnamefont
  {Essig}}, \bibinfo {author} {\bibfnamefont {J.}~\bibnamefont {Mardon}}, \
  and\ \bibinfo {author} {\bibfnamefont {T.}~\bibnamefont {Volansky}},\ }\href
  {\doibase 10.1103/PhysRevD.85.076007} {\bibfield  {journal} {\bibinfo
  {journal} {Phys. Rev.}\ }\textbf {\bibinfo {volume} {D85}},\ \bibinfo {pages}
  {076007} (\bibinfo {year} {2012}{\natexlab{a}})},\ \Eprint
  {http://arxiv.org/abs/1108.5383} {arXiv:1108.5383 [hep-ph]} \BibitemShut
  {NoStop}%
\bibitem [{\citenamefont {Essig}\ \emph {et~al.}(2016)\citenamefont {Essig},
  \citenamefont {Fernandez-Serra}, \citenamefont {Mardon}, \citenamefont
  {Soto}, \citenamefont {Volansky},\ and\ \citenamefont {Yu}}]{Essig:2015cda}%
  \BibitemOpen
  \bibfield  {author} {\bibinfo {author} {\bibfnamefont {R.}~\bibnamefont
  {Essig}}, \bibinfo {author} {\bibfnamefont {M.}~\bibnamefont
  {Fernandez-Serra}}, \bibinfo {author} {\bibfnamefont {J.}~\bibnamefont
  {Mardon}}, \bibinfo {author} {\bibfnamefont {A.}~\bibnamefont {Soto}},
  \bibinfo {author} {\bibfnamefont {T.}~\bibnamefont {Volansky}}, \ and\
  \bibinfo {author} {\bibfnamefont {T.-T.}\ \bibnamefont {Yu}},\ }\href
  {\doibase 10.1007/JHEP05(2016)046} {\bibfield  {journal} {\bibinfo  {journal}
  {JHEP}\ }\textbf {\bibinfo {volume} {05}},\ \bibinfo {pages} {046} (\bibinfo
  {year} {2016})},\ \Eprint {http://arxiv.org/abs/1509.01598} {arXiv:1509.01598
  [hep-ph]} \BibitemShut {NoStop}%
\bibitem [{\citenamefont {Graham}\ \emph {et~al.}(2012)\citenamefont {Graham},
  \citenamefont {Kaplan}, \citenamefont {Rajendran},\ and\ \citenamefont
  {Walters}}]{Graham:2012su}%
  \BibitemOpen
  \bibfield  {author} {\bibinfo {author} {\bibfnamefont {P.~W.}\ \bibnamefont
  {Graham}}, \bibinfo {author} {\bibfnamefont {D.~E.}\ \bibnamefont {Kaplan}},
  \bibinfo {author} {\bibfnamefont {S.}~\bibnamefont {Rajendran}}, \ and\
  \bibinfo {author} {\bibfnamefont {M.~T.}\ \bibnamefont {Walters}},\ }\href
  {\doibase 10.1016/j.dark.2012.09.001} {\bibfield  {journal} {\bibinfo
  {journal} {Phys. Dark Univ.}\ }\textbf {\bibinfo {volume} {1}},\ \bibinfo
  {pages} {32} (\bibinfo {year} {2012})},\ \Eprint
  {http://arxiv.org/abs/1203.2531} {arXiv:1203.2531 [hep-ph]} \BibitemShut
  {NoStop}%
\bibitem [{\citenamefont {Lee}\ \emph {et~al.}(2015)\citenamefont {Lee},
  \citenamefont {Lisanti}, \citenamefont {Mishra-Sharma},\ and\ \citenamefont
  {Safdi}}]{Lee:2015qva}%
  \BibitemOpen
  \bibfield  {author} {\bibinfo {author} {\bibfnamefont {S.~K.}\ \bibnamefont
  {Lee}}, \bibinfo {author} {\bibfnamefont {M.}~\bibnamefont {Lisanti}},
  \bibinfo {author} {\bibfnamefont {S.}~\bibnamefont {Mishra-Sharma}}, \ and\
  \bibinfo {author} {\bibfnamefont {B.~R.}\ \bibnamefont {Safdi}},\ }\href
  {\doibase 10.1103/PhysRevD.92.083517} {\bibfield  {journal} {\bibinfo
  {journal} {Phys. Rev.}\ }\textbf {\bibinfo {volume} {D92}},\ \bibinfo {pages}
  {083517} (\bibinfo {year} {2015})},\ \Eprint
  {http://arxiv.org/abs/1508.07361} {arXiv:1508.07361 [hep-ph]} \BibitemShut
  {NoStop}%
\bibitem [{\citenamefont {Essig}\ \emph
  {et~al.}(2012{\natexlab{b}})\citenamefont {Essig}, \citenamefont
  {Manalaysay}, \citenamefont {Mardon}, \citenamefont {Sorensen},\ and\
  \citenamefont {Volansky}}]{Essig:2012yx}%
  \BibitemOpen
  \bibfield  {author} {\bibinfo {author} {\bibfnamefont {R.}~\bibnamefont
  {Essig}}, \bibinfo {author} {\bibfnamefont {A.}~\bibnamefont {Manalaysay}},
  \bibinfo {author} {\bibfnamefont {J.}~\bibnamefont {Mardon}}, \bibinfo
  {author} {\bibfnamefont {P.}~\bibnamefont {Sorensen}}, \ and\ \bibinfo
  {author} {\bibfnamefont {T.}~\bibnamefont {Volansky}},\ }\href {\doibase
  10.1103/PhysRevLett.109.021301} {\bibfield  {journal} {\bibinfo  {journal}
  {Phys. Rev. Lett.}\ }\textbf {\bibinfo {volume} {109}},\ \bibinfo {pages}
  {021301} (\bibinfo {year} {2012}{\natexlab{b}})},\ \Eprint
  {http://arxiv.org/abs/1206.2644} {arXiv:1206.2644 [astro-ph.CO]} \BibitemShut
  {NoStop}%
\bibitem [{\citenamefont {Hochberg}\ \emph
  {et~al.}(2016{\natexlab{a}})\citenamefont {Hochberg}, \citenamefont {Zhao},\
  and\ \citenamefont {Zurek}}]{Hochberg:2015pha}%
  \BibitemOpen
  \bibfield  {author} {\bibinfo {author} {\bibfnamefont {Y.}~\bibnamefont
  {Hochberg}}, \bibinfo {author} {\bibfnamefont {Y.}~\bibnamefont {Zhao}}, \
  and\ \bibinfo {author} {\bibfnamefont {K.~M.}\ \bibnamefont {Zurek}},\ }\href
  {\doibase 10.1103/PhysRevLett.116.011301} {\bibfield  {journal} {\bibinfo
  {journal} {Phys. Rev. Lett.}\ }\textbf {\bibinfo {volume} {116}},\ \bibinfo
  {pages} {011301} (\bibinfo {year} {2016}{\natexlab{a}})},\ \Eprint
  {http://arxiv.org/abs/1504.07237} {arXiv:1504.07237 [hep-ph]} \BibitemShut
  {NoStop}%
\bibitem [{\citenamefont {Hochberg}\ \emph
  {et~al.}(2015{\natexlab{a}})\citenamefont {Hochberg}, \citenamefont {Pyle},
  \citenamefont {Zhao},\ and\ \citenamefont {Zurek}}]{Hochberg:2015fth}%
  \BibitemOpen
  \bibfield  {author} {\bibinfo {author} {\bibfnamefont {Y.}~\bibnamefont
  {Hochberg}}, \bibinfo {author} {\bibfnamefont {M.}~\bibnamefont {Pyle}},
  \bibinfo {author} {\bibfnamefont {Y.}~\bibnamefont {Zhao}}, \ and\ \bibinfo
  {author} {\bibfnamefont {K.~M.}\ \bibnamefont {Zurek}},\ }\href@noop {} {\
  (\bibinfo {year} {2015}{\natexlab{a}})},\ \Eprint
  {http://arxiv.org/abs/1512.04533} {arXiv:1512.04533 [hep-ph]} \BibitemShut
  {NoStop}%
\bibitem [{\citenamefont {Schutz}\ and\ \citenamefont
  {Zurek}(2016)}]{Schutz:2016tid}%
  \BibitemOpen
  \bibfield  {author} {\bibinfo {author} {\bibfnamefont {K.}~\bibnamefont
  {Schutz}}\ and\ \bibinfo {author} {\bibfnamefont {K.~M.}\ \bibnamefont
  {Zurek}},\ }\href@noop {} {\  (\bibinfo {year} {2016})},\ \Eprint
  {http://arxiv.org/abs/1604.08206} {arXiv:1604.08206 [hep-ph]} \BibitemShut
  {NoStop}%
\bibitem [{\citenamefont {Hochberg}\ \emph
  {et~al.}(2016{\natexlab{b}})\citenamefont {Hochberg}, \citenamefont {Kahn},
  \citenamefont {Lisanti}, \citenamefont {Tully},\ and\ \citenamefont
  {Zurek}}]{Hochberg:2016ntt}%
  \BibitemOpen
  \bibfield  {author} {\bibinfo {author} {\bibfnamefont {Y.}~\bibnamefont
  {Hochberg}}, \bibinfo {author} {\bibfnamefont {Y.}~\bibnamefont {Kahn}},
  \bibinfo {author} {\bibfnamefont {M.}~\bibnamefont {Lisanti}}, \bibinfo
  {author} {\bibfnamefont {C.~G.}\ \bibnamefont {Tully}}, \ and\ \bibinfo
  {author} {\bibfnamefont {K.~M.}\ \bibnamefont {Zurek}},\ }\href@noop {} {\
  (\bibinfo {year} {2016}{\natexlab{b}})},\ \Eprint
  {http://arxiv.org/abs/1606.08849} {arXiv:1606.08849 [hep-ph]} \BibitemShut
  {NoStop}%
\bibitem [{\citenamefont {Essig}\ \emph
  {et~al.}(2013{\natexlab{a}})\citenamefont {Essig}, \citenamefont {Jaros},
  \citenamefont {Wester}, \citenamefont {Adrian}, \citenamefont {Andreas} \emph
  {et~al.}}]{Essig:2013lka}%
  \BibitemOpen
  \bibfield  {author} {\bibinfo {author} {\bibfnamefont {R.}~\bibnamefont
  {Essig}}, \bibinfo {author} {\bibfnamefont {J.~A.}\ \bibnamefont {Jaros}},
  \bibinfo {author} {\bibfnamefont {W.}~\bibnamefont {Wester}}, \bibinfo
  {author} {\bibfnamefont {P.~H.}\ \bibnamefont {Adrian}}, \bibinfo {author}
  {\bibfnamefont {S.}~\bibnamefont {Andreas}},  \emph {et~al.},\ }\href@noop {}
  {\  (\bibinfo {year} {2013}{\natexlab{a}})},\ \Eprint
  {http://arxiv.org/abs/1311.0029} {arXiv:1311.0029 [hep-ph]} \BibitemShut
  {NoStop}%
\bibitem [{\citenamefont {Bird}\ \emph {et~al.}(2004)\citenamefont {Bird},
  \citenamefont {Jackson}, \citenamefont {Kowalewski},\ and\ \citenamefont
  {Pospelov}}]{Bird:2004ts}%
  \BibitemOpen
  \bibfield  {author} {\bibinfo {author} {\bibfnamefont {C.}~\bibnamefont
  {Bird}}, \bibinfo {author} {\bibfnamefont {P.}~\bibnamefont {Jackson}},
  \bibinfo {author} {\bibfnamefont {R.~V.}\ \bibnamefont {Kowalewski}}, \ and\
  \bibinfo {author} {\bibfnamefont {M.}~\bibnamefont {Pospelov}},\ }\href
  {\doibase 10.1103/PhysRevLett.93.201803} {\bibfield  {journal} {\bibinfo
  {journal} {Phys.Rev.Lett.}\ }\textbf {\bibinfo {volume} {93}},\ \bibinfo
  {pages} {201803} (\bibinfo {year} {2004})},\ \Eprint
  {http://arxiv.org/abs/hep-ph/0401195} {arXiv:hep-ph/0401195 [hep-ph]}
  \BibitemShut {NoStop}%
\bibitem [{\citenamefont {Borodatchenkova}\ \emph {et~al.}(2006)\citenamefont
  {Borodatchenkova}, \citenamefont {Choudhury},\ and\ \citenamefont
  {Drees}}]{Borodatchenkova:2005ct}%
  \BibitemOpen
  \bibfield  {author} {\bibinfo {author} {\bibfnamefont {N.}~\bibnamefont
  {Borodatchenkova}}, \bibinfo {author} {\bibfnamefont {D.}~\bibnamefont
  {Choudhury}}, \ and\ \bibinfo {author} {\bibfnamefont {M.}~\bibnamefont
  {Drees}},\ }\href {\doibase 10.1103/PhysRevLett.96.141802} {\bibfield
  {journal} {\bibinfo  {journal} {Phys.Rev.Lett.}\ }\textbf {\bibinfo {volume}
  {96}},\ \bibinfo {pages} {141802} (\bibinfo {year} {2006})},\ \Eprint
  {http://arxiv.org/abs/hep-ph/0510147} {arXiv:hep-ph/0510147 [hep-ph]}
  \BibitemShut {NoStop}%
\bibitem [{\citenamefont {McElrath}(2005)}]{McElrath:2005bp}%
  \BibitemOpen
  \bibfield  {author} {\bibinfo {author} {\bibfnamefont {B.}~\bibnamefont
  {McElrath}},\ }\href {\doibase 10.1103/PhysRevD.72.103508} {\bibfield
  {journal} {\bibinfo  {journal} {Phys.Rev.}\ }\textbf {\bibinfo {volume}
  {D72}},\ \bibinfo {pages} {103508} (\bibinfo {year} {2005})},\ \Eprint
  {http://arxiv.org/abs/hep-ph/0506151} {arXiv:hep-ph/0506151 [hep-ph]}
  \BibitemShut {NoStop}%
\bibitem [{\citenamefont {Fayet}(2006)}]{Fayet:2006sp}%
  \BibitemOpen
  \bibfield  {author} {\bibinfo {author} {\bibfnamefont {P.}~\bibnamefont
  {Fayet}},\ }\href {\doibase 10.1103/PhysRevD.74.054034} {\bibfield  {journal}
  {\bibinfo  {journal} {Phys.Rev.}\ }\textbf {\bibinfo {volume} {D74}},\
  \bibinfo {pages} {054034} (\bibinfo {year} {2006})},\ \Eprint
  {http://arxiv.org/abs/hep-ph/0607318} {arXiv:hep-ph/0607318 [hep-ph]}
  \BibitemShut {NoStop}%
\bibitem [{\citenamefont {Bird}\ \emph {et~al.}(2006)\citenamefont {Bird},
  \citenamefont {Kowalewski},\ and\ \citenamefont {Pospelov}}]{Bird:2006jd}%
  \BibitemOpen
  \bibfield  {author} {\bibinfo {author} {\bibfnamefont {C.}~\bibnamefont
  {Bird}}, \bibinfo {author} {\bibfnamefont {R.~V.}\ \bibnamefont
  {Kowalewski}}, \ and\ \bibinfo {author} {\bibfnamefont {M.}~\bibnamefont
  {Pospelov}},\ }\href {\doibase 10.1142/S0217732306019852} {\bibfield
  {journal} {\bibinfo  {journal} {Mod.Phys.Lett.}\ }\textbf {\bibinfo {volume}
  {A21}},\ \bibinfo {pages} {457} (\bibinfo {year} {2006})},\ \Eprint
  {http://arxiv.org/abs/hep-ph/0601090} {arXiv:hep-ph/0601090 [hep-ph]}
  \BibitemShut {NoStop}%
\bibitem [{\citenamefont {Kahn}\ \emph {et~al.}(2008)\citenamefont {Kahn},
  \citenamefont {Schmitt},\ and\ \citenamefont {Tait}}]{Kahn:2007ru}%
  \BibitemOpen
  \bibfield  {author} {\bibinfo {author} {\bibfnamefont {Y.}~\bibnamefont
  {Kahn}}, \bibinfo {author} {\bibfnamefont {M.}~\bibnamefont {Schmitt}}, \
  and\ \bibinfo {author} {\bibfnamefont {T.~M.}\ \bibnamefont {Tait}},\ }\href
  {\doibase 10.1103/PhysRevD.78.115002} {\bibfield  {journal} {\bibinfo
  {journal} {Phys.Rev.}\ }\textbf {\bibinfo {volume} {D78}},\ \bibinfo {pages}
  {115002} (\bibinfo {year} {2008})},\ \Eprint {http://arxiv.org/abs/0712.0007}
  {arXiv:0712.0007 [hep-ph]} \BibitemShut {NoStop}%
\bibitem [{\citenamefont {Fayet}(2007)}]{Fayet:2007ua}%
  \BibitemOpen
  \bibfield  {author} {\bibinfo {author} {\bibfnamefont {P.}~\bibnamefont
  {Fayet}},\ }\href {\doibase 10.1103/PhysRevD.75.115017} {\bibfield  {journal}
  {\bibinfo  {journal} {Phys.Rev.}\ }\textbf {\bibinfo {volume} {D75}},\
  \bibinfo {pages} {115017} (\bibinfo {year} {2007})},\ \Eprint
  {http://arxiv.org/abs/hep-ph/0702176} {arXiv:hep-ph/0702176 [HEP-PH]}
  \BibitemShut {NoStop}%
\bibitem [{\citenamefont {Essig}\ \emph {et~al.}(2009)\citenamefont {Essig},
  \citenamefont {Schuster},\ and\ \citenamefont {Toro}}]{Essig:2009nc}%
  \BibitemOpen
  \bibfield  {author} {\bibinfo {author} {\bibfnamefont {R.}~\bibnamefont
  {Essig}}, \bibinfo {author} {\bibfnamefont {P.}~\bibnamefont {Schuster}}, \
  and\ \bibinfo {author} {\bibfnamefont {N.}~\bibnamefont {Toro}},\ }\href
  {\doibase 10.1103/PhysRevD.80.015003} {\bibfield  {journal} {\bibinfo
  {journal} {Phys.Rev.}\ }\textbf {\bibinfo {volume} {D80}},\ \bibinfo {pages}
  {015003} (\bibinfo {year} {2009})},\ \Eprint {http://arxiv.org/abs/0903.3941}
  {arXiv:0903.3941 [hep-ph]} \BibitemShut {NoStop}%
\bibitem [{\citenamefont {Bjorken}\ \emph {et~al.}(2009)\citenamefont
  {Bjorken}, \citenamefont {Essig}, \citenamefont {Schuster},\ and\
  \citenamefont {Toro}}]{Bjorken:2009mm}%
  \BibitemOpen
  \bibfield  {author} {\bibinfo {author} {\bibfnamefont {J.~D.}\ \bibnamefont
  {Bjorken}}, \bibinfo {author} {\bibfnamefont {R.}~\bibnamefont {Essig}},
  \bibinfo {author} {\bibfnamefont {P.}~\bibnamefont {Schuster}}, \ and\
  \bibinfo {author} {\bibfnamefont {N.}~\bibnamefont {Toro}},\ }\href {\doibase
  10.1103/PhysRevD.80.075018} {\bibfield  {journal} {\bibinfo  {journal} {Phys.
  Rev.}\ }\textbf {\bibinfo {volume} {D80}},\ \bibinfo {pages} {075018}
  (\bibinfo {year} {2009})}\BibitemShut {NoStop}%
\bibitem [{\citenamefont {Reece}\ and\ \citenamefont
  {Wang}(2009)}]{Reece:2009un}%
  \BibitemOpen
  \bibfield  {author} {\bibinfo {author} {\bibfnamefont {M.}~\bibnamefont
  {Reece}}\ and\ \bibinfo {author} {\bibfnamefont {L.-T.}\ \bibnamefont
  {Wang}},\ }\href {\doibase 10.1088/1126-6708/2009/07/051} {\bibfield
  {journal} {\bibinfo  {journal} {JHEP}\ }\textbf {\bibinfo {volume} {0907}},\
  \bibinfo {pages} {051} (\bibinfo {year} {2009})},\ \Eprint
  {http://arxiv.org/abs/0904.1743} {arXiv:0904.1743 [hep-ph]} \BibitemShut
  {NoStop}%
\bibitem [{\citenamefont {Fayet}(2010)}]{Fayet:2009tv}%
  \BibitemOpen
  \bibfield  {author} {\bibinfo {author} {\bibfnamefont {P.}~\bibnamefont
  {Fayet}},\ }\href {\doibase 10.1103/PhysRevD.81.054025} {\bibfield  {journal}
  {\bibinfo  {journal} {Phys.Rev.}\ }\textbf {\bibinfo {volume} {D81}},\
  \bibinfo {pages} {054025} (\bibinfo {year} {2010})},\ \Eprint
  {http://arxiv.org/abs/0910.2587} {arXiv:0910.2587 [hep-ph]} \BibitemShut
  {NoStop}%
\bibitem [{\citenamefont {Yeghiyan}(2009)}]{Yeghiyan:2009xc}%
  \BibitemOpen
  \bibfield  {author} {\bibinfo {author} {\bibfnamefont {G.~K.}\ \bibnamefont
  {Yeghiyan}},\ }\href {\doibase 10.1103/PhysRevD.80.115019} {\bibfield
  {journal} {\bibinfo  {journal} {Phys.Rev.}\ }\textbf {\bibinfo {volume}
  {D80}},\ \bibinfo {pages} {115019} (\bibinfo {year} {2009})},\ \Eprint
  {http://arxiv.org/abs/0909.4919} {arXiv:0909.4919 [hep-ph]} \BibitemShut
  {NoStop}%
\bibitem [{\citenamefont {Badin}\ and\ \citenamefont
  {Petrov}(2010)}]{Badin:2010uh}%
  \BibitemOpen
  \bibfield  {author} {\bibinfo {author} {\bibfnamefont {A.}~\bibnamefont
  {Badin}}\ and\ \bibinfo {author} {\bibfnamefont {A.~A.}\ \bibnamefont
  {Petrov}},\ }\href {\doibase 10.1103/PhysRevD.82.034005} {\bibfield
  {journal} {\bibinfo  {journal} {Phys.Rev.}\ }\textbf {\bibinfo {volume}
  {D82}},\ \bibinfo {pages} {034005} (\bibinfo {year} {2010})},\ \Eprint
  {http://arxiv.org/abs/1005.1277} {arXiv:1005.1277 [hep-ph]} \BibitemShut
  {NoStop}%
\bibitem [{\citenamefont {Echenard}(2012)}]{Echenard:2012iq}%
  \BibitemOpen
  \bibfield  {author} {\bibinfo {author} {\bibfnamefont {B.}~\bibnamefont
  {Echenard}},\ }\href {\doibase 10.1142/S0217732312300169} {\bibfield
  {journal} {\bibinfo  {journal} {Mod.Phys.Lett.}\ }\textbf {\bibinfo {volume}
  {A27}},\ \bibinfo {pages} {1230016} (\bibinfo {year} {2012})},\ \Eprint
  {http://arxiv.org/abs/1205.3505} {arXiv:1205.3505 [hep-ex]} \BibitemShut
  {NoStop}%
\bibitem [{\citenamefont {March-Russell}\ \emph {et~al.}(2012)\citenamefont
  {March-Russell}, \citenamefont {Unwin},\ and\ \citenamefont
  {West}}]{MarchRussell:2012hi}%
  \BibitemOpen
  \bibfield  {author} {\bibinfo {author} {\bibfnamefont {J.}~\bibnamefont
  {March-Russell}}, \bibinfo {author} {\bibfnamefont {J.}~\bibnamefont
  {Unwin}}, \ and\ \bibinfo {author} {\bibfnamefont {S.~M.}\ \bibnamefont
  {West}},\ }\href {\doibase 10.1007/JHEP08(2012)029} {\bibfield  {journal}
  {\bibinfo  {journal} {JHEP}\ }\textbf {\bibinfo {volume} {1208}},\ \bibinfo
  {pages} {029} (\bibinfo {year} {2012})},\ \Eprint
  {http://arxiv.org/abs/1203.4854} {arXiv:1203.4854 [hep-ph]} \BibitemShut
  {NoStop}%
\bibitem [{\citenamefont {Essig}\ \emph
  {et~al.}(2013{\natexlab{b}})\citenamefont {Essig}, \citenamefont {Mardon},
  \citenamefont {Papucci}, \citenamefont {Volansky},\ and\ \citenamefont
  {Zhong}}]{Essig:2013vha}%
  \BibitemOpen
  \bibfield  {author} {\bibinfo {author} {\bibfnamefont {R.}~\bibnamefont
  {Essig}}, \bibinfo {author} {\bibfnamefont {J.}~\bibnamefont {Mardon}},
  \bibinfo {author} {\bibfnamefont {M.}~\bibnamefont {Papucci}}, \bibinfo
  {author} {\bibfnamefont {T.}~\bibnamefont {Volansky}}, \ and\ \bibinfo
  {author} {\bibfnamefont {Y.-M.}\ \bibnamefont {Zhong}},\ }\href {\doibase
  10.1007/JHEP11(2013)167} {\bibfield  {journal} {\bibinfo  {journal} {JHEP}\
  }\textbf {\bibinfo {volume} {1311}},\ \bibinfo {pages} {167} (\bibinfo {year}
  {2013}{\natexlab{b}})},\ \Eprint {http://arxiv.org/abs/1309.5084}
  {arXiv:1309.5084 [hep-ph]} \BibitemShut {NoStop}%
\bibitem [{\citenamefont {Essig}\ \emph
  {et~al.}(2013{\natexlab{c}})\citenamefont {Essig}, \citenamefont {Kuflik},
  \citenamefont {McDermott}, \citenamefont {Volansky},\ and\ \citenamefont
  {Zurek}}]{Essig:2013goa}%
  \BibitemOpen
  \bibfield  {author} {\bibinfo {author} {\bibfnamefont {R.}~\bibnamefont
  {Essig}}, \bibinfo {author} {\bibfnamefont {E.}~\bibnamefont {Kuflik}},
  \bibinfo {author} {\bibfnamefont {S.~D.}\ \bibnamefont {McDermott}}, \bibinfo
  {author} {\bibfnamefont {T.}~\bibnamefont {Volansky}}, \ and\ \bibinfo
  {author} {\bibfnamefont {K.~M.}\ \bibnamefont {Zurek}},\ }\href {\doibase
  10.1007/JHEP11(2013)193} {\bibfield  {journal} {\bibinfo  {journal} {JHEP}\
  }\textbf {\bibinfo {volume} {1311}},\ \bibinfo {pages} {193} (\bibinfo {year}
  {2013}{\natexlab{c}})},\ \Eprint {http://arxiv.org/abs/1309.4091}
  {arXiv:1309.4091 [hep-ph]} \BibitemShut {NoStop}%
\bibitem [{\citenamefont {Boehm}\ \emph {et~al.}(2013)\citenamefont {Boehm},
  \citenamefont {Dolan},\ and\ \citenamefont {McCabe}}]{Boehm:2013jpa}%
  \BibitemOpen
  \bibfield  {author} {\bibinfo {author} {\bibfnamefont {C.}~\bibnamefont
  {Boehm}}, \bibinfo {author} {\bibfnamefont {M.~J.}\ \bibnamefont {Dolan}}, \
  and\ \bibinfo {author} {\bibfnamefont {C.}~\bibnamefont {McCabe}},\ }\href
  {\doibase 10.1088/1475-7516/2013/08/041} {\bibfield  {journal} {\bibinfo
  {journal} {JCAP}\ }\textbf {\bibinfo {volume} {1308}},\ \bibinfo {pages}
  {041} (\bibinfo {year} {2013})},\ \Eprint {http://arxiv.org/abs/1303.6270}
  {arXiv:1303.6270 [hep-ph]} \BibitemShut {NoStop}%
\bibitem [{\citenamefont {Nollett}\ and\ \citenamefont
  {Steigman}(2014)}]{Nollett:2013pwa}%
  \BibitemOpen
  \bibfield  {author} {\bibinfo {author} {\bibfnamefont {K.~M.}\ \bibnamefont
  {Nollett}}\ and\ \bibinfo {author} {\bibfnamefont {G.}~\bibnamefont
  {Steigman}},\ }\href {\doibase 10.1103/PhysRevD.89.083508} {\bibfield
  {journal} {\bibinfo  {journal} {Phys. Rev.}\ }\textbf {\bibinfo {volume}
  {D89}},\ \bibinfo {pages} {083508} (\bibinfo {year} {2014})},\ \Eprint
  {http://arxiv.org/abs/1312.5725} {arXiv:1312.5725 [astro-ph.CO]} \BibitemShut
  {NoStop}%
\bibitem [{\citenamefont {Andreas}\ \emph {et~al.}(2013)\citenamefont {Andreas}
  \emph {et~al.}}]{Andreas:2013lya}%
  \BibitemOpen
  \bibfield  {author} {\bibinfo {author} {\bibfnamefont {S.}~\bibnamefont
  {Andreas}} \emph {et~al.},\ }\href@noop {} {\  (\bibinfo {year} {2013})},\
  \Eprint {http://arxiv.org/abs/1312.3309} {arXiv:1312.3309 [hep-ex]}
  \BibitemShut {NoStop}%
\bibitem [{\citenamefont {Izaguirre}\ \emph {et~al.}(2013)\citenamefont
  {Izaguirre}, \citenamefont {Krnjaic}, \citenamefont {Schuster},\ and\
  \citenamefont {Toro}}]{Izaguirre:2013uxa}%
  \BibitemOpen
  \bibfield  {author} {\bibinfo {author} {\bibfnamefont {E.}~\bibnamefont
  {Izaguirre}}, \bibinfo {author} {\bibfnamefont {G.}~\bibnamefont {Krnjaic}},
  \bibinfo {author} {\bibfnamefont {P.}~\bibnamefont {Schuster}}, \ and\
  \bibinfo {author} {\bibfnamefont {N.}~\bibnamefont {Toro}},\ }\href {\doibase
  10.1103/PhysRevD.88.114015} {\bibfield  {journal} {\bibinfo  {journal}
  {Phys.Rev.}\ }\textbf {\bibinfo {volume} {D88}},\ \bibinfo {pages} {114015}
  (\bibinfo {year} {2013})},\ \Eprint {http://arxiv.org/abs/1307.6554}
  {arXiv:1307.6554 [hep-ph]} \BibitemShut {NoStop}%
\bibitem [{\citenamefont {Battaglieri}\ \emph {et~al.}(2014)\citenamefont
  {Battaglieri} \emph {et~al.}}]{Battaglieri:2014qoa}%
  \BibitemOpen
  \bibfield  {author} {\bibinfo {author} {\bibfnamefont {M.}~\bibnamefont
  {Battaglieri}} \emph {et~al.} (\bibinfo {collaboration} {BDX
  Collaboration}),\ }\href@noop {} {\  (\bibinfo {year} {2014})},\ \Eprint
  {http://arxiv.org/abs/1406.3028} {arXiv:1406.3028 [physics.ins-det]}
  \BibitemShut {NoStop}%
\bibitem [{\citenamefont {Izaguirre}\ \emph {et~al.}(2014)\citenamefont
  {Izaguirre}, \citenamefont {Krnjaic}, \citenamefont {Schuster},\ and\
  \citenamefont {Toro}}]{Izaguirre:2014bca}%
  \BibitemOpen
  \bibfield  {author} {\bibinfo {author} {\bibfnamefont {E.}~\bibnamefont
  {Izaguirre}}, \bibinfo {author} {\bibfnamefont {G.}~\bibnamefont {Krnjaic}},
  \bibinfo {author} {\bibfnamefont {P.}~\bibnamefont {Schuster}}, \ and\
  \bibinfo {author} {\bibfnamefont {N.}~\bibnamefont {Toro}},\ }\href@noop {}
  {\  (\bibinfo {year} {2014})},\ \Eprint {http://arxiv.org/abs/1411.1404}
  {arXiv:1411.1404 [hep-ph]} \BibitemShut {NoStop}%
\bibitem [{\citenamefont {Batell}\ \emph {et~al.}(2014)\citenamefont {Batell},
  \citenamefont {Essig},\ and\ \citenamefont {Surujon}}]{Batell:2014mga}%
  \BibitemOpen
  \bibfield  {author} {\bibinfo {author} {\bibfnamefont {B.}~\bibnamefont
  {Batell}}, \bibinfo {author} {\bibfnamefont {R.}~\bibnamefont {Essig}}, \
  and\ \bibinfo {author} {\bibfnamefont {Z.}~\bibnamefont {Surujon}},\ }\href
  {\doibase 10.1103/PhysRevLett.113.171802} {\bibfield  {journal} {\bibinfo
  {journal} {Phys.Rev.Lett.}\ }\textbf {\bibinfo {volume} {113}},\ \bibinfo
  {pages} {171802} (\bibinfo {year} {2014})},\ \Eprint
  {http://arxiv.org/abs/1406.2698} {arXiv:1406.2698 [hep-ph]} \BibitemShut
  {NoStop}%
\bibitem [{\citenamefont {Kahn}\ \emph {et~al.}(2015)\citenamefont {Kahn},
  \citenamefont {Krnjaic}, \citenamefont {Thaler},\ and\ \citenamefont
  {Toups}}]{Kahn:2014sra}%
  \BibitemOpen
  \bibfield  {author} {\bibinfo {author} {\bibfnamefont {Y.}~\bibnamefont
  {Kahn}}, \bibinfo {author} {\bibfnamefont {G.}~\bibnamefont {Krnjaic}},
  \bibinfo {author} {\bibfnamefont {J.}~\bibnamefont {Thaler}}, \ and\ \bibinfo
  {author} {\bibfnamefont {M.}~\bibnamefont {Toups}},\ }\href {\doibase
  10.1103/PhysRevD.91.055006} {\bibfield  {journal} {\bibinfo  {journal} {Phys.
  Rev.}\ }\textbf {\bibinfo {volume} {D91}},\ \bibinfo {pages} {055006}
  (\bibinfo {year} {2015})},\ \Eprint {http://arxiv.org/abs/1411.1055}
  {arXiv:1411.1055 [hep-ph]} \BibitemShut {NoStop}%
\bibitem [{\citenamefont {Krnjaic}(2015)}]{Krnjaic:2015mbs}%
  \BibitemOpen
  \bibfield  {author} {\bibinfo {author} {\bibfnamefont {G.}~\bibnamefont
  {Krnjaic}},\ }\href@noop {} {\  (\bibinfo {year} {2015})},\ \Eprint
  {http://arxiv.org/abs/1512.04119} {arXiv:1512.04119 [hep-ph]} \BibitemShut
  {NoStop}%
\bibitem [{\citenamefont {Batell}\ \emph {et~al.}(2009)\citenamefont {Batell},
  \citenamefont {Pospelov},\ and\ \citenamefont {Ritz}}]{Batell:2009di}%
  \BibitemOpen
  \bibfield  {author} {\bibinfo {author} {\bibfnamefont {B.}~\bibnamefont
  {Batell}}, \bibinfo {author} {\bibfnamefont {M.}~\bibnamefont {Pospelov}}, \
  and\ \bibinfo {author} {\bibfnamefont {A.}~\bibnamefont {Ritz}},\ }\href
  {\doibase 10.1103/PhysRevD.80.095024} {\bibfield  {journal} {\bibinfo
  {journal} {Phys. Rev.}\ }\textbf {\bibinfo {volume} {D80}},\ \bibinfo {pages}
  {095024} (\bibinfo {year} {2009})},\ \Eprint {http://arxiv.org/abs/0906.5614}
  {arXiv:0906.5614 [hep-ph]} \BibitemShut {NoStop}%
\bibitem [{\citenamefont {Izaguirre}\ \emph {et~al.}(2015)\citenamefont
  {Izaguirre}, \citenamefont {Krnjaic}, \citenamefont {Schuster},\ and\
  \citenamefont {Toro}}]{Izaguirre:2015yja}%
  \BibitemOpen
  \bibfield  {author} {\bibinfo {author} {\bibfnamefont {E.}~\bibnamefont
  {Izaguirre}}, \bibinfo {author} {\bibfnamefont {G.}~\bibnamefont {Krnjaic}},
  \bibinfo {author} {\bibfnamefont {P.}~\bibnamefont {Schuster}}, \ and\
  \bibinfo {author} {\bibfnamefont {N.}~\bibnamefont {Toro}},\ }\href@noop {}
  {\  (\bibinfo {year} {2015})},\ \Eprint {http://arxiv.org/abs/1505.00011}
  {arXiv:1505.00011 [hep-ph]} \BibitemShut {NoStop}%
\bibitem [{\citenamefont {Boehm}\ and\ \citenamefont
  {Fayet}(2004)}]{Boehm:2003hm}%
  \BibitemOpen
  \bibfield  {author} {\bibinfo {author} {\bibfnamefont {C.}~\bibnamefont
  {Boehm}}\ and\ \bibinfo {author} {\bibfnamefont {P.}~\bibnamefont {Fayet}},\
  }\href {\doibase 10.1016/j.nuclphysb.2004.01.015} {\bibfield  {journal}
  {\bibinfo  {journal} {Nucl.Phys.}\ }\textbf {\bibinfo {volume} {B683}},\
  \bibinfo {pages} {219} (\bibinfo {year} {2004})},\ \Eprint
  {http://arxiv.org/abs/hep-ph/0305261} {arXiv:hep-ph/0305261 [hep-ph]}
  \BibitemShut {NoStop}%
\bibitem [{\citenamefont {Strassler}\ and\ \citenamefont
  {Zurek}(2007)}]{Strassler:2006im}%
  \BibitemOpen
  \bibfield  {author} {\bibinfo {author} {\bibfnamefont {M.~J.}\ \bibnamefont
  {Strassler}}\ and\ \bibinfo {author} {\bibfnamefont {K.~M.}\ \bibnamefont
  {Zurek}},\ }\href {\doibase 10.1016/j.physletb.2007.06.055} {\bibfield
  {journal} {\bibinfo  {journal} {Phys.Lett.}\ }\textbf {\bibinfo {volume}
  {B651}},\ \bibinfo {pages} {374} (\bibinfo {year} {2007})},\ \Eprint
  {http://arxiv.org/abs/hep-ph/0604261} {arXiv:hep-ph/0604261 [hep-ph]}
  \BibitemShut {NoStop}%
\bibitem [{\citenamefont {Arkani-Hamed}\ \emph {et~al.}(2009)\citenamefont
  {Arkani-Hamed}, \citenamefont {Finkbeiner}, \citenamefont {Slatyer},\ and\
  \citenamefont {Weiner}}]{ArkaniHamed:2008qn}%
  \BibitemOpen
  \bibfield  {author} {\bibinfo {author} {\bibfnamefont {N.}~\bibnamefont
  {Arkani-Hamed}}, \bibinfo {author} {\bibfnamefont {D.~P.}\ \bibnamefont
  {Finkbeiner}}, \bibinfo {author} {\bibfnamefont {T.~R.}\ \bibnamefont
  {Slatyer}}, \ and\ \bibinfo {author} {\bibfnamefont {N.}~\bibnamefont
  {Weiner}},\ }\href {\doibase 10.1103/PhysRevD.79.015014} {\bibfield
  {journal} {\bibinfo  {journal} {Phys.Rev.}\ }\textbf {\bibinfo {volume}
  {D79}},\ \bibinfo {pages} {015014} (\bibinfo {year} {2009})},\ \Eprint
  {http://arxiv.org/abs/0810.0713} {arXiv:0810.0713 [hep-ph]} \BibitemShut
  {NoStop}%
\bibitem [{\citenamefont {Pospelov}\ and\ \citenamefont
  {Ritz}(2009)}]{Pospelov:2008jd}%
  \BibitemOpen
  \bibfield  {author} {\bibinfo {author} {\bibfnamefont {M.}~\bibnamefont
  {Pospelov}}\ and\ \bibinfo {author} {\bibfnamefont {A.}~\bibnamefont
  {Ritz}},\ }\href {\doibase 10.1016/j.physletb.2008.12.012} {\bibfield
  {journal} {\bibinfo  {journal} {Phys. Lett.}\ }\textbf {\bibinfo {volume}
  {B671}},\ \bibinfo {pages} {391} (\bibinfo {year} {2009})}\BibitemShut
  {NoStop}%
\bibitem [{\citenamefont {Hooper}\ and\ \citenamefont
  {Zurek}(2008)}]{Hooper:2008im}%
  \BibitemOpen
  \bibfield  {author} {\bibinfo {author} {\bibfnamefont {D.}~\bibnamefont
  {Hooper}}\ and\ \bibinfo {author} {\bibfnamefont {K.~M.}\ \bibnamefont
  {Zurek}},\ }\href {\doibase 10.1103/PhysRevD.77.087302} {\bibfield  {journal}
  {\bibinfo  {journal} {Phys.Rev.}\ }\textbf {\bibinfo {volume} {D77}},\
  \bibinfo {pages} {087302} (\bibinfo {year} {2008})},\ \Eprint
  {http://arxiv.org/abs/0801.3686} {arXiv:0801.3686 [hep-ph]} \BibitemShut
  {NoStop}%
\bibitem [{\citenamefont {Feng}\ and\ \citenamefont
  {Kumar}(2008)}]{Feng:2008ya}%
  \BibitemOpen
  \bibfield  {author} {\bibinfo {author} {\bibfnamefont {J.~L.}\ \bibnamefont
  {Feng}}\ and\ \bibinfo {author} {\bibfnamefont {J.}~\bibnamefont {Kumar}},\
  }\href {\doibase 10.1103/PhysRevLett.101.231301} {\bibfield  {journal}
  {\bibinfo  {journal} {Phys.Rev.Lett.}\ }\textbf {\bibinfo {volume} {101}},\
  \bibinfo {pages} {231301} (\bibinfo {year} {2008})},\ \Eprint
  {http://arxiv.org/abs/0803.4196} {arXiv:0803.4196 [hep-ph]} \BibitemShut
  {NoStop}%
\bibitem [{\citenamefont {Morrissey}\ \emph {et~al.}(2009)\citenamefont
  {Morrissey}, \citenamefont {Poland},\ and\ \citenamefont
  {Zurek}}]{Morrissey:2009ur}%
  \BibitemOpen
  \bibfield  {author} {\bibinfo {author} {\bibfnamefont {D.~E.}\ \bibnamefont
  {Morrissey}}, \bibinfo {author} {\bibfnamefont {D.}~\bibnamefont {Poland}}, \
  and\ \bibinfo {author} {\bibfnamefont {K.~M.}\ \bibnamefont {Zurek}},\ }\href
  {\doibase 10.1088/1126-6708/2009/07/050} {\bibfield  {journal} {\bibinfo
  {journal} {JHEP}\ }\textbf {\bibinfo {volume} {0907}},\ \bibinfo {pages}
  {050} (\bibinfo {year} {2009})},\ \Eprint {http://arxiv.org/abs/0904.2567}
  {arXiv:0904.2567 [hep-ph]} \BibitemShut {NoStop}%
\bibitem [{\citenamefont {Essig}\ \emph {et~al.}(2010)\citenamefont {Essig},
  \citenamefont {Kaplan}, \citenamefont {Schuster},\ and\ \citenamefont
  {Toro}}]{Essig:2010ye}%
  \BibitemOpen
  \bibfield  {author} {\bibinfo {author} {\bibfnamefont {R.}~\bibnamefont
  {Essig}}, \bibinfo {author} {\bibfnamefont {J.}~\bibnamefont {Kaplan}},
  \bibinfo {author} {\bibfnamefont {P.}~\bibnamefont {Schuster}}, \ and\
  \bibinfo {author} {\bibfnamefont {N.}~\bibnamefont {Toro}},\ }\href@noop {}
  {\  (\bibinfo {year} {2010})},\ \Eprint {http://arxiv.org/abs/1004.0691}
  {arXiv:1004.0691 [hep-ph]} \BibitemShut {NoStop}%
\bibitem [{\citenamefont {Cohen}\ \emph {et~al.}(2010)\citenamefont {Cohen},
  \citenamefont {Phalen}, \citenamefont {Pierce},\ and\ \citenamefont
  {Zurek}}]{Cohen:2010kn}%
  \BibitemOpen
  \bibfield  {author} {\bibinfo {author} {\bibfnamefont {T.}~\bibnamefont
  {Cohen}}, \bibinfo {author} {\bibfnamefont {D.~J.}\ \bibnamefont {Phalen}},
  \bibinfo {author} {\bibfnamefont {A.}~\bibnamefont {Pierce}}, \ and\ \bibinfo
  {author} {\bibfnamefont {K.~M.}\ \bibnamefont {Zurek}},\ }\href {\doibase
  10.1103/PhysRevD.82.056001} {\bibfield  {journal} {\bibinfo  {journal}
  {Phys.Rev.}\ }\textbf {\bibinfo {volume} {D82}},\ \bibinfo {pages} {056001}
  (\bibinfo {year} {2010})},\ \Eprint {http://arxiv.org/abs/1005.1655}
  {arXiv:1005.1655 [hep-ph]} \BibitemShut {NoStop}%
\bibitem [{\citenamefont {Lin}\ \emph {et~al.}(2012)\citenamefont {Lin},
  \citenamefont {Yu},\ and\ \citenamefont {Zurek}}]{Lin:2011gj}%
  \BibitemOpen
  \bibfield  {author} {\bibinfo {author} {\bibfnamefont {T.}~\bibnamefont
  {Lin}}, \bibinfo {author} {\bibfnamefont {H.-B.}\ \bibnamefont {Yu}}, \ and\
  \bibinfo {author} {\bibfnamefont {K.~M.}\ \bibnamefont {Zurek}},\ }\href
  {\doibase 10.1103/PhysRevD.85.063503} {\bibfield  {journal} {\bibinfo
  {journal} {Phys.Rev.}\ }\textbf {\bibinfo {volume} {D85}},\ \bibinfo {pages}
  {063503} (\bibinfo {year} {2012})},\ \Eprint {http://arxiv.org/abs/1111.0293}
  {arXiv:1111.0293 [hep-ph]} \BibitemShut {NoStop}%
\bibitem [{\citenamefont {Chu}\ \emph {et~al.}(2012)\citenamefont {Chu},
  \citenamefont {Hambye},\ and\ \citenamefont {Tytgat}}]{Chu:2011be}%
  \BibitemOpen
  \bibfield  {author} {\bibinfo {author} {\bibfnamefont {X.}~\bibnamefont
  {Chu}}, \bibinfo {author} {\bibfnamefont {T.}~\bibnamefont {Hambye}}, \ and\
  \bibinfo {author} {\bibfnamefont {M.~H.}\ \bibnamefont {Tytgat}},\ }\href
  {\doibase 10.1088/1475-7516/2012/05/034} {\bibfield  {journal} {\bibinfo
  {journal} {JCAP}\ }\textbf {\bibinfo {volume} {1205}},\ \bibinfo {pages}
  {034} (\bibinfo {year} {2012})},\ \Eprint {http://arxiv.org/abs/1112.0493}
  {arXiv:1112.0493 [hep-ph]} \BibitemShut {NoStop}%
\bibitem [{\citenamefont {Hochberg}\ \emph {et~al.}(2014)\citenamefont
  {Hochberg}, \citenamefont {Kuflik}, \citenamefont {Volansky},\ and\
  \citenamefont {Wacker}}]{Hochberg:2014dra}%
  \BibitemOpen
  \bibfield  {author} {\bibinfo {author} {\bibfnamefont {Y.}~\bibnamefont
  {Hochberg}}, \bibinfo {author} {\bibfnamefont {E.}~\bibnamefont {Kuflik}},
  \bibinfo {author} {\bibfnamefont {T.}~\bibnamefont {Volansky}}, \ and\
  \bibinfo {author} {\bibfnamefont {J.~G.}\ \bibnamefont {Wacker}},\ }\href
  {\doibase 10.1103/PhysRevLett.113.171301} {\bibfield  {journal} {\bibinfo
  {journal} {Phys.Rev.Lett.}\ }\textbf {\bibinfo {volume} {113}},\ \bibinfo
  {pages} {171301} (\bibinfo {year} {2014})},\ \Eprint
  {http://arxiv.org/abs/1402.5143} {arXiv:1402.5143 [hep-ph]} \BibitemShut
  {NoStop}%
\bibitem [{\citenamefont {Hochberg}\ \emph
  {et~al.}(2015{\natexlab{b}})\citenamefont {Hochberg}, \citenamefont {Kuflik},
  \citenamefont {Murayama}, \citenamefont {Volansky},\ and\ \citenamefont
  {Wacker}}]{Hochberg:2014kqa}%
  \BibitemOpen
  \bibfield  {author} {\bibinfo {author} {\bibfnamefont {Y.}~\bibnamefont
  {Hochberg}}, \bibinfo {author} {\bibfnamefont {E.}~\bibnamefont {Kuflik}},
  \bibinfo {author} {\bibfnamefont {H.}~\bibnamefont {Murayama}}, \bibinfo
  {author} {\bibfnamefont {T.}~\bibnamefont {Volansky}}, \ and\ \bibinfo
  {author} {\bibfnamefont {J.~G.}\ \bibnamefont {Wacker}},\ }\href {\doibase
  10.1103/PhysRevLett.115.021301} {\bibfield  {journal} {\bibinfo  {journal}
  {Phys. Rev. Lett.}\ }\textbf {\bibinfo {volume} {115}},\ \bibinfo {pages}
  {021301} (\bibinfo {year} {2015}{\natexlab{b}})},\ \Eprint
  {http://arxiv.org/abs/1411.3727} {arXiv:1411.3727 [hep-ph]} \BibitemShut
  {NoStop}%
\bibitem [{\citenamefont {Cushman}\ \emph {et~al.}(2013)\citenamefont {Cushman}
  \emph {et~al.}}]{Cushman:2013zza}%
  \BibitemOpen
  \bibfield  {author} {\bibinfo {author} {\bibfnamefont {P.}~\bibnamefont
  {Cushman}} \emph {et~al.},\ }in\ \href {http://arxiv.org/pdf/1310.8327.pdf}
  {\emph {\bibinfo {booktitle} {{Community Summer Study 2013: Snowmass on the
  Mississippi (CSS2013) Minneapolis, MN, USA, July 29-August 6, 2013}}}}\
  (\bibinfo {year} {2013})\ \Eprint {http://arxiv.org/abs/1310.8327}
  {arXiv:1310.8327 [hep-ex]} \BibitemShut {NoStop}%
\bibitem [{\citenamefont {Angle}\ \emph {et~al.}(2011)\citenamefont {Angle}
  \emph {et~al.}}]{Angle:2011th}%
  \BibitemOpen
  \bibfield  {author} {\bibinfo {author} {\bibfnamefont {J.}~\bibnamefont
  {Angle}} \emph {et~al.} (\bibinfo {collaboration} {XENON10}),\ }\href
  {\doibase 10.1103/PhysRevLett.110.249901, 10.1103/PhysRevLett.107.051301}
  {\bibfield  {journal} {\bibinfo  {journal} {Phys. Rev. Lett.}\ }\textbf
  {\bibinfo {volume} {107}},\ \bibinfo {pages} {051301} (\bibinfo {year}
  {2011})},\ \bibinfo {note} {[Erratum: Phys. Rev. Lett.110,249901(2013)]},\
  \Eprint {http://arxiv.org/abs/1104.3088} {arXiv:1104.3088 [astro-ph.CO]}
  \BibitemShut {NoStop}%
\bibitem [{\citenamefont {Agnese}\ \emph {et~al.}(2016)\citenamefont {Agnese}
  \emph {et~al.}}]{Agnese:2015nto}%
  \BibitemOpen
  \bibfield  {author} {\bibinfo {author} {\bibfnamefont {R.}~\bibnamefont
  {Agnese}} \emph {et~al.} (\bibinfo {collaboration} {SuperCDMS}),\ }\href
  {\doibase 10.1103/PhysRevLett.116.071301} {\bibfield  {journal} {\bibinfo
  {journal} {Phys. Rev. Lett.}\ }\textbf {\bibinfo {volume} {116}},\ \bibinfo
  {pages} {071301} (\bibinfo {year} {2016})},\ \Eprint
  {http://arxiv.org/abs/1509.02448} {arXiv:1509.02448 [astro-ph.CO]}
  \BibitemShut {NoStop}%
\bibitem [{\citenamefont {Fernandez~Moroni}\ \emph {et~al.}(2012)\citenamefont
  {Fernandez~Moroni}, \citenamefont {Estrada}, \citenamefont {Cancelo},
  \citenamefont {Holland}, \citenamefont {Paolini},\ and\ \citenamefont
  {Diehl}}]{Moroni:2011xs}%
  \BibitemOpen
  \bibfield  {author} {\bibinfo {author} {\bibfnamefont {G.}~\bibnamefont
  {Fernandez~Moroni}}, \bibinfo {author} {\bibfnamefont {J.}~\bibnamefont
  {Estrada}}, \bibinfo {author} {\bibfnamefont {G.}~\bibnamefont {Cancelo}},
  \bibinfo {author} {\bibfnamefont {S.~E.}\ \bibnamefont {Holland}}, \bibinfo
  {author} {\bibfnamefont {E.~E.}\ \bibnamefont {Paolini}}, \ and\ \bibinfo
  {author} {\bibfnamefont {H.~T.}\ \bibnamefont {Diehl}},\ }\href {\doibase
  10.1007/s10686-012-9298-x} {\bibfield  {journal} {\bibinfo  {journal} {Exper.
  Astron.}\ }\textbf {\bibinfo {volume} {34}},\ \bibinfo {pages} {43} (\bibinfo
  {year} {2012})},\ \Eprint {http://arxiv.org/abs/1106.1839} {arXiv:1106.1839
  [astro-ph.IM]} \BibitemShut {NoStop}%
\bibitem [{Note1()}]{Note1}%
  \BibitemOpen
  \bibinfo {note} {Note that~\cite {Starkman:1994gf} proposed the search of one
  or more photons from Weak-scale dark matter through atomic
  excitations.}\BibitemShut {Stop}%
\bibitem [{\citenamefont {Pyle}()}]{Pyle}%
  \BibitemOpen
  \bibfield  {author} {\bibinfo {author} {\bibfnamefont {M.}~\bibnamefont
  {Pyle}},\ }\href@noop {} {\bibinfo  {journal} {private communication}\
  }\BibitemShut {NoStop}%
\bibitem [{\citenamefont {Aprile}\ \emph
  {et~al.}(2016{\natexlab{a}})\citenamefont {Aprile} \emph
  {et~al.}}]{Aprile:2016wwo}%
  \BibitemOpen
\bibfield  {journal} {  }\bibfield  {author} {\bibinfo {author} {\bibfnamefont
  {E.}~\bibnamefont {Aprile}} \emph {et~al.} (\bibinfo {collaboration}
  {XENON100}),\ }\href@noop {} {\  (\bibinfo {year} {2016}{\natexlab{a}})},\
  \Eprint {http://arxiv.org/abs/1605.06262} {arXiv:1605.06262 [astro-ph.CO]}
  \BibitemShut {NoStop}%
\bibitem [{\citenamefont {Luke}\ \emph {et~al.}(1990)\citenamefont {Luke},
  \citenamefont {Beeman}, \citenamefont {Goulding}, \citenamefont {Labov},\
  and\ \citenamefont {Silver}}]{Luke:1990ir}%
  \BibitemOpen
  \bibfield  {author} {\bibinfo {author} {\bibfnamefont {P.}~\bibnamefont
  {Luke}}, \bibinfo {author} {\bibfnamefont {J.}~\bibnamefont {Beeman}},
  \bibinfo {author} {\bibfnamefont {F.}~\bibnamefont {Goulding}}, \bibinfo
  {author} {\bibfnamefont {S.}~\bibnamefont {Labov}}, \ and\ \bibinfo {author}
  {\bibfnamefont {E.}~\bibnamefont {Silver}},\ }\href {\doibase
  10.1016/0168-9002(90)91510-I} {\bibfield  {journal} {\bibinfo  {journal}
  {Nucl.Instrum.Meth.}\ }\textbf {\bibinfo {volume} {A289}},\ \bibinfo {pages}
  {406} (\bibinfo {year} {1990})}\BibitemShut {NoStop}%
\bibitem [{\citenamefont {Neganov}\ and\ \citenamefont
  {Trofimov}(1985)}]{Neganov:1985}%
  \BibitemOpen
  \bibfield  {author} {\bibinfo {author} {\bibfnamefont {B.}~\bibnamefont
  {Neganov}}\ and\ \bibinfo {author} {\bibfnamefont {V.}~\bibnamefont
  {Trofimov}},\ }\href@noop {} {\bibfield  {journal} {\bibinfo  {journal}
  {Otkrytiya, Izobret}\ }\textbf {\bibinfo {volume} {146}},\ \bibinfo {pages}
  {215} (\bibinfo {year} {1985})}\BibitemShut {NoStop}%
\bibitem [{\citenamefont {Golwala}(2016)}]{GolwalaTalk}%
  \BibitemOpen
  \bibfield  {author} {\bibinfo {author} {\bibfnamefont {S.}~\bibnamefont
  {Golwala}} (\bibinfo {collaboration} {SuperCDMS}),\ }\href
  {https://conferences.pa.ucla.edu/dm16/talks/golwala.pdf} {\enquote {\bibinfo
  {title} {{SuperCDMS SNOLAB: Goals, Design, and Status}},}\ } (\bibinfo {year}
  {2016}),\ \bibinfo {note} {talk given at UCLA DM 2016}\BibitemShut {NoStop}%
\bibitem [{\citenamefont {Mazin}\ \emph {et~al.}(2012)\citenamefont {Mazin},
  \citenamefont {Bumble}, \citenamefont {Meeker}, \citenamefont {O'Brien},
  \citenamefont {McHugh},\ and\ \citenamefont {Langman}}]{Mazin:12}%
  \BibitemOpen
  \bibfield  {author} {\bibinfo {author} {\bibfnamefont {B.~A.}\ \bibnamefont
  {Mazin}}, \bibinfo {author} {\bibfnamefont {B.}~\bibnamefont {Bumble}},
  \bibinfo {author} {\bibfnamefont {S.~R.}\ \bibnamefont {Meeker}}, \bibinfo
  {author} {\bibfnamefont {K.}~\bibnamefont {O'Brien}}, \bibinfo {author}
  {\bibfnamefont {S.}~\bibnamefont {McHugh}}, \ and\ \bibinfo {author}
  {\bibfnamefont {E.}~\bibnamefont {Langman}},\ }\href {\doibase
  10.1364/OE.20.001503} {\bibfield  {journal} {\bibinfo  {journal} {Opt.
  Express}\ }\textbf {\bibinfo {volume} {20}},\ \bibinfo {pages} {1503}
  (\bibinfo {year} {2012})}\BibitemShut {NoStop}%
\bibitem [{\citenamefont {Irwin}\ and\ \citenamefont {Hilton}(2005)}]{TES}%
  \BibitemOpen
  \bibfield  {author} {\bibinfo {author} {\bibfnamefont {K.}~\bibnamefont
  {Irwin}}\ and\ \bibinfo {author} {\bibfnamefont {G.}~\bibnamefont {Hilton}},\
  }\href@noop {} {\bibfield  {journal} {\bibinfo  {journal} {Topics in Applied
  Physics}\ }\textbf {\bibinfo {volume} {99, 63}} (\bibinfo {year}
  {2005})}\BibitemShut {NoStop}%
\bibitem [{\citenamefont {Petroff}\ \emph {et~al.}(1987)\citenamefont
  {Petroff}, \citenamefont {Stapelbroek},\ and\ \citenamefont
  {Kleinhans}}]{TES3}%
  \BibitemOpen
  \bibfield  {author} {\bibinfo {author} {\bibfnamefont {M.~D.}\ \bibnamefont
  {Petroff}}, \bibinfo {author} {\bibfnamefont {M.~G.}\ \bibnamefont
  {Stapelbroek}}, \ and\ \bibinfo {author} {\bibfnamefont {W.~A.}\ \bibnamefont
  {Kleinhans}},\ }\href {\doibase http://dx.doi.org/10.1063/1.98404} {\bibfield
   {journal} {\bibinfo  {journal} {Applied Physics Letters}\ }\textbf {\bibinfo
  {volume} {51}},\ \bibinfo {pages} {406} (\bibinfo {year} {1987})}\BibitemShut
  {NoStop}%
\bibitem [{\citenamefont {Santavicca}\ \emph {et~al.}(2010)\citenamefont
  {Santavicca}, \citenamefont {Reulet}, \citenamefont {Karasik}, \citenamefont
  {Pereverzev}, \citenamefont {Olaya}, \citenamefont {Gershenson},
  \citenamefont {Frunzio},\ and\ \citenamefont {Prober}}]{TES4}%
  \BibitemOpen
  \bibfield  {author} {\bibinfo {author} {\bibfnamefont {D.~F.}\ \bibnamefont
  {Santavicca}}, \bibinfo {author} {\bibfnamefont {B.}~\bibnamefont {Reulet}},
  \bibinfo {author} {\bibfnamefont {B.~S.}\ \bibnamefont {Karasik}}, \bibinfo
  {author} {\bibfnamefont {S.~V.}\ \bibnamefont {Pereverzev}}, \bibinfo
  {author} {\bibfnamefont {D.}~\bibnamefont {Olaya}}, \bibinfo {author}
  {\bibfnamefont {M.~E.}\ \bibnamefont {Gershenson}}, \bibinfo {author}
  {\bibfnamefont {L.}~\bibnamefont {Frunzio}}, \ and\ \bibinfo {author}
  {\bibfnamefont {D.~E.}\ \bibnamefont {Prober}},\ }\href {\doibase
  http://dx.doi.org/10.1063/1.3336008} {\bibfield  {journal} {\bibinfo
  {journal} {Applied Physics Letters}\ }\textbf {\bibinfo {volume} {96}},\
  \bibinfo {eid} {083505} (\bibinfo {year} {2010}),\
  http://dx.doi.org/10.1063/1.3336008}\BibitemShut {NoStop}%
\bibitem [{\citenamefont {Karasik}\ \emph {et~al.}(2012)\citenamefont
  {Karasik}, \citenamefont {Pereverzev}, \citenamefont {Soibel}, \citenamefont
  {Santavicca}, \citenamefont {Prober}, \citenamefont {Olaya},\ and\
  \citenamefont {Gershenson}}]{TES1}%
  \BibitemOpen
  \bibfield  {author} {\bibinfo {author} {\bibfnamefont {B.~S.}\ \bibnamefont
  {Karasik}}, \bibinfo {author} {\bibfnamefont {S.~V.}\ \bibnamefont
  {Pereverzev}}, \bibinfo {author} {\bibfnamefont {A.}~\bibnamefont {Soibel}},
  \bibinfo {author} {\bibfnamefont {D.~F.}\ \bibnamefont {Santavicca}},
  \bibinfo {author} {\bibfnamefont {D.~E.}\ \bibnamefont {Prober}}, \bibinfo
  {author} {\bibfnamefont {D.}~\bibnamefont {Olaya}}, \ and\ \bibinfo {author}
  {\bibfnamefont {M.~E.}\ \bibnamefont {Gershenson}},\ }\href {\doibase
  http://dx.doi.org/10.1063/1.4739839} {\bibfield  {journal} {\bibinfo
  {journal} {Applied Physics Letters}\ }\textbf {\bibinfo {volume} {101}},\
  \bibinfo {eid} {052601} (\bibinfo {year} {2012}),\
  http://dx.doi.org/10.1063/1.4739839}\BibitemShut {NoStop}%
\bibitem [{\citenamefont {Goldie}\ \emph {et~al.}(2011)\citenamefont {Goldie},
  \citenamefont {Velichko}, \citenamefont {Glowacka},\ and\ \citenamefont
  {Withington}}]{Goldie}%
  \BibitemOpen
  \bibfield  {author} {\bibinfo {author} {\bibfnamefont {D.~J.}\ \bibnamefont
  {Goldie}}, \bibinfo {author} {\bibfnamefont {A.~V.}\ \bibnamefont
  {Velichko}}, \bibinfo {author} {\bibfnamefont {D.~M.}\ \bibnamefont
  {Glowacka}}, \ and\ \bibinfo {author} {\bibfnamefont {S.}~\bibnamefont
  {Withington}},\ }\href {\doibase http://dx.doi.org/10.1063/1.3561432}
  {\bibfield  {journal} {\bibinfo  {journal} {Journal of Applied Physics}\
  }\textbf {\bibinfo {volume} {109}},\ \bibinfo {eid} {084507} (\bibinfo {year}
  {2011}),\ http://dx.doi.org/10.1063/1.3561432}\BibitemShut {NoStop}%
\bibitem [{\citenamefont {Miller}\ \emph {et~al.}(2003)\citenamefont {Miller},
  \citenamefont {Nam}, \citenamefont {Martinis},\ and\ \citenamefont
  {Sergienko}}]{Miller}%
  \BibitemOpen
  \bibfield  {author} {\bibinfo {author} {\bibfnamefont {A.~J.}\ \bibnamefont
  {Miller}}, \bibinfo {author} {\bibfnamefont {S.~W.}\ \bibnamefont {Nam}},
  \bibinfo {author} {\bibfnamefont {J.~M.}\ \bibnamefont {Martinis}}, \ and\
  \bibinfo {author} {\bibfnamefont {A.~V.}\ \bibnamefont {Sergienko}},\ }\href
  {\doibase http://dx.doi.org/10.1063/1.1596723} {\bibfield  {journal}
  {\bibinfo  {journal} {Applied Physics Letters}\ }\textbf {\bibinfo {volume}
  {83}},\ \bibinfo {pages} {791} (\bibinfo {year} {2003})}\BibitemShut
  {NoStop}%
\bibitem [{\citenamefont {Angloher}\ \emph
  {et~al.}(2016{\natexlab{a}})\citenamefont {Angloher} \emph
  {et~al.}}]{Angloher:2015ewa}%
  \BibitemOpen
  \bibfield  {author} {\bibinfo {author} {\bibfnamefont {G.}~\bibnamefont
  {Angloher}} \emph {et~al.} (\bibinfo {collaboration} {CRESST}),\ }\href
  {\doibase 10.1140/epjc/s10052-016-3877-3} {\bibfield  {journal} {\bibinfo
  {journal} {Eur. Phys. J.}\ }\textbf {\bibinfo {volume} {C76}},\ \bibinfo
  {pages} {25} (\bibinfo {year} {2016}{\natexlab{a}})},\ \Eprint
  {http://arxiv.org/abs/1509.01515} {arXiv:1509.01515 [astro-ph.CO]}
  \BibitemShut {NoStop}%
\bibitem [{\citenamefont {Angloher}\ \emph
  {et~al.}(2016{\natexlab{b}})\citenamefont {Angloher} \emph
  {et~al.}}]{Angloher:2016hbv}%
  \BibitemOpen
  \bibfield  {author} {\bibinfo {author} {\bibfnamefont {G.}~\bibnamefont
  {Angloher}} \emph {et~al.},\ }\href@noop {} {\  (\bibinfo {year}
  {2016}{\natexlab{b}})},\ \Eprint {http://arxiv.org/abs/1602.08884}
  {arXiv:1602.08884 [physics.ins-det]} \BibitemShut {NoStop}%
\bibitem [{\citenamefont {Biroth}\ \emph {et~al.}(2015)\citenamefont {Biroth},
  \citenamefont {Achenbach}, \citenamefont {Downie},\ and\ \citenamefont
  {Thomas}}]{Biroth}%
  \BibitemOpen
  \bibfield  {author} {\bibinfo {author} {\bibfnamefont {M.}~\bibnamefont
  {Biroth}}, \bibinfo {author} {\bibfnamefont {P.}~\bibnamefont {Achenbach}},
  \bibinfo {author} {\bibfnamefont {E.}~\bibnamefont {Downie}}, \ and\ \bibinfo
  {author} {\bibfnamefont {A.}~\bibnamefont {Thomas}},\ }\href@noop {}
  {\bibfield  {journal} {\bibinfo  {journal} {Nuclear Instruments \& Methods in
  Physics Research Section A-Accelerators Spectrometers Detectors and
  Associated Equipment}\ }\textbf {\bibinfo {volume} {787}},\ \bibinfo {pages}
  {68} (\bibinfo {year} {2015})}\BibitemShut {NoStop}%
\bibitem [{\citenamefont {Achenbach}\ \emph {et~al.}(2016)\citenamefont
  {Achenbach}, \citenamefont {Biroth}, \citenamefont {Downie},\ and\
  \citenamefont {Thomas}}]{Achenbach}%
  \BibitemOpen
  \bibfield  {author} {\bibinfo {author} {\bibfnamefont {P.}~\bibnamefont
  {Achenbach}}, \bibinfo {author} {\bibfnamefont {M.}~\bibnamefont {Biroth}},
  \bibinfo {author} {\bibfnamefont {E.}~\bibnamefont {Downie}}, \ and\ \bibinfo
  {author} {\bibfnamefont {A.}~\bibnamefont {Thomas}},\ }\href@noop {}
  {\bibfield  {journal} {\bibinfo  {journal} {Nuclear Instruments \& Methods in
  Physics Research Section A-Accelerators Spectrometers Detectors and
  Associated Equipment}\ }\textbf {\bibinfo {volume} {824}},\ \bibinfo {pages}
  {74} (\bibinfo {year} {2016})}\BibitemShut {NoStop}%
\bibitem [{\citenamefont {Drukier}\ \emph {et~al.}(1986)\citenamefont
  {Drukier}, \citenamefont {Freese},\ and\ \citenamefont
  {Spergel}}]{Drukier:1986tm}%
  \BibitemOpen
  \bibfield  {author} {\bibinfo {author} {\bibfnamefont {A.~K.}\ \bibnamefont
  {Drukier}}, \bibinfo {author} {\bibfnamefont {K.}~\bibnamefont {Freese}}, \
  and\ \bibinfo {author} {\bibfnamefont {D.~N.}\ \bibnamefont {Spergel}},\
  }\href {\doibase 10.1103/PhysRevD.33.3495} {\bibfield  {journal} {\bibinfo
  {journal} {Phys. Rev.}\ }\textbf {\bibinfo {volume} {D33}},\ \bibinfo {pages}
  {3495} (\bibinfo {year} {1986})}\BibitemShut {NoStop}%
\bibitem [{\citenamefont {Tucker-Smith}\ and\ \citenamefont
  {Weiner}(2001)}]{TuckerSmith:2001hy}%
  \BibitemOpen
  \bibfield  {author} {\bibinfo {author} {\bibfnamefont {D.}~\bibnamefont
  {Tucker-Smith}}\ and\ \bibinfo {author} {\bibfnamefont {N.}~\bibnamefont
  {Weiner}},\ }\href {\doibase 10.1103/PhysRevD.64.043502} {\bibfield
  {journal} {\bibinfo  {journal} {Phys. Rev.}\ }\textbf {\bibinfo {volume}
  {D64}},\ \bibinfo {pages} {043502} (\bibinfo {year} {2001})},\ \Eprint
  {http://arxiv.org/abs/hep-ph/0101138} {arXiv:hep-ph/0101138 [hep-ph]}
  \BibitemShut {NoStop}%
\bibitem [{Note2()}]{Note2}%
  \BibitemOpen
  \bibinfo {note} {We acknowledge Matthew Pyle for insightful
  discussions.}\BibitemShut {Stop}%
\bibitem [{\citenamefont {Moszynski}\ \emph
  {et~al.}(2005{\natexlab{a}})\citenamefont {Moszynski}, \citenamefont
  {Balcerzyk}, \citenamefont {Czarnacki}, \citenamefont {Kapusta},
  \citenamefont {Klamra}, \citenamefont {Schotanus}, \citenamefont {Syntfeld},
  \citenamefont {Szawlowski},\ and\ \citenamefont {Kozlov}}]{Moszynski2005357}%
  \BibitemOpen
  \bibfield  {author} {\bibinfo {author} {\bibfnamefont {M.}~\bibnamefont
  {Moszynski}}, \bibinfo {author} {\bibfnamefont {M.}~\bibnamefont
  {Balcerzyk}}, \bibinfo {author} {\bibfnamefont {W.}~\bibnamefont
  {Czarnacki}}, \bibinfo {author} {\bibfnamefont {M.}~\bibnamefont {Kapusta}},
  \bibinfo {author} {\bibfnamefont {W.}~\bibnamefont {Klamra}}, \bibinfo
  {author} {\bibfnamefont {P.}~\bibnamefont {Schotanus}}, \bibinfo {author}
  {\bibfnamefont {A.}~\bibnamefont {Syntfeld}}, \bibinfo {author}
  {\bibfnamefont {M.}~\bibnamefont {Szawlowski}}, \ and\ \bibinfo {author}
  {\bibfnamefont {V.}~\bibnamefont {Kozlov}},\ }\href {\doibase
  http://dx.doi.org/10.1016/j.nima.2004.08.043} {\bibfield  {journal} {\bibinfo
   {journal} {Nuclear Instruments and Methods in Physics Research Section A:
  Accelerators, Spectrometers, Detectors and Associated Equipment}\ }\textbf
  {\bibinfo {volume} {537}},\ \bibinfo {pages} {357 } (\bibinfo {year}
  {2005}{\natexlab{a}})},\ \bibinfo {note} {proceedings of the 7th
  International Conference on Inorganic Scintillators and their Use in
  Scientific and Industrial Applications}\BibitemShut {NoStop}%
\bibitem [{\citenamefont {ADAMS}\ and\ \citenamefont
  {DAMS}(1970)}]{tagkey1970iv}%
  \BibitemOpen
  \bibinfo {editor} {\bibfnamefont {C.~C.}\ \bibnamefont {ADAMS}}\ and\
  \bibinfo {editor} {\bibfnamefont {R.}~\bibnamefont {DAMS}},\ eds.,\ \href
  {\doibase http://dx.doi.org/10.1016/B978-0-08-006888-6.50002-X} {\emph
  {\bibinfo {title} {Applied Gamma-Ray Spectrometry}}},\ \bibinfo {edition}
  {second edition completely revised and enlarged}\ ed.,\ \bibinfo {series}
  {International Series of Monographs on Analytical Chemistry}, Vol.~\bibinfo
  {volume} {2}\ (\bibinfo  {publisher} {Pergamon},\ \bibinfo {year} {1970})\
  pp.\ \bibinfo {pages} {iv --}\BibitemShut {NoStop}%
\bibitem [{\citenamefont {Hofstadter}(1949{\natexlab{a}})}]{PhysRev.75.796}%
  \BibitemOpen
  \bibfield  {author} {\bibinfo {author} {\bibfnamefont {R.}~\bibnamefont
  {Hofstadter}},\ }\href {\doibase 10.1103/PhysRev.75.796} {\bibfield
  {journal} {\bibinfo  {journal} {Phys. Rev.}\ }\textbf {\bibinfo {volume}
  {75}},\ \bibinfo {pages} {796} (\bibinfo {year}
  {1949}{\natexlab{a}})}\BibitemShut {NoStop}%
\bibitem [{\citenamefont {Hofstadter}(1949{\natexlab{b}})}]{PhysRev.75.1611.2}%
  \BibitemOpen
  \bibfield  {author} {\bibinfo {author} {\bibfnamefont {R.}~\bibnamefont
  {Hofstadter}},\ }\href {\doibase 10.1103/PhysRev.75.1611.2} {\bibfield
  {journal} {\bibinfo  {journal} {Phys. Rev.}\ }\textbf {\bibinfo {volume}
  {75}},\ \bibinfo {pages} {1611} (\bibinfo {year}
  {1949}{\natexlab{b}})}\BibitemShut {NoStop}%
\bibitem [{Sai()}]{SaintGobain}%
  \BibitemOpen
  \href@noop {} {\enquote {\bibinfo {title} {Saint-gobain crystals},}\
  }\bibinfo {howpublished} {\url{http://www.crystals.saint-gobain.com/}},\
  \bibinfo {note} {accessed: 2016-02-17}\BibitemShut {NoStop}%
\bibitem [{\citenamefont {Moszyriski}\ \emph {et~al.}(2008)\citenamefont
  {Moszyriski}, \citenamefont {Nassalski}, \citenamefont {Syntfeld-Kazuch},
  \citenamefont {Swiderski},\ and\ \citenamefont {Szczesniak}}]{4545171}%
  \BibitemOpen
  \bibfield  {author} {\bibinfo {author} {\bibfnamefont {M.}~\bibnamefont
  {Moszyriski}}, \bibinfo {author} {\bibfnamefont {A.}~\bibnamefont
  {Nassalski}}, \bibinfo {author} {\bibfnamefont {A.}~\bibnamefont
  {Syntfeld-Kazuch}}, \bibinfo {author} {\bibfnamefont {L.}~\bibnamefont
  {Swiderski}}, \ and\ \bibinfo {author} {\bibfnamefont {T.}~\bibnamefont
  {Szczesniak}},\ }\href {\doibase 10.1109/TNS.2007.908580} {\bibfield
  {journal} {\bibinfo  {journal} {Nuclear Science, IEEE Transactions on}\
  }\textbf {\bibinfo {volume} {55}},\ \bibinfo {pages} {1062} (\bibinfo {year}
  {2008})}\BibitemShut {NoStop}%
\bibitem [{\citenamefont {{Sailer}}\ \emph {et~al.}(2012)\citenamefont
  {{Sailer}}, \citenamefont {{Lubsandorzhiev}}, \citenamefont {{Strandhagen}},\
  and\ \citenamefont {{Jochum}}}]{2012EPJC72.2061S}%
  \BibitemOpen
  \bibfield  {author} {\bibinfo {author} {\bibfnamefont {C.}~\bibnamefont
  {{Sailer}}}, \bibinfo {author} {\bibfnamefont {B.}~\bibnamefont
  {{Lubsandorzhiev}}}, \bibinfo {author} {\bibfnamefont {C.}~\bibnamefont
  {{Strandhagen}}}, \ and\ \bibinfo {author} {\bibfnamefont {J.}~\bibnamefont
  {{Jochum}}},\ }\href {\doibase 10.1140/epjc/s10052-012-2061-7} {\bibfield
  {journal} {\bibinfo  {journal} {European Physical Journal C}\ }\textbf
  {\bibinfo {volume} {72}},\ \bibinfo {eid} {2061} (\bibinfo {year} {2012})},\
  \Eprint {http://arxiv.org/abs/1203.1172} {arXiv:1203.1172 [physics.ins-det]}
  \BibitemShut {NoStop}%
\bibitem [{\citenamefont {Mikhailik}\ \emph {et~al.}(2015)\citenamefont
  {Mikhailik}, \citenamefont {Kapustyanyk}, \citenamefont {Tsybulskyi},
  \citenamefont {Rudyk},\ and\ \citenamefont {Kraus}}]{PSSB:PSSB201451464}%
  \BibitemOpen
  \bibfield  {author} {\bibinfo {author} {\bibfnamefont {V.~B.}\ \bibnamefont
  {Mikhailik}}, \bibinfo {author} {\bibfnamefont {V.}~\bibnamefont
  {Kapustyanyk}}, \bibinfo {author} {\bibfnamefont {V.}~\bibnamefont
  {Tsybulskyi}}, \bibinfo {author} {\bibfnamefont {V.}~\bibnamefont {Rudyk}}, \
  and\ \bibinfo {author} {\bibfnamefont {H.}~\bibnamefont {Kraus}},\ }\href
  {\doibase 10.1002/pssb.201451464} {\bibfield  {journal} {\bibinfo  {journal}
  {physica status solidi (b)}\ }\textbf {\bibinfo {volume} {252}},\ \bibinfo
  {pages} {804} (\bibinfo {year} {2015})}\BibitemShut {NoStop}%
\bibitem [{\citenamefont {Cusano}(1964)}]{Cusano}%
  \BibitemOpen
  \bibfield  {author} {\bibinfo {author} {\bibfnamefont {D.}~\bibnamefont
  {Cusano}},\ }\href {\doibase http://dx.doi.org/10.1016/0038-1098(64)90259-5}
  {\bibfield  {journal} {\bibinfo  {journal} {Solid State Communications}\
  }\textbf {\bibinfo {volume} {2}},\ \bibinfo {pages} {353 } (\bibinfo {year}
  {1964})}\BibitemShut {NoStop}%
\bibitem [{\citenamefont {Knapitsch}\ and\ \citenamefont
  {Lecoq}(2014)}]{Knapitsch}%
  \BibitemOpen
  \bibfield  {author} {\bibinfo {author} {\bibfnamefont {A.}~\bibnamefont
  {Knapitsch}}\ and\ \bibinfo {author} {\bibfnamefont {P.}~\bibnamefont
  {Lecoq}},\ }\href {\doibase 10.1142/S0217751X14300701} {\bibfield  {journal}
  {\bibinfo  {journal} {International Journal of Modern Physics A}\ }\textbf
  {\bibinfo {volume} {29}},\ \bibinfo {pages} {1430070} (\bibinfo {year}
  {2014})}\BibitemShut {NoStop}%
\bibitem [{\citenamefont {Moszynski}\ \emph {et~al.}(2002)\citenamefont
  {Moszynski}, \citenamefont {Balcerzyk}, \citenamefont {Czarnacki},
  \citenamefont {Kapusta}, \citenamefont {Klamra}, \citenamefont {Schotanus},
  \citenamefont {Syntfeld},\ and\ \citenamefont {Szawlowski}}]{Moszynski2003}%
  \BibitemOpen
  \bibfield  {author} {\bibinfo {author} {\bibfnamefont {M.}~\bibnamefont
  {Moszynski}}, \bibinfo {author} {\bibfnamefont {M.}~\bibnamefont
  {Balcerzyk}}, \bibinfo {author} {\bibfnamefont {W.}~\bibnamefont
  {Czarnacki}}, \bibinfo {author} {\bibfnamefont {M.}~\bibnamefont {Kapusta}},
  \bibinfo {author} {\bibfnamefont {W.}~\bibnamefont {Klamra}}, \bibinfo
  {author} {\bibfnamefont {P.}~\bibnamefont {Schotanus}}, \bibinfo {author}
  {\bibfnamefont {A.}~\bibnamefont {Syntfeld}}, \ and\ \bibinfo {author}
  {\bibfnamefont {M.}~\bibnamefont {Szawlowski}},\ }in\ \href {\doibase
  10.1109/NSSMIC.2002.1239330} {\emph {\bibinfo {booktitle} {Nuclear Science
  Symposium Conference Record, 2002 IEEE}}},\ Vol.~\bibinfo {volume} {1}\
  (\bibinfo {year} {2002})\ pp.\ \bibinfo {pages} {346--351 vol.1}\BibitemShut
  {NoStop}%
\bibitem [{\citenamefont {Amsler}\ \emph {et~al.}(2002)\citenamefont {Amsler},
  \citenamefont {Grögler}, \citenamefont {Joffrain}, \citenamefont
  {Lindelöf}, \citenamefont {Marchesotti}, \citenamefont {Niederberger},
  \citenamefont {Pruys}, \citenamefont {Regenfus}, \citenamefont {Riedler},\
  and\ \citenamefont {Rotondi}}]{Amsler2002494}%
  \BibitemOpen
  \bibfield  {author} {\bibinfo {author} {\bibfnamefont {C.}~\bibnamefont
  {Amsler}}, \bibinfo {author} {\bibfnamefont {D.}~\bibnamefont {Grögler}},
  \bibinfo {author} {\bibfnamefont {W.}~\bibnamefont {Joffrain}}, \bibinfo
  {author} {\bibfnamefont {D.}~\bibnamefont {Lindelöf}}, \bibinfo {author}
  {\bibfnamefont {M.}~\bibnamefont {Marchesotti}}, \bibinfo {author}
  {\bibfnamefont {P.}~\bibnamefont {Niederberger}}, \bibinfo {author}
  {\bibfnamefont {H.}~\bibnamefont {Pruys}}, \bibinfo {author} {\bibfnamefont
  {C.}~\bibnamefont {Regenfus}}, \bibinfo {author} {\bibfnamefont
  {P.}~\bibnamefont {Riedler}}, \ and\ \bibinfo {author} {\bibfnamefont
  {A.}~\bibnamefont {Rotondi}},\ }\href {\doibase
  http://dx.doi.org/10.1016/S0168-9002(01)01239-6} {\bibfield  {journal}
  {\bibinfo  {journal} {Nuclear Instruments and Methods in Physics Research
  Section A: Accelerators, Spectrometers, Detectors and Associated Equipment}\
  }\textbf {\bibinfo {volume} {480}},\ \bibinfo {pages} {494 } (\bibinfo {year}
  {2002})}\BibitemShut {NoStop}%
\bibitem [{\citenamefont {Trilling}(1970)}]{Trilling1970}%
  \BibitemOpen
  \bibfield  {author} {\bibinfo {author} {\bibfnamefont {G.}~\bibnamefont
  {Trilling}},\ }\href {\doibase 10.2172/4082446} {\emph {\bibinfo {title}
  {Bubble chamber physics in the seventies}}}\ (\bibinfo {year}
  {1970})\BibitemShut {NoStop}%
\bibitem [{\citenamefont {Zdesenko}\ \emph {et~al.}(2005)\citenamefont
  {Zdesenko}, \citenamefont {III}, \citenamefont {Brudanin}, \citenamefont
  {Danevich}, \citenamefont {Nagorny}, \citenamefont {Solsky},\ and\
  \citenamefont {Tretyak}}]{Zdesenko2005657}%
  \BibitemOpen
  \bibfield  {author} {\bibinfo {author} {\bibfnamefont {Y.}~\bibnamefont
  {Zdesenko}}, \bibinfo {author} {\bibfnamefont {F.~A.}\ \bibnamefont {III}},
  \bibinfo {author} {\bibfnamefont {V.}~\bibnamefont {Brudanin}}, \bibinfo
  {author} {\bibfnamefont {F.}~\bibnamefont {Danevich}}, \bibinfo {author}
  {\bibfnamefont {S.}~\bibnamefont {Nagorny}}, \bibinfo {author} {\bibfnamefont
  {I.}~\bibnamefont {Solsky}}, \ and\ \bibinfo {author} {\bibfnamefont
  {V.}~\bibnamefont {Tretyak}},\ }\href {\doibase
  http://dx.doi.org/10.1016/j.nima.2004.09.030} {\bibfield  {journal} {\bibinfo
   {journal} {Nuclear Instruments and Methods in Physics Research Section A:
  Accelerators, Spectrometers, Detectors and Associated Equipment}\ }\textbf
  {\bibinfo {volume} {538}},\ \bibinfo {pages} {657 } (\bibinfo {year}
  {2005})}\BibitemShut {NoStop}%
\bibitem [{\citenamefont {Aprile}\ \emph {et~al.}(2009)\citenamefont {Aprile},
  \citenamefont {Baudis}, \citenamefont {Choi}, \citenamefont {Giboni},
  \citenamefont {Lim}, \citenamefont {Manalaysay}, \citenamefont {Monzani},
  \citenamefont {Plante}, \citenamefont {Santorelli},\ and\ \citenamefont
  {Yamashita}}]{Aprile:2008rc}%
  \BibitemOpen
  \bibfield  {author} {\bibinfo {author} {\bibfnamefont {E.}~\bibnamefont
  {Aprile}}, \bibinfo {author} {\bibfnamefont {L.}~\bibnamefont {Baudis}},
  \bibinfo {author} {\bibfnamefont {B.}~\bibnamefont {Choi}}, \bibinfo {author}
  {\bibfnamefont {K.~L.}\ \bibnamefont {Giboni}}, \bibinfo {author}
  {\bibfnamefont {K.}~\bibnamefont {Lim}}, \bibinfo {author} {\bibfnamefont
  {A.}~\bibnamefont {Manalaysay}}, \bibinfo {author} {\bibfnamefont {M.~E.}\
  \bibnamefont {Monzani}}, \bibinfo {author} {\bibfnamefont {G.}~\bibnamefont
  {Plante}}, \bibinfo {author} {\bibfnamefont {R.}~\bibnamefont {Santorelli}},
  \ and\ \bibinfo {author} {\bibfnamefont {M.}~\bibnamefont {Yamashita}},\
  }\href {\doibase 10.1103/PhysRevC.79.045807} {\bibfield  {journal} {\bibinfo
  {journal} {Phys. Rev.}\ }\textbf {\bibinfo {volume} {C79}},\ \bibinfo {pages}
  {045807} (\bibinfo {year} {2009})},\ \Eprint {http://arxiv.org/abs/0810.0274}
  {arXiv:0810.0274 [astro-ph]} \BibitemShut {NoStop}%
\bibitem [{\citenamefont {Chepel}\ and\ \citenamefont
  {Araujo}(2013)}]{Chepel:2012sj}%
  \BibitemOpen
  \bibfield  {author} {\bibinfo {author} {\bibfnamefont {V.}~\bibnamefont
  {Chepel}}\ and\ \bibinfo {author} {\bibfnamefont {H.}~\bibnamefont
  {Araujo}},\ }\href {\doibase 10.1088/1748-0221/8/04/R04001} {\bibfield
  {journal} {\bibinfo  {journal} {JINST}\ }\textbf {\bibinfo {volume} {8}},\
  \bibinfo {pages} {R04001} (\bibinfo {year} {2013})},\ \Eprint
  {http://arxiv.org/abs/1207.2292} {arXiv:1207.2292 [physics.ins-det]}
  \BibitemShut {NoStop}%
\bibitem [{\citenamefont {Dehmer}\ and\ \citenamefont {Pratt}(1982)}]{Ar}%
  \BibitemOpen
  \bibfield  {author} {\bibinfo {author} {\bibfnamefont {P.~M.}\ \bibnamefont
  {Dehmer}}\ and\ \bibinfo {author} {\bibfnamefont {S.~T.}\ \bibnamefont
  {Pratt}},\ }\href {\doibase http://dx.doi.org/10.1063/1.443056} {\bibfield
  {journal} {\bibinfo  {journal} {The Journal of Chemical Physics}\ }\textbf
  {\bibinfo {volume} {76}},\ \bibinfo {pages} {843} (\bibinfo {year}
  {1982})}\BibitemShut {NoStop}%
\bibitem [{\citenamefont {Guo}\ and\ \citenamefont
  {McKinsey}(2013)}]{Guo:2013dt}%
  \BibitemOpen
  \bibfield  {author} {\bibinfo {author} {\bibfnamefont {W.}~\bibnamefont
  {Guo}}\ and\ \bibinfo {author} {\bibfnamefont {D.~N.}\ \bibnamefont
  {McKinsey}},\ }\href {\doibase 10.1103/PhysRevD.87.115001} {\bibfield
  {journal} {\bibinfo  {journal} {Phys. Rev.}\ }\textbf {\bibinfo {volume}
  {D87}},\ \bibinfo {pages} {115001} (\bibinfo {year} {2013})},\ \Eprint
  {http://arxiv.org/abs/1302.0534} {arXiv:1302.0534 [astro-ph.IM]} \BibitemShut
  {NoStop}%
\bibitem [{\citenamefont {de~Haas}\ and\ \citenamefont
  {Dorenbos}(2008)}]{deHaas2008}%
  \BibitemOpen
  \bibfield  {author} {\bibinfo {author} {\bibfnamefont {J.~T.~M.}\
  \bibnamefont {de~Haas}}\ and\ \bibinfo {author} {\bibfnamefont
  {P.}~\bibnamefont {Dorenbos}},\ }\href {\doibase 10.1109/TNS.2008.922819}
  {\bibfield  {journal} {\bibinfo  {journal} {IEEE Transactions on Nuclear
  Science}\ }\textbf {\bibinfo {volume} {55}},\ \bibinfo {pages} {1086}
  (\bibinfo {year} {2008})}\BibitemShut {NoStop}%
\bibitem [{\citenamefont {of~Standards}\ and\ \citenamefont {(NIST)}()}]{NIST}%
  \BibitemOpen
  \bibfield  {author} {\bibinfo {author} {\bibfnamefont {T.~N.~I.}\
  \bibnamefont {of~Standards}}\ and\ \bibinfo {author} {\bibfnamefont
  {T.}~\bibnamefont {(NIST)}},\ }\href@noop {} {\enquote {\bibinfo {title}
  {Handbook of basic atomic spectroscopic data},}\ }\BibitemShut {NoStop}%
\bibitem [{\citenamefont {Moszynski}\ \emph
  {et~al.}(2005{\natexlab{b}})\citenamefont {Moszynski}, \citenamefont
  {Balcerzyk}, \citenamefont {Czarnacki}, \citenamefont {Nassalski},
  \citenamefont {Szczesniak}, \citenamefont {Kraus}, \citenamefont
  {Mikhailik},\ and\ \citenamefont {Solskii}}]{Moszynski}%
  \BibitemOpen
  \bibfield  {author} {\bibinfo {author} {\bibfnamefont {M.}~\bibnamefont
  {Moszynski}}, \bibinfo {author} {\bibfnamefont {M.}~\bibnamefont
  {Balcerzyk}}, \bibinfo {author} {\bibfnamefont {W.}~\bibnamefont
  {Czarnacki}}, \bibinfo {author} {\bibfnamefont {A.}~\bibnamefont
  {Nassalski}}, \bibinfo {author} {\bibfnamefont {T.}~\bibnamefont
  {Szczesniak}}, \bibinfo {author} {\bibfnamefont {H.}~\bibnamefont {Kraus}},
  \bibinfo {author} {\bibfnamefont {V.~B.}\ \bibnamefont {Mikhailik}}, \ and\
  \bibinfo {author} {\bibfnamefont {I.~M.}\ \bibnamefont {Solskii}},\
  }\href@noop {} {\bibfield  {journal} {\bibinfo  {journal} {Nuclear
  Instruments \& Methods in Physics Research Section A-Accelerators
  Spectrometers Detectors and Associated Equipment}\ }\textbf {\bibinfo
  {volume} {553}},\ \bibinfo {pages} {578} (\bibinfo {year}
  {2005}{\natexlab{b}})}\BibitemShut {NoStop}%
\bibitem [{\citenamefont {Angloher}\ \emph {et~al.}(2009)\citenamefont
  {Angloher} \emph {et~al.}}]{Angloher:2008jj}%
  \BibitemOpen
  \bibfield  {author} {\bibinfo {author} {\bibfnamefont {G.}~\bibnamefont
  {Angloher}} \emph {et~al.},\ }\href {\doibase
  10.1016/j.astropartphys.2009.02.007} {\bibfield  {journal} {\bibinfo
  {journal} {Astropart. Phys.}\ }\textbf {\bibinfo {volume} {31}},\ \bibinfo
  {pages} {270} (\bibinfo {year} {2009})},\ \Eprint
  {http://arxiv.org/abs/0809.1829} {arXiv:0809.1829 [astro-ph]} \BibitemShut
  {NoStop}%
\bibitem [{\citenamefont {Akerib}\ \emph {et~al.}(2016)\citenamefont {Akerib}
  \emph {et~al.}}]{Akerib:2016lao}%
  \BibitemOpen
  \bibfield  {author} {\bibinfo {author} {\bibfnamefont {D.~S.}\ \bibnamefont
  {Akerib}} \emph {et~al.} (\bibinfo {collaboration} {LUX}),\ }\href {\doibase
  10.1103/PhysRevLett.116.161302} {\bibfield  {journal} {\bibinfo  {journal}
  {Phys. Rev. Lett.}\ }\textbf {\bibinfo {volume} {116}},\ \bibinfo {pages}
  {161302} (\bibinfo {year} {2016})},\ \Eprint
  {http://arxiv.org/abs/1602.03489} {arXiv:1602.03489 [hep-ex]} \BibitemShut
  {NoStop}%
\bibitem [{\citenamefont {Abe}\ \emph {et~al.}(2013)\citenamefont {Abe} \emph
  {et~al.}}]{Abe:2013tc}%
  \BibitemOpen
  \bibfield  {author} {\bibinfo {author} {\bibfnamefont {K.}~\bibnamefont
  {Abe}} \emph {et~al.},\ }\href {\doibase 10.1016/j.nima.2013.03.059}
  {\bibfield  {journal} {\bibinfo  {journal} {Nucl. Instrum. Meth.}\ }\textbf
  {\bibinfo {volume} {A716}},\ \bibinfo {pages} {78} (\bibinfo {year}
  {2013})},\ \Eprint {http://arxiv.org/abs/1301.2815} {arXiv:1301.2815
  [physics.ins-det]} \BibitemShut {NoStop}%
\bibitem [{\citenamefont {Aprile}\ \emph
  {et~al.}(2016{\natexlab{b}})\citenamefont {Aprile} \emph
  {et~al.}}]{Aprile:2015uzo}%
  \BibitemOpen
  \bibfield  {author} {\bibinfo {author} {\bibfnamefont {E.}~\bibnamefont
  {Aprile}} \emph {et~al.} (\bibinfo {collaboration} {XENON}),\ }\href
  {\doibase 10.1088/1475-7516/2016/04/027} {\bibfield  {journal} {\bibinfo
  {journal} {JCAP}\ }\textbf {\bibinfo {volume} {1604}},\ \bibinfo {pages}
  {027} (\bibinfo {year} {2016}{\natexlab{b}})},\ \Eprint
  {http://arxiv.org/abs/1512.07501} {arXiv:1512.07501 [physics.ins-det]}
  \BibitemShut {NoStop}%
\bibitem [{\citenamefont {Akerib}\ \emph {et~al.}(2015)\citenamefont {Akerib}
  \emph {et~al.}}]{Akerib:2015cja}%
  \BibitemOpen
  \bibfield  {author} {\bibinfo {author} {\bibfnamefont {D.~S.}\ \bibnamefont
  {Akerib}} \emph {et~al.} (\bibinfo {collaboration} {LZ}),\ }\href@noop {} {\
  (\bibinfo {year} {2015})},\ \Eprint {http://arxiv.org/abs/1509.02910}
  {arXiv:1509.02910 [physics.ins-det]} \BibitemShut {NoStop}%
\bibitem [{\citenamefont {Aalseth}\ \emph {et~al.}(2015)\citenamefont {Aalseth}
  \emph {et~al.}}]{Aalseth:2015mba}%
  \BibitemOpen
  \bibfield  {author} {\bibinfo {author} {\bibfnamefont {C.~E.}\ \bibnamefont
  {Aalseth}} \emph {et~al.},\ }\href {\doibase 10.1155/2015/541362} {\bibfield
  {journal} {\bibinfo  {journal} {Adv. High Energy Phys.}\ }\textbf {\bibinfo
  {volume} {2015}},\ \bibinfo {pages} {541362} (\bibinfo {year}
  {2015})}\BibitemShut {NoStop}%
\bibitem [{\citenamefont {Bernabei}\ \emph {et~al.}(2010)\citenamefont
  {Bernabei} \emph {et~al.}}]{Bernabei:2010mq}%
  \BibitemOpen
  \bibfield  {author} {\bibinfo {author} {\bibfnamefont {R.}~\bibnamefont
  {Bernabei}} \emph {et~al.} (\bibinfo {collaboration} {DAMA, LIBRA}),\ }\href
  {\doibase 10.1140/epjc/s10052-010-1303-9} {\bibfield  {journal} {\bibinfo
  {journal} {Eur. Phys. J.}\ }\textbf {\bibinfo {volume} {C67}},\ \bibinfo
  {pages} {39} (\bibinfo {year} {2010})},\ \Eprint
  {http://arxiv.org/abs/1002.1028} {arXiv:1002.1028 [astro-ph.GA]} \BibitemShut
  {NoStop}%
\bibitem [{\citenamefont {Froborg}(2016)}]{Froborg:2016ova}%
  \BibitemOpen
  \bibfield  {author} {\bibinfo {author} {\bibfnamefont {F.}~\bibnamefont
  {Froborg}} (\bibinfo {collaboration} {SABRE}),\ }in\ \href
  {https://inspirehep.net/record/1416177/files/arXiv:1601.05307.pdf} {\emph
  {\bibinfo {booktitle} {{14th International Conference on Topics in
  Astroparticle and Underground Physics (TAUP 2015) Torino, Italy, September
  7-11, 2015}}}}\ (\bibinfo {year} {2016})\ \Eprint
  {http://arxiv.org/abs/1601.05307} {arXiv:1601.05307 [physics.ins-det]}
  \BibitemShut {NoStop}%
\bibitem [{\citenamefont {Barbosa~de Souza}\ \emph {et~al.}(2016)\citenamefont
  {Barbosa~de Souza} \emph {et~al.}}]{deSouza:2016fxg}%
  \BibitemOpen
  \bibfield  {author} {\bibinfo {author} {\bibfnamefont {E.}~\bibnamefont
  {Barbosa~de Souza}} \emph {et~al.} (\bibinfo {collaboration} {DM-Ice}),\
  }\href@noop {} {\bibfield  {journal} {\bibinfo  {journal} {Submitted to:
  Phys. Rev. Lett.}\ } (\bibinfo {year} {2016})},\ \Eprint
  {http://arxiv.org/abs/1602.05939} {arXiv:1602.05939 [physics.ins-det]}
  \BibitemShut {NoStop}%
\bibitem [{\citenamefont {Lee}\ \emph {et~al.}(2007)\citenamefont {Lee} \emph
  {et~al.}}]{Lee.:2007qn}%
  \BibitemOpen
  \bibfield  {author} {\bibinfo {author} {\bibfnamefont {H.~S.}\ \bibnamefont
  {Lee}} \emph {et~al.} (\bibinfo {collaboration} {KIMS}),\ }\href {\doibase
  10.1103/PhysRevLett.99.091301} {\bibfield  {journal} {\bibinfo  {journal}
  {Phys. Rev. Lett.}\ }\textbf {\bibinfo {volume} {99}},\ \bibinfo {pages}
  {091301} (\bibinfo {year} {2007})},\ \Eprint {http://arxiv.org/abs/0704.0423}
  {arXiv:0704.0423 [astro-ph]} \BibitemShut {NoStop}%
\bibitem [{Note3()}]{Note3}%
  \BibitemOpen
  \bibinfo {note} {DM-electron scattering in e.g.~xenon TPCs could produce two
  photons in a multi-step de-excitation process. However the efficiency to
  detect a photon is low (e.g. $\sim 10\%$ in LUX). Moreover, the PMTs are not
  sensitive to the second photon, which is in the infrared.}\BibitemShut
  {Stop}%
\bibitem [{\citenamefont {Derenzo}\ \emph {et~al.}(2006)\citenamefont
  {Derenzo}, \citenamefont {Bourret-Courchesne}, \citenamefont {James},
  \citenamefont {Klintenberg}, \citenamefont {Porter-Chapman}, \citenamefont
  {Wang},\ and\ \citenamefont {Weber}}]{Derenzo}%
  \BibitemOpen
  \bibfield  {author} {\bibinfo {author} {\bibfnamefont {S.~E.}\ \bibnamefont
  {Derenzo}}, \bibinfo {author} {\bibfnamefont {E.}~\bibnamefont
  {Bourret-Courchesne}}, \bibinfo {author} {\bibfnamefont {F.~J.}\ \bibnamefont
  {James}}, \bibinfo {author} {\bibfnamefont {M.~K.}\ \bibnamefont
  {Klintenberg}}, \bibinfo {author} {\bibfnamefont {Y.}~\bibnamefont
  {Porter-Chapman}}, \bibinfo {author} {\bibfnamefont {J.}~\bibnamefont
  {Wang}}, \ and\ \bibinfo {author} {\bibfnamefont {M.~J.}\ \bibnamefont
  {Weber}},\ }\href@noop {} {\bibfield  {journal} {\bibinfo  {journal} {IEEE
  Nuclear Science Symposium Conference Record}\ ,\ \bibinfo {pages} {1132}}
  (\bibinfo {year} {2006})}\BibitemShut {NoStop}%
\bibitem [{\citenamefont {Steger}\ \emph {et~al.}(2011)\citenamefont {Steger},
  \citenamefont {Yang}, \citenamefont {Sekiguchi}, \citenamefont {Saeedi},
  \citenamefont {Thewalt}, \citenamefont {Henry}, \citenamefont {Johnston},
  \citenamefont {Riemann}, \citenamefont {Abrosimov}, \citenamefont
  {Churbanov}, \citenamefont {Gusev}, \citenamefont {Kaliteevskii},
  \citenamefont {Godisov}, \citenamefont {Becker},\ and\ \citenamefont
  {Pohl}}]{Steger}%
  \BibitemOpen
  \bibfield  {author} {\bibinfo {author} {\bibfnamefont {M.}~\bibnamefont
  {Steger}}, \bibinfo {author} {\bibfnamefont {A.}~\bibnamefont {Yang}},
  \bibinfo {author} {\bibfnamefont {T.}~\bibnamefont {Sekiguchi}}, \bibinfo
  {author} {\bibfnamefont {K.}~\bibnamefont {Saeedi}}, \bibinfo {author}
  {\bibfnamefont {M.~L.~W.}\ \bibnamefont {Thewalt}}, \bibinfo {author}
  {\bibfnamefont {M.~O.}\ \bibnamefont {Henry}}, \bibinfo {author}
  {\bibfnamefont {K.}~\bibnamefont {Johnston}}, \bibinfo {author}
  {\bibfnamefont {H.}~\bibnamefont {Riemann}}, \bibinfo {author} {\bibfnamefont
  {N.~V.}\ \bibnamefont {Abrosimov}}, \bibinfo {author} {\bibfnamefont {M.~F.}\
  \bibnamefont {Churbanov}}, \bibinfo {author} {\bibfnamefont {A.~V.}\
  \bibnamefont {Gusev}}, \bibinfo {author} {\bibfnamefont {A.~K.}\ \bibnamefont
  {Kaliteevskii}}, \bibinfo {author} {\bibfnamefont {O.~N.}\ \bibnamefont
  {Godisov}}, \bibinfo {author} {\bibfnamefont {P.}~\bibnamefont {Becker}}, \
  and\ \bibinfo {author} {\bibfnamefont {H.-J.}\ \bibnamefont {Pohl}},\ }\href
  {\doibase http://dx.doi.org/10.1063/1.3651774} {\bibfield  {journal}
  {\bibinfo  {journal} {Journal of Applied Physics}\ }\textbf {\bibinfo
  {volume} {110}},\ \bibinfo {eid} {081301} (\bibinfo {year} {2011}),\
  http://dx.doi.org/10.1063/1.3651774}\BibitemShut {NoStop}%
\bibitem [{\citenamefont {Davies}(1994)}]{Davies1994}%
  \BibitemOpen
  \bibfield  {author} {\bibinfo {author} {\bibfnamefont {G.}~\bibnamefont
  {Davies}},\ }\href {http://stacks.iop.org/1402-4896/1994/i=T54/a=001}
  {\bibfield  {journal} {\bibinfo  {journal} {Physica Scripta}\ }\textbf
  {\bibinfo {volume} {1994}},\ \bibinfo {pages} {7} (\bibinfo {year}
  {1994})}\BibitemShut {NoStop}%
\bibitem [{\citenamefont {Davies}\ \emph {et~al.}(1992)\citenamefont {Davies},
  \citenamefont {Lightowlers}, \citenamefont {Itoh}, \citenamefont {Hansen},
  \citenamefont {Haller},\ and\ \citenamefont {Ozhogin}}]{Davies1992}%
  \BibitemOpen
  \bibfield  {author} {\bibinfo {author} {\bibfnamefont {G.}~\bibnamefont
  {Davies}}, \bibinfo {author} {\bibfnamefont {E.~C.}\ \bibnamefont
  {Lightowlers}}, \bibinfo {author} {\bibfnamefont {K.}~\bibnamefont {Itoh}},
  \bibinfo {author} {\bibfnamefont {W.~L.}\ \bibnamefont {Hansen}}, \bibinfo
  {author} {\bibfnamefont {E.~E.}\ \bibnamefont {Haller}}, \ and\ \bibinfo
  {author} {\bibfnamefont {V.}~\bibnamefont {Ozhogin}},\ }\href
  {http://stacks.iop.org/0268-1242/7/i=10/a=010} {\bibfield  {journal}
  {\bibinfo  {journal} {Semiconductor Science and Technology}\ }\textbf
  {\bibinfo {volume} {7}},\ \bibinfo {pages} {1271} (\bibinfo {year}
  {1992})}\BibitemShut {NoStop}%
\bibitem [{\citenamefont {Perdew}\ \emph {et~al.}(1996)\citenamefont {Perdew},
  \citenamefont {Burke},\ and\ \citenamefont
  {Ernzerhof}}]{PhysRevLett.77.3865}%
  \BibitemOpen
  \bibfield  {author} {\bibinfo {author} {\bibfnamefont {J.~P.}\ \bibnamefont
  {Perdew}}, \bibinfo {author} {\bibfnamefont {K.}~\bibnamefont {Burke}}, \
  and\ \bibinfo {author} {\bibfnamefont {M.}~\bibnamefont {Ernzerhof}},\ }\href
  {\doibase 10.1103/PhysRevLett.77.3865} {\bibfield  {journal} {\bibinfo
  {journal} {Phys. Rev. Lett.}\ }\textbf {\bibinfo {volume} {77}},\ \bibinfo
  {pages} {3865} (\bibinfo {year} {1996})}\BibitemShut {NoStop}%
\bibitem [{\citenamefont {Hamann}\ \emph {et~al.}(1979)\citenamefont {Hamann},
  \citenamefont {Schl\"uter},\ and\ \citenamefont
  {Chiang}}]{PhysRevLett.43.1494}%
  \BibitemOpen
  \bibfield  {author} {\bibinfo {author} {\bibfnamefont {D.~R.}\ \bibnamefont
  {Hamann}}, \bibinfo {author} {\bibfnamefont {M.}~\bibnamefont {Schl\"uter}},
  \ and\ \bibinfo {author} {\bibfnamefont {C.}~\bibnamefont {Chiang}},\ }\href
  {\doibase 10.1103/PhysRevLett.43.1494} {\bibfield  {journal} {\bibinfo
  {journal} {Phys. Rev. Lett.}\ }\textbf {\bibinfo {volume} {43}},\ \bibinfo
  {pages} {1494} (\bibinfo {year} {1979})}\BibitemShut {NoStop}%
\bibitem [{\citenamefont {Levine}\ and\ \citenamefont
  {Allan}(1989)}]{PhysRevLett.63.1719}%
  \BibitemOpen
  \bibfield  {author} {\bibinfo {author} {\bibfnamefont {Z.}~\bibnamefont
  {Levine}}\ and\ \bibinfo {author} {\bibfnamefont {D.}~\bibnamefont {Allan}},\
  }\href {\doibase 10.1103/PhysRevLett.63.1719} {\bibfield  {journal} {\bibinfo
   {journal} {Phys. Rev. Lett.}\ }\textbf {\bibinfo {volume} {63}},\ \bibinfo
  {pages} {1719} (\bibinfo {year} {1989})}\BibitemShut {NoStop}%
\bibitem [{\citenamefont {Levine}\ and\ \citenamefont
  {Allan}(1991)}]{PhysRevB.43.4187}%
  \BibitemOpen
  \bibfield  {author} {\bibinfo {author} {\bibfnamefont {Z.}~\bibnamefont
  {Levine}}\ and\ \bibinfo {author} {\bibfnamefont {D.}~\bibnamefont {Allan}},\
  }\href {\doibase 10.1103/PhysRevB.43.4187} {\bibfield  {journal} {\bibinfo
  {journal} {Phys. Rev. B}\ }\textbf {\bibinfo {volume} {43}},\ \bibinfo
  {pages} {4187} (\bibinfo {year} {1991})}\BibitemShut {NoStop}%
\bibitem [{\citenamefont {{Klein}}(1968)}]{Klein:1968}%
  \BibitemOpen
  \bibfield  {author} {\bibinfo {author} {\bibfnamefont {C.~A.}\ \bibnamefont
  {{Klein}}},\ }\href {\doibase 10.1063/1.1656484} {\bibfield  {journal}
  {\bibinfo  {journal} {Journal of Applied Physics}\ }\textbf {\bibinfo
  {volume} {39}},\ \bibinfo {pages} {2029} (\bibinfo {year}
  {1968})}\BibitemShut {NoStop}%
\bibitem [{\citenamefont {Knoll}(2010)}]{Knoll:1300754}%
  \BibitemOpen
  \bibfield  {author} {\bibinfo {author} {\bibfnamefont {G.~F.}\ \bibnamefont
  {Knoll}},\ }\href {http://cds.cern.ch/record/1300754} {\emph {\bibinfo
  {title} {{Radiation detection and measurement; 4th ed.}}}}\ (\bibinfo
  {publisher} {Wiley},\ \bibinfo {address} {New York, NY},\ \bibinfo {year}
  {2010})\BibitemShut {NoStop}%
\bibitem [{\citenamefont {Knox}(1963)}]{knox1963theory}%
  \BibitemOpen
  \bibfield  {author} {\bibinfo {author} {\bibfnamefont {R.}~\bibnamefont
  {Knox}},\ }\href {https://books.google.com/books?id=GuQ9AQAAIAAJ} {\emph
  {\bibinfo {title} {Theory of excitons}}},\ Solid state physics: Supplement\
  (\bibinfo  {publisher} {Academic Press},\ \bibinfo {year} {1963})\BibitemShut
  {NoStop}%
\bibitem [{\citenamefont {Song}\ and\ \citenamefont
  {Williams}(1996)}]{song1996}%
  \BibitemOpen
  \bibfield  {author} {\bibinfo {author} {\bibfnamefont {A.}~\bibnamefont
  {Song}}\ and\ \bibinfo {author} {\bibfnamefont {R.}~\bibnamefont
  {Williams}},\ }\href {https://books.google.com/books?id=90QsAAAAYAAJ} {\emph
  {\bibinfo {title} {Self-trapped excitons}}},\ Springer series in solid-state
  sciences\ (\bibinfo  {publisher} {Springer},\ \bibinfo {year}
  {1996})\BibitemShut {NoStop}%
\bibitem [{\citenamefont {Cho}\ \emph {et~al.}(2012)\citenamefont {Cho},
  \citenamefont {Dean}, \citenamefont {Fischer}, \citenamefont {Herbert},
  \citenamefont {Lagois},\ and\ \citenamefont {Yu}}]{cho2012}%
  \BibitemOpen
  \bibfield  {author} {\bibinfo {author} {\bibfnamefont {K.}~\bibnamefont
  {Cho}}, \bibinfo {author} {\bibfnamefont {P.}~\bibnamefont {Dean}}, \bibinfo
  {author} {\bibfnamefont {B.}~\bibnamefont {Fischer}}, \bibinfo {author}
  {\bibfnamefont {D.}~\bibnamefont {Herbert}}, \bibinfo {author} {\bibfnamefont
  {J.}~\bibnamefont {Lagois}}, \ and\ \bibinfo {author} {\bibfnamefont
  {P.}~\bibnamefont {Yu}},\ }\href
  {https://books.google.com/books?id=WyLrCAAAQBAJ} {\emph {\bibinfo {title}
  {Excitons}}},\ Topics in Current Physics\ (\bibinfo  {publisher} {Springer
  Berlin Heidelberg},\ \bibinfo {year} {2012})\BibitemShut {NoStop}%
\bibitem [{\citenamefont {Kubo}\ and\ \citenamefont
  {Hanamura}(2012)}]{kubo2012}%
  \BibitemOpen
  \bibfield  {author} {\bibinfo {author} {\bibfnamefont {R.}~\bibnamefont
  {Kubo}}\ and\ \bibinfo {author} {\bibfnamefont {E.}~\bibnamefont
  {Hanamura}},\ }\href {https://books.google.com/books?id=ISrwCAAAQBAJ} {\emph
  {\bibinfo {title} {Relaxation of Elementary Excitations: Proceedings of the
  Taniguchi International Symposium, Susono-shi, Japan, October 12--16,
  1979}}},\ Springer Series in Solid-State Sciences\ (\bibinfo  {publisher}
  {Springer Berlin Heidelberg},\ \bibinfo {year} {2012})\BibitemShut {NoStop}%
\bibitem [{Alm()}]{AlmazOptics}%
  \BibitemOpen
  \href@noop {} {\enquote {\bibinfo {title} {Almaz optics},}\ }\bibinfo
  {howpublished} {\url{http://www.almazoptics.com}},\ \bibinfo {note}
  {accessed: 2016-06-16}\BibitemShut {NoStop}%
\bibitem [{\citenamefont {Setyawan}\ \emph {et~al.}(2009)\citenamefont
  {Setyawan}, \citenamefont {Gaume}, \citenamefont {Feigelson},\ and\
  \citenamefont {Curtarolo}}]{5280482}%
  \BibitemOpen
  \bibfield  {author} {\bibinfo {author} {\bibfnamefont {W.}~\bibnamefont
  {Setyawan}}, \bibinfo {author} {\bibfnamefont {R.~M.}\ \bibnamefont {Gaume}},
  \bibinfo {author} {\bibfnamefont {R.~S.}\ \bibnamefont {Feigelson}}, \ and\
  \bibinfo {author} {\bibfnamefont {S.}~\bibnamefont {Curtarolo}},\ }\href
  {\doibase 10.1109/TNS.2009.2027019} {\bibfield  {journal} {\bibinfo
  {journal} {IEEE Transactions on Nuclear Science}\ }\textbf {\bibinfo {volume}
  {56}},\ \bibinfo {pages} {2989} (\bibinfo {year} {2009})}\BibitemShut
  {NoStop}%
\bibitem [{\citenamefont {Kamiyoshi}\ and\ \citenamefont
  {Nigara}(1970)}]{PSSA:PSSA19700030320}%
  \BibitemOpen
  \bibfield  {author} {\bibinfo {author} {\bibfnamefont {K.}~\bibnamefont
  {Kamiyoshi}}\ and\ \bibinfo {author} {\bibfnamefont {Y.}~\bibnamefont
  {Nigara}},\ }\href {\doibase 10.1002/pssa.19700030320} {\bibfield  {journal}
  {\bibinfo  {journal} {physica status solidi (a)}\ }\textbf {\bibinfo {volume}
  {3}},\ \bibinfo {pages} {735} (\bibinfo {year} {1970})}\BibitemShut {NoStop}%
\bibitem [{\citenamefont {Singh}(1992)}]{Singh1993}%
  \BibitemOpen
  \bibfield  {author} {\bibinfo {author} {\bibfnamefont {J.}~\bibnamefont
  {Singh}},\ }\href@noop {} {\emph {\bibinfo {title} {Physics of semiconductors
  and their heterostructures}}}\ (\bibinfo  {publisher} {McGraw-Hill},\
  \bibinfo {year} {1992})\BibitemShut {NoStop}%
\end{thebibliography}%

\clearpage

\onecolumngrid

\begin{center}
{\large{\bf Supplemental Material: \\ \vskip 1mm
Direct Detection of sub-GeV Dark Matter with Scintillating Targets}}
\end{center}

\vskip 0.5cm

\twocolumngrid

In these supplementary materials, we provide a few more details that are not essential for understanding the letter.  
In particular, we discuss the scintillation mechanisms of various materials mentioned in Table~\ref{tab:materials}, as well as give a brief discussion on whether the effect of excitons should be included in the calculation of the DM-electron scattering rates.  
For completeness, we also provide plots showing our calculated band structures and density of states for the five elements shown in Fig.~\ref{fig:reach}, as well as the recoil spectra for GaAs, NaI, and CsI. 

\vskip -4mm
\section{BRIEF REVIEW OF SCINTILLATION MECHANISMS} \vskip -3mm

\begin{figure}[b]
\includegraphics[width=0.38\textwidth]{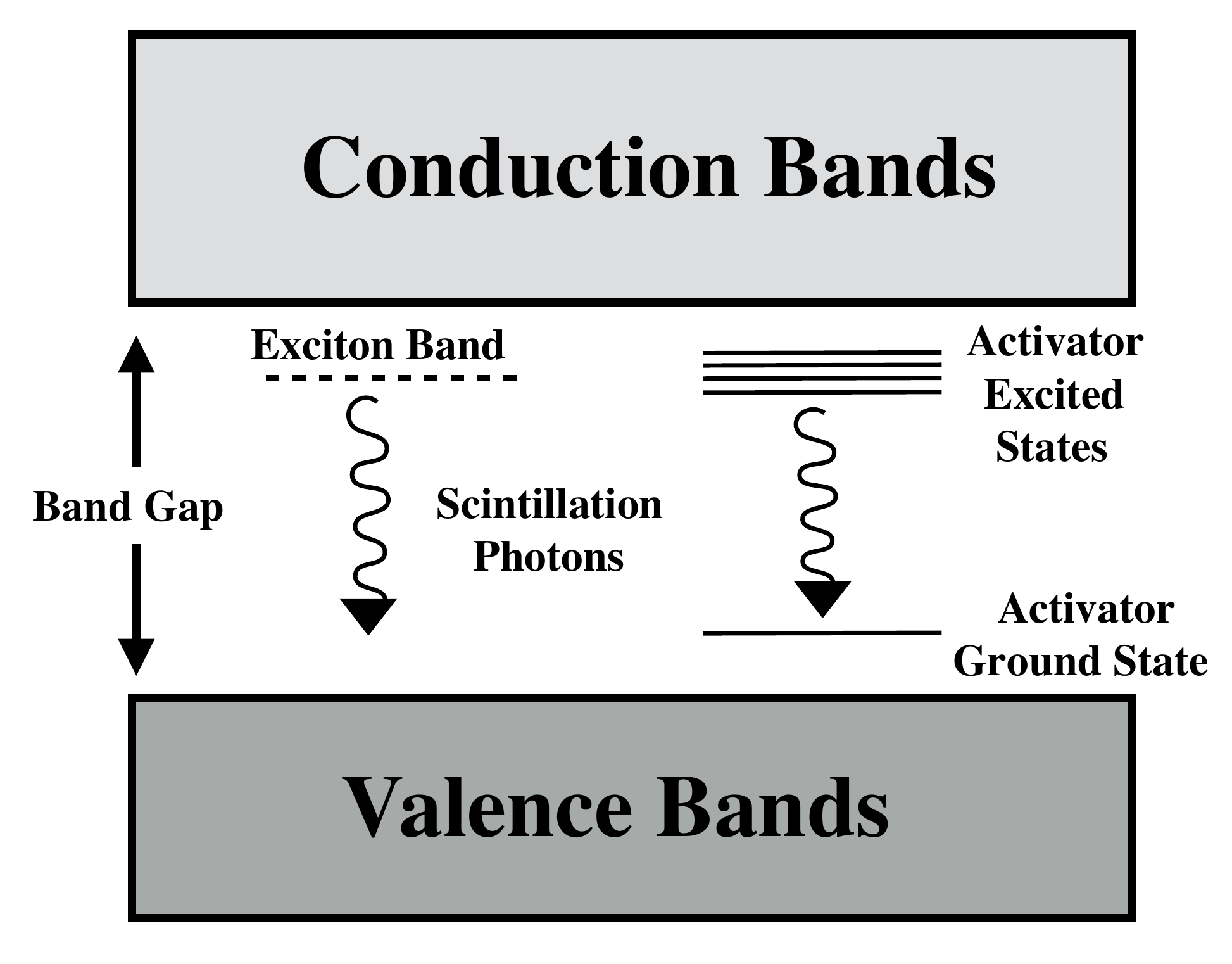}
\caption{Illustration of the different mechanisms for light emission in a scintillating crystal.}
\label{fig:scintillator}
\end{figure}

We review briefly the scintillation mechanisms of the materials listed in Table~\ref{tab:materials} of the letter.  
In general, for a material to be a scintillator, it must contain luminescent centers. These centers can be either extrinsic (e.g.~dopants and 
impurities) or intrinsic (e.g.~defects of the lattice or excitons), and give rise to a transition between a higher- and a lower-energy state. 
Moreover, the energy levels involved in the transition must be contained in a forbidden energy region (e.g.~the band gap for semiconductors and insulators, or excimer states in gases) to avoid re-absorption of the emitted light or photoionization of the center (Fig.~\ref{fig:scintillator}).

Pure CsI and NaI at cryogenic temperatures scintillate via the formation of self-trapped excitons, where an exciton (an electron-hole bound state) becomes self-trapped by deforming the lattice structure around it.  At cryogenic temperatures the system lies at the minimum energy in lattice configuration space, and the system can only return to the ground state by emission of a photon.  At higher temperatures, thermally induced lattice vibrations allow the system to return to the ground state by phonon emission resulting in a low radiative efficiency.  At room temperature, this thermal quenching is over- come by doping the material with e.g. thallium. In these cases, 
Tl$^+$ traps the excitons and provides an efficient luminescence center.

Direct-gap semiconductors, like GaAs, have the advantage that an excited electron can recombine with a hole without requiring a change in crystal momentum. In practice, however, dopants are used to enhance the radiative quantum efficiency, by providing radiative centers, and to reduce non-radiative recombination from impurities and native defects.

Indirect-gap semiconductors, like Si and Ge, require dopants to allow radiative recombination at cryogenic temperatures through the formation of a bound exciton that can radiate without the need for a change in crystal momentum.

Plastic scintillators consist of a base polymer that contains delocalized $\pi$-orbital electrons and a small concentration of fluorescent molecules. Excited $\pi$-orbital electrons will diffuse through the base polymer and excite fluorescent molecules. 
These excitations have radiative lifetimes of $1-2$~nanoseconds. This process is efficient both at room and cryogenic temperatures.

In tungstate scintillators, valence-band electrons on the oxygen ions can be excited to conduction band states on the tungsten ions.  
In PbWO$_4$, the excited state is thermally quenched so that at room temperature the luminosity is low and the decay time is short. 
CaWO$_4$ and CdWO$_4$ are more efficient at room temperature and their decay times are $\sim$10~microseconds.

\vskip -4mm
\section{EFFECT OF EXCITONS ON DARK MATTER-ELECTRON SCATTERING-RATE CALCULATION} \vskip -3mm

Our calculation of the DM-electron scattering rate neglects the effect of {\it excitons}.  
In this section, we discuss why we expect this to be a good approximation for the low-band-gap materials (Ge, Si, and GaAs),  but that 
there may be an $\mathcal{O}(1)$ correction for the large-band-gap insulators (NaI and CsI). 

Semiconductors or insulating crystals are characterized by a finite band gap, $E_g$, between the top of the valence band and the bottom of the conduction band.  
These bands form an energy continuum for the excitation of an electron from the valence to 
the conduction band, which can be viewed as the creation of a free-electron-free-hole pair. 
In our calculation of the DM-electron scattering rate, we included the contribution of this continuum of states.

The small electrostatic Coulomb attraction between the negatively charged electron and positively charged hole creates an 
exciton, a bound electron-hole pair (see e.g.~\cite{knox1963theory,song1996,cho2012,kubo2012} and references therein).  
As we will see below, this Coulomb-bound electron-hole pair can be modeled with Rydberg-like states with energies $E_g-E_{B,n}$, where $E_{B,n}$ is the binding energy and $n$ labels the Rydberg-like energy level.  
The energy of these excitons is therefore in the ``forbidden'' band-gap region, so that the density of states is nonzero even at energies slightly below the conduction band.  
Moreover, the bound electron-hole pair has ionized states with a continuous energy due to their relative motion.  
It turns out that excitons therefore also moderately increase the density of states just above the band gap compared to a calculation that 
neglects them. 
Including exciton effects in the DM-electron scattering-rate calculation could thus be important for two reasons. First, a nonzero density of states below the band gap means that the actual mass threshold is slightly lower. Second, any calculation that neglects exciton effects might underestimate slightly the scattering rate. 

Excitons are extensively studied in solid state physics and play an important role in determining the properties of various materials.  
For example, it is well known that excitons are crucial in understanding the spectrum for the absorption of light, as they allow for photons with an energy just below $E_g$ to be absorbed by an electron. Similarly, excitons can play an essential role in determining the scintillation properties of a material. For example, an electron excited from the valence to the conduction band can quickly relax to the bottom of the conduction band and then into an exciton state by emitting phonons.  
The radiative decay of the exciton then yields a photon whose energy is just below that of the band gap. This typically allows the photon to traverse the material without being absorbed again, i.e.~the material scintillates.  

\begin{table}[t!]
\begin{center}
\begin{tabular}{|l|c|c|c|c|c|}
\hline
     & $\varepsilon$ & $m^*_e/m_e$ & $m^*_h/m_e$ & $\Delta E_{B,n=1}$ [eV] & $a_{n=1}/a$ \\ \hline
CsI~\cite{AlmazOptics,5280482}    & 5.65 &   0.312  & 2.270  & 0.117 & 2.37 \\ \hline
NaI~\cite{PSSA:PSSA19700030320,5280482}    &   7.28  &   0.287   &  2.397  &  0.066 & 2.20\\ \hline
GaAs~\cite{AlmazOptics,Singh1993} &  12.85 & 0.067  &  0.45 & 0.005 & 20.3        \\ \hline 
Ge~\cite{AlmazOptics,Singh1993}     &  16 &   0.2      &  0.28  & 0.006   &  12.7 \\ \hline
Si~\cite{AlmazOptics,Singh1993}       &   13 &  0.33  & 0.49 &  0.016 & 6.38 \\ \hline
\end{tabular}
\caption{\label{tab:exciton-info}
Dielectric constant ($\varepsilon$), effective electron mass ($m^*_e$), effective hole mass ($m^*_h$), 1s-exciton binding energy, 
and 1s-exciton radius (in units of the lattice constant $a$ in Table~\ref{tab:qedark-parameters}) for various materials. 
\vspace{-5mm}}
\end{center}
\end{table}

\begin{figure*}[t]
\vspace{4mm}
\includegraphics[width=0.43\textwidth]{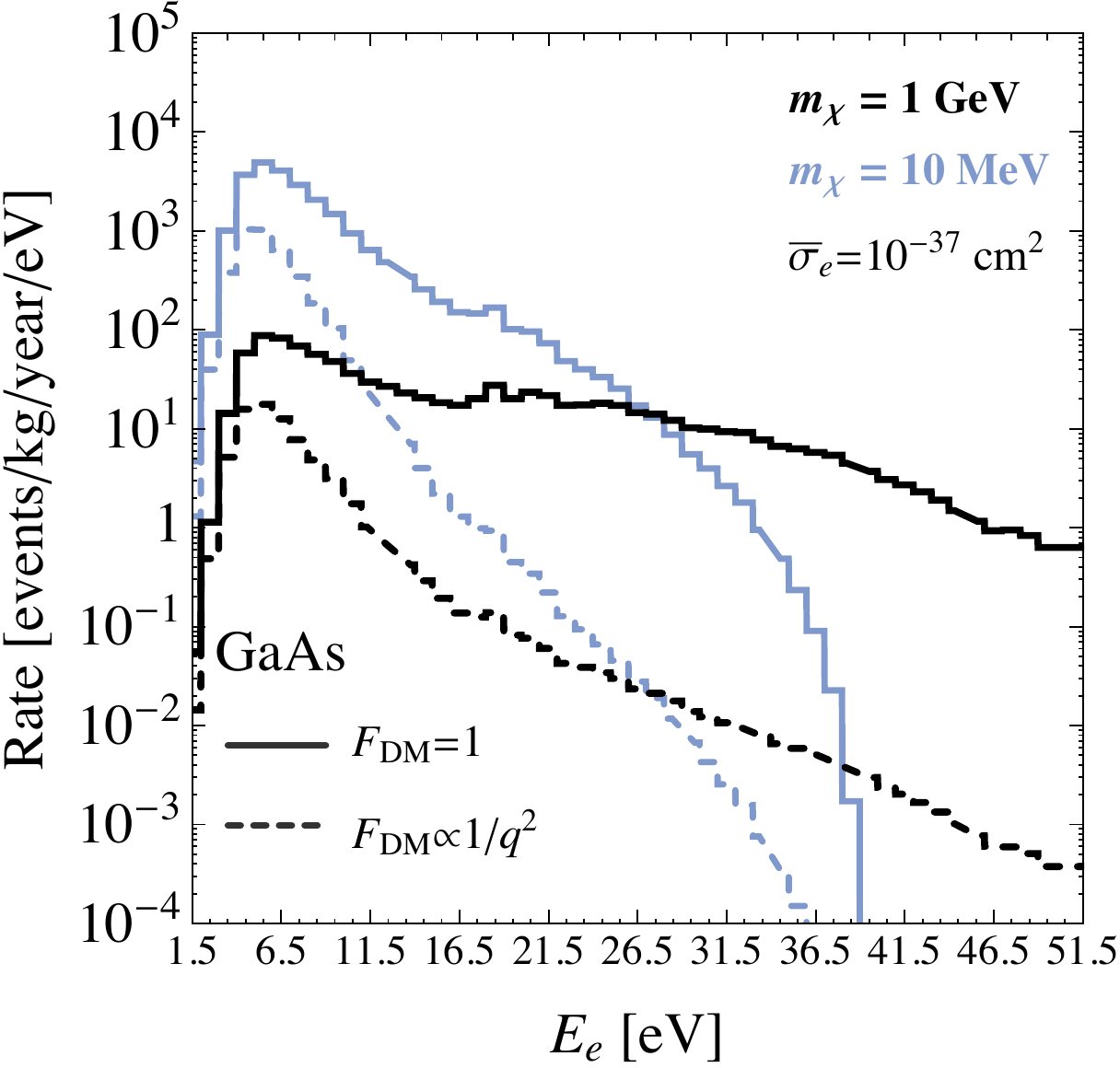}
~~~~\includegraphics[width=0.43\textwidth]{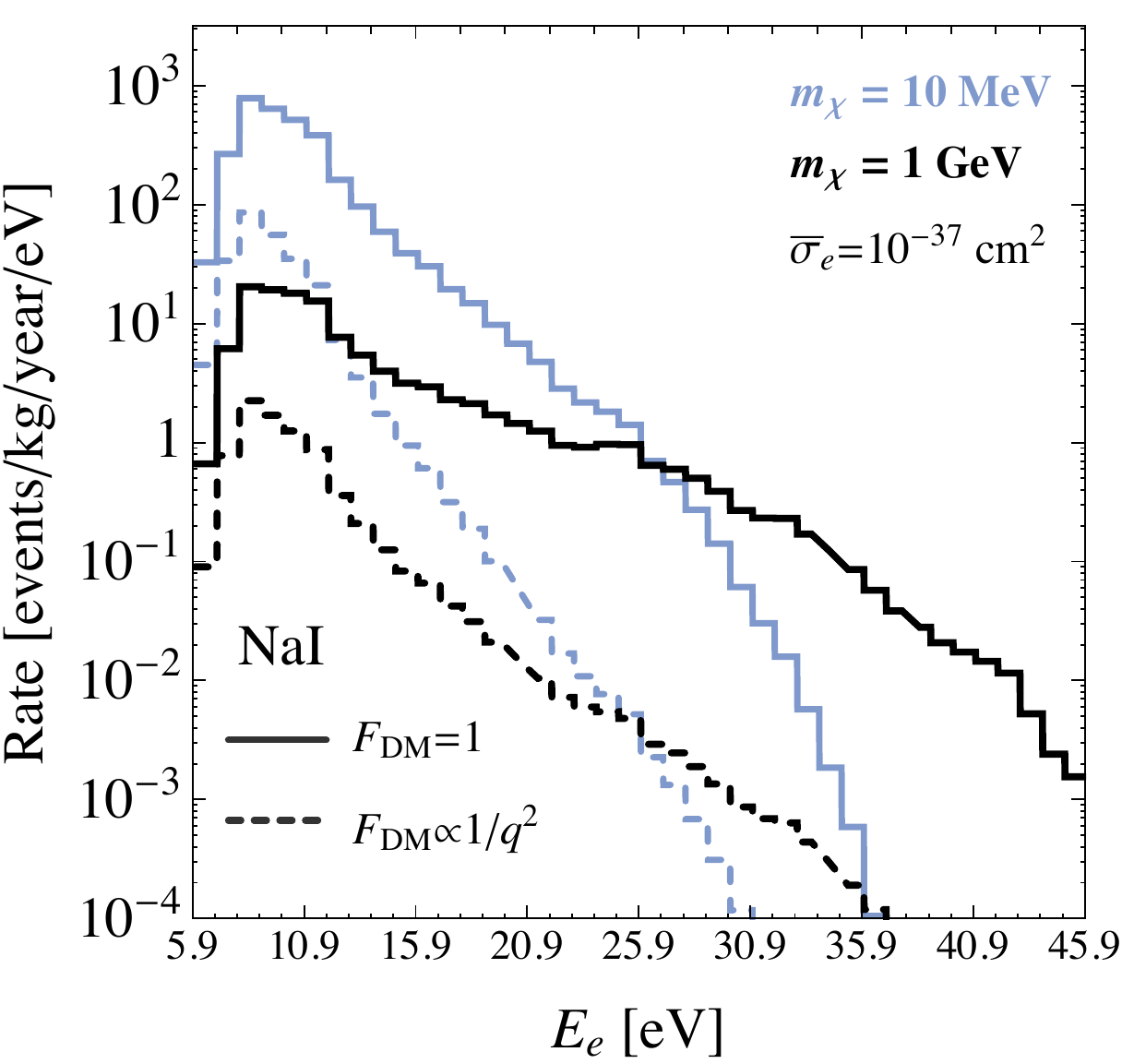}\\
\includegraphics[width=0.43\textwidth]{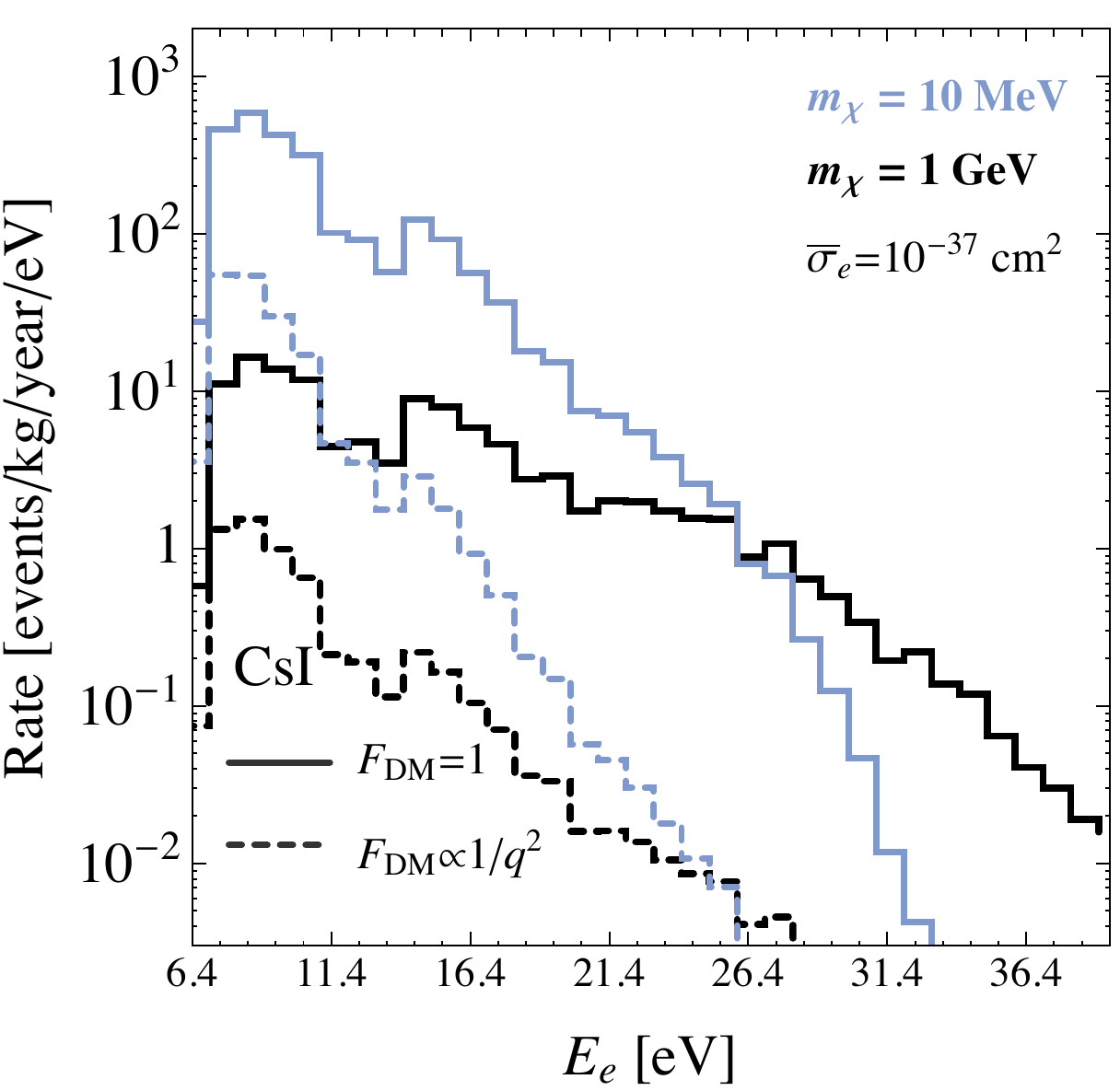}
~~~~\includegraphics[width=0.43\textwidth]{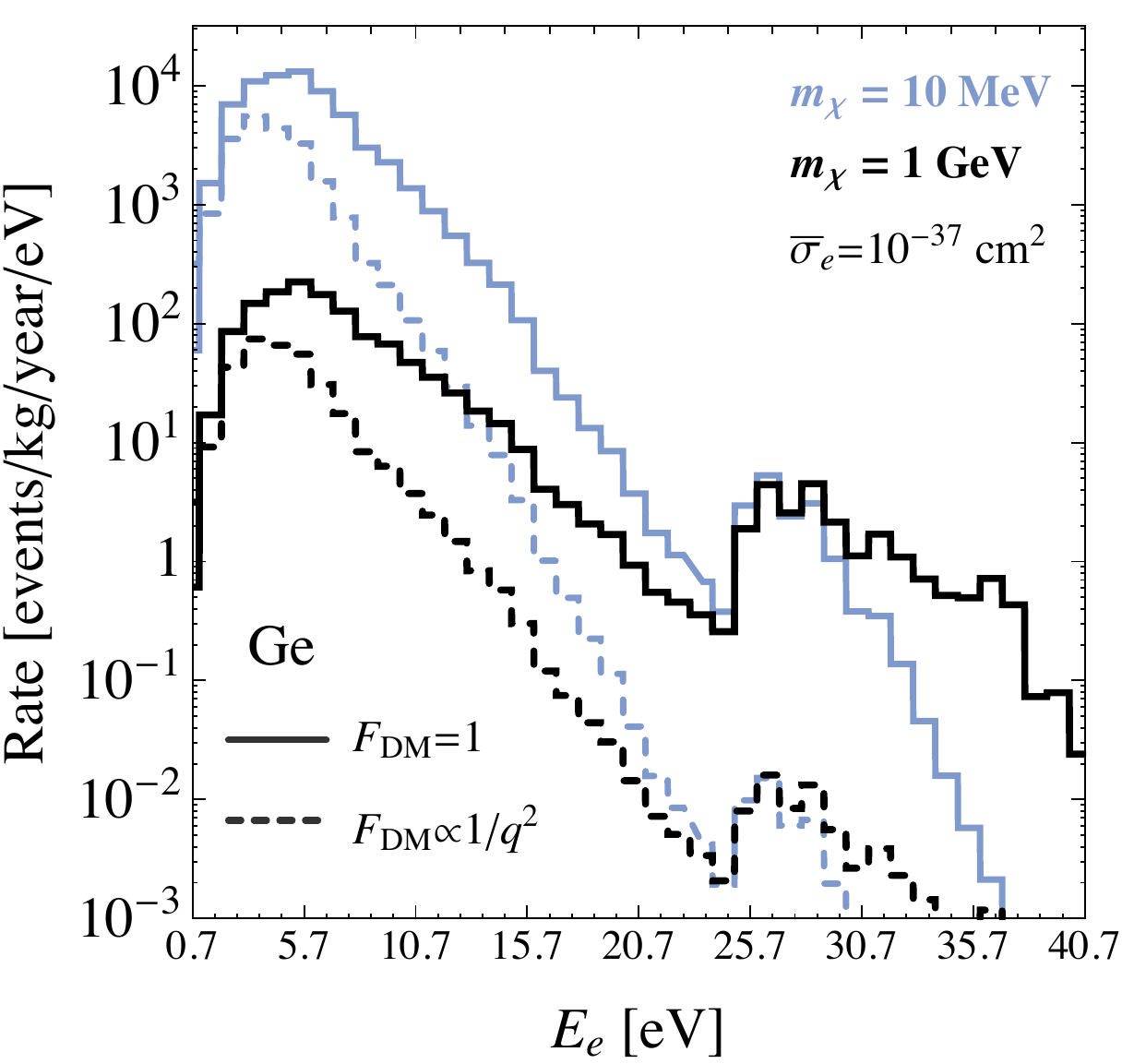}\\%
\begin{minipage}[t]{0.5\textwidth}
\mbox{}\\[-0.1\baselineskip]
~~~~~~~\includegraphics[width=0.9\textwidth]{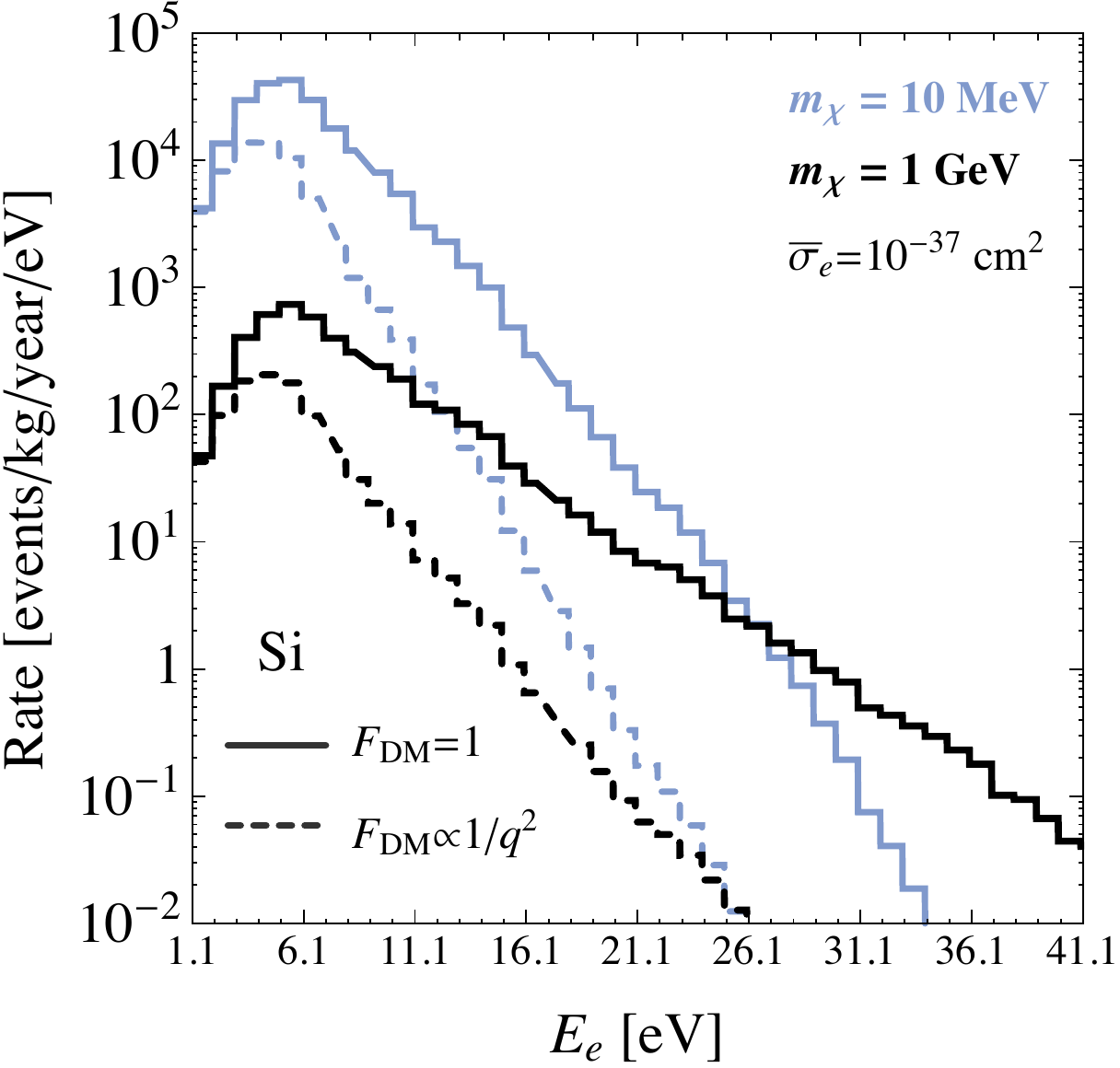}
\end{minipage}
\hfill
\begin{minipage}[t]{0.45\textwidth}
\mbox{}\\[0.3\baselineskip]
\caption{Electron recoil spectra from DM-electron scattering in GaAs, NaI, CsI, Ge, and Si 
as a function of total deposited energy $E_e$, for $m_\chi=10$~MeV (blue lines) and 1~GeV (black lines) and 
DM form factors $F_{\rm DM}=1$ (solid lines) and $F_{\rm DM}=(\alpha m_e/q)^2$ (dashed lines). 
We fix $\overline\sigma_e=10^{-37}\rm{cm}^2$ and assume an exposure of 1~kg-year.  
The $E_e$-axis begins at the band-gap energies $E_g$.
\label{fig:spectra}
}
\end{minipage}
\end{figure*}

\begin{figure*}[t]
\vspace{4mm}
\includegraphics[width=0.48\textwidth]{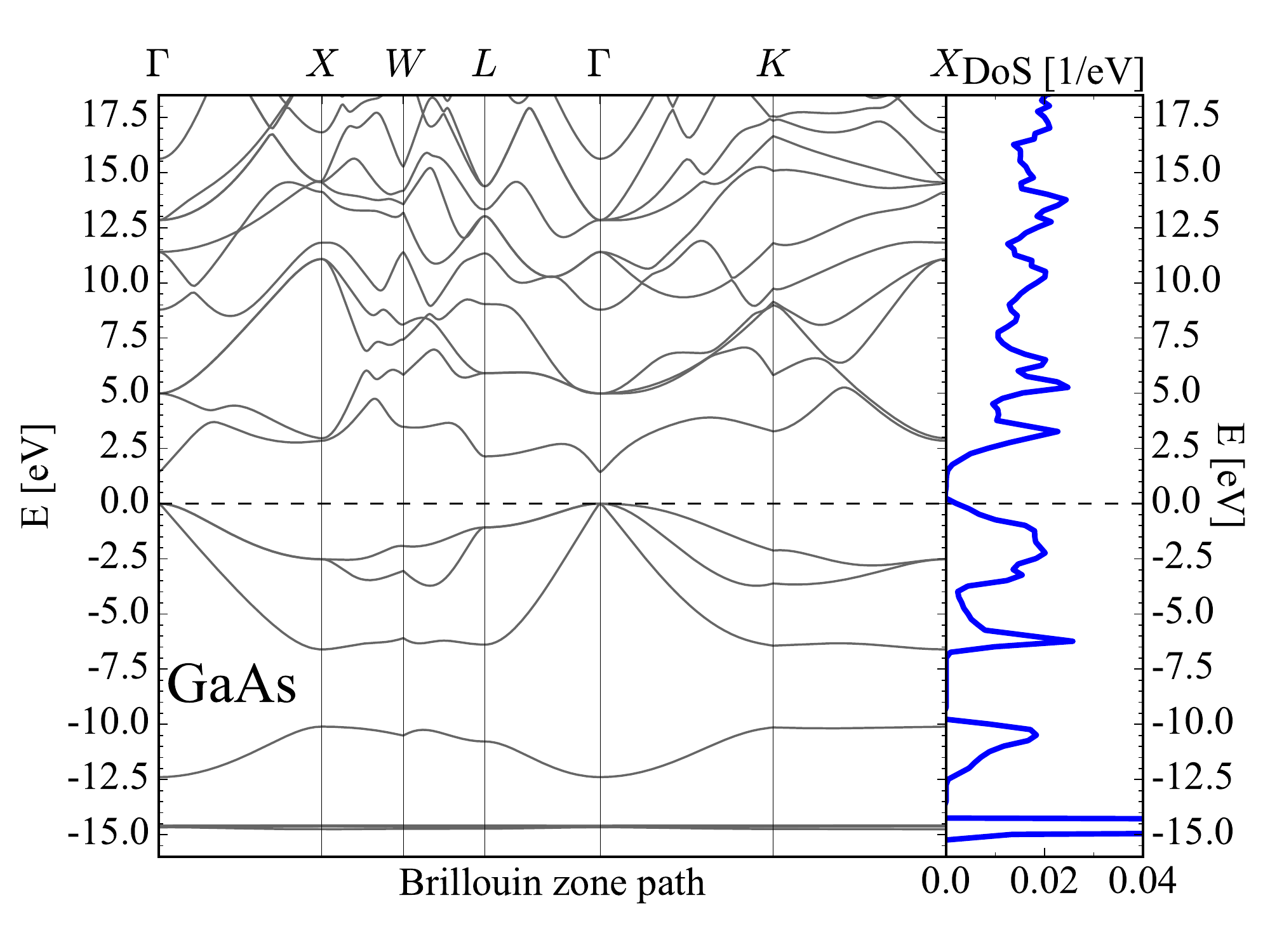}
\includegraphics[width=0.48\textwidth]{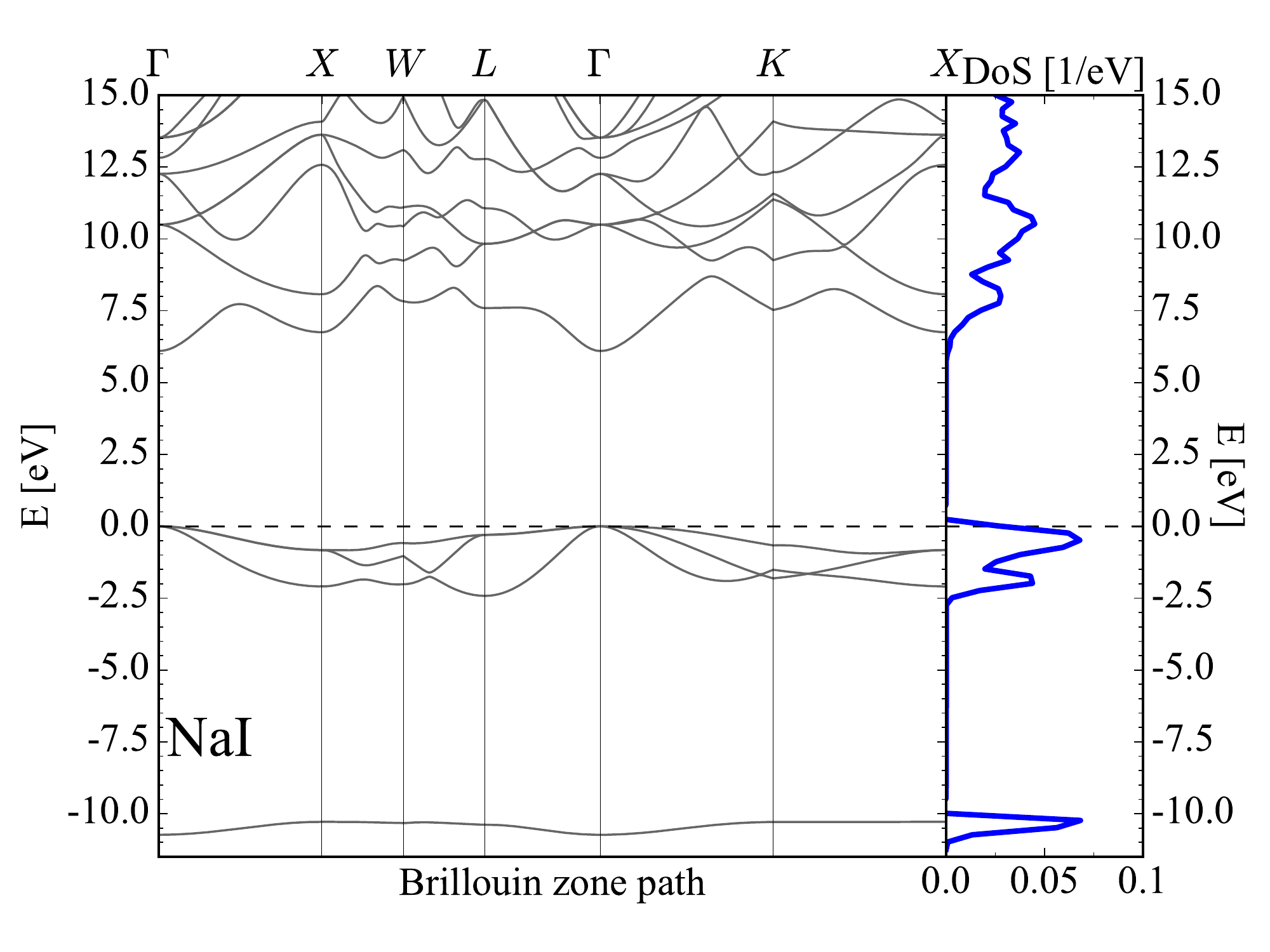}\\
\includegraphics[width=0.48\textwidth]{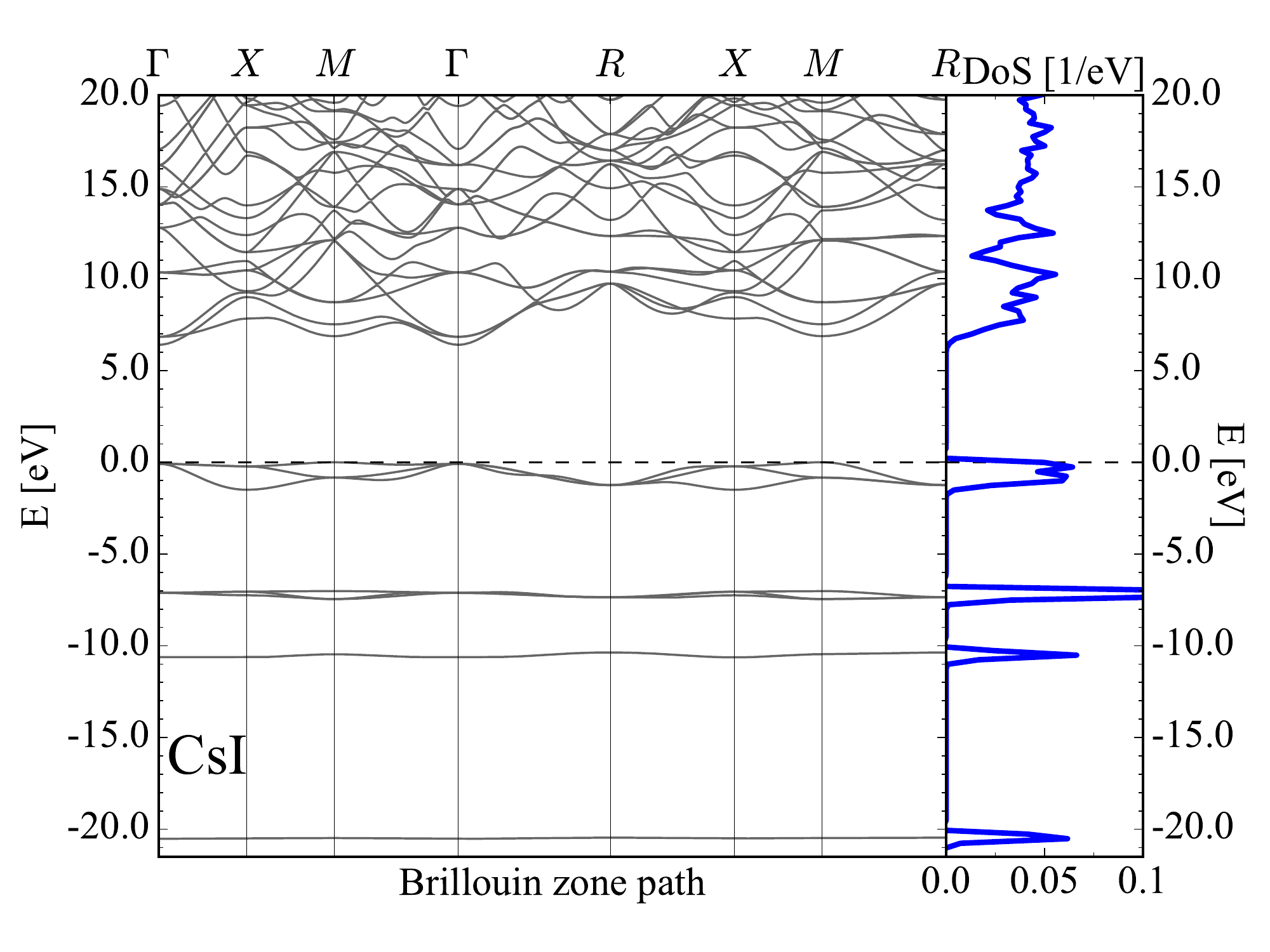}
\includegraphics[width=0.48\textwidth]{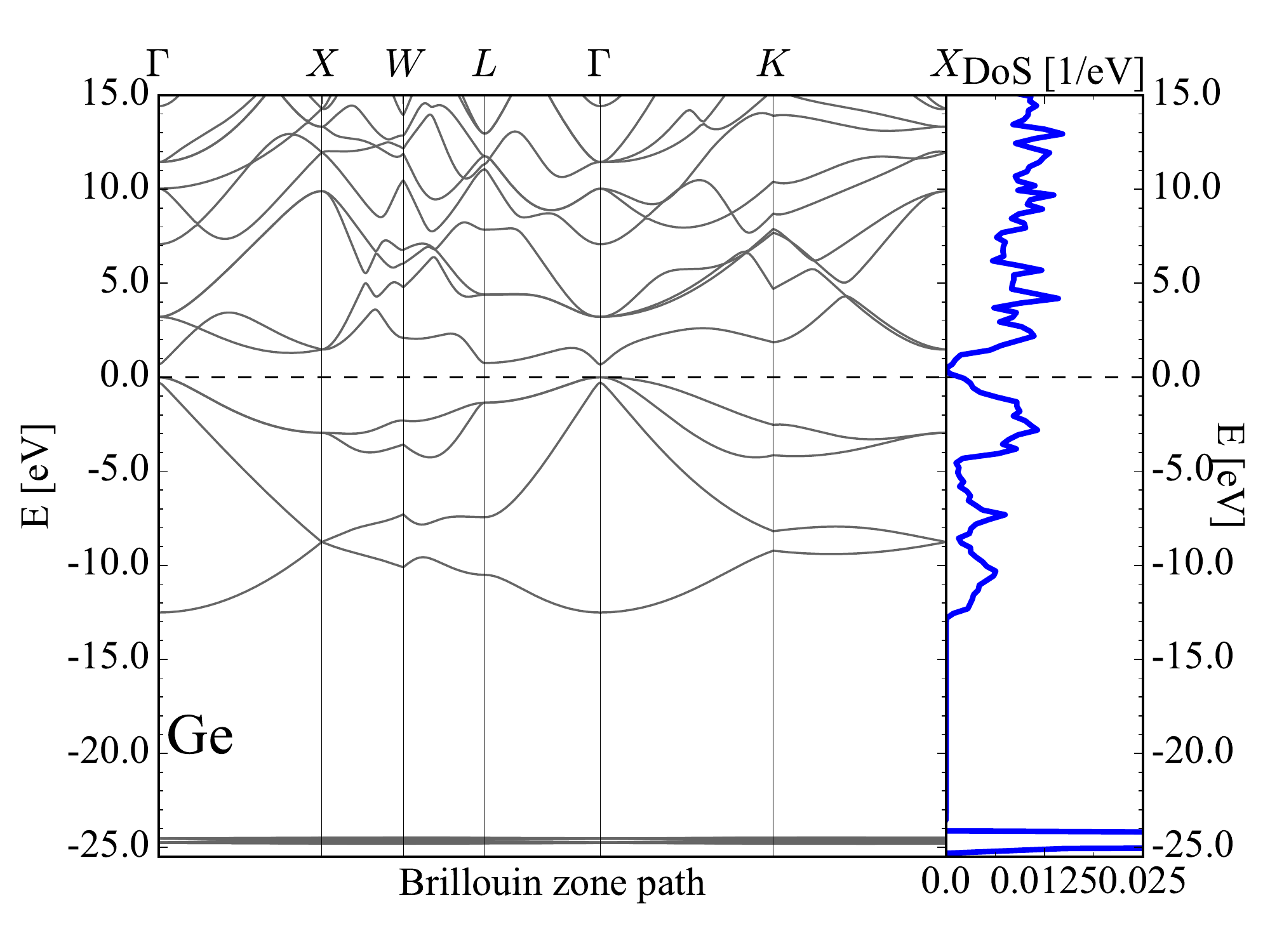} \\
\begin{minipage}[t]{0.5\textwidth}
\mbox{}\\[-0.3\baselineskip]
\includegraphics[width=0.98\textwidth]{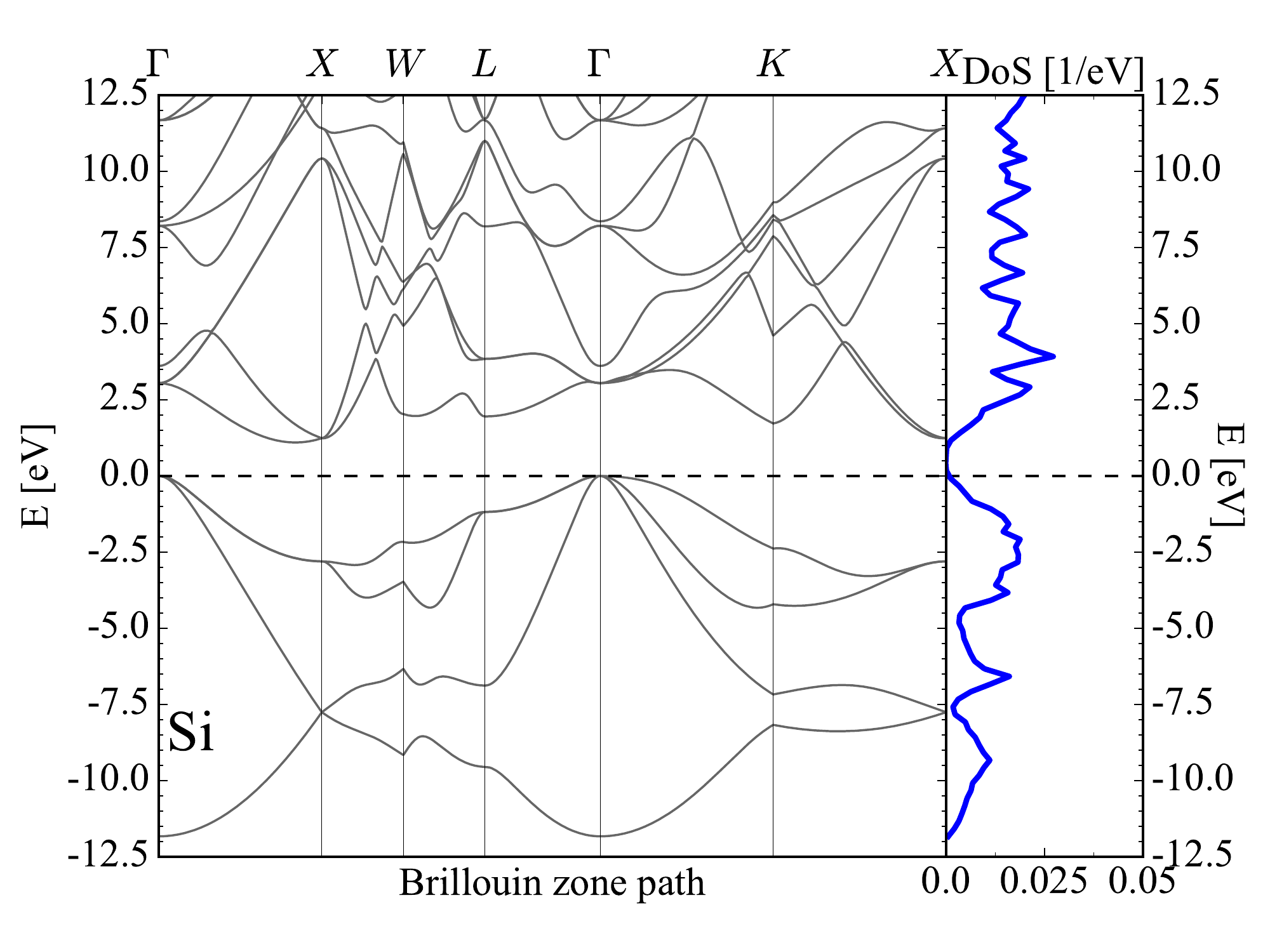}
\end{minipage}
\hfill
\begin{minipage}[t]{0.45\textwidth}
\mbox{}\\[2\baselineskip]
\caption{
Calculated band structure (black lines) and density of states (DoS, blue lines) of the electronic states for gallium arsenide (GaAs), 
sodium iodine (NaI), cesium iodine (CsI), germanium (Ge), and silicon (Si). 
We show all valence electron states included in our DM-electron-scattering-rate calculation as well as the bottom of the conduction band. 
The DoS was calculated by smearing the energy with a Gaussian function of width $\delta E = 0.25$~eV.  
\label{fig:DOS}
}
\end{minipage}
\end{figure*}

We can estimate how far below the conduction band the density of states will be nonzero from exciton effects by 
using a hydrogen-like model for the electron-hole pair.  
In particular, the exciton binding energies $E_{B,n}$ can be approximated by a modified Rydberg energy, namely 
\begin{equation}\label{eq:exciton-binding}
\Delta E_{B,n} = \frac{\alpha^2 \, \mu_{eh}}{2\,\varepsilon^2\,n^2} \, \,,
\end{equation}
where $\varepsilon$ is the dielectric constant of the crystal, $n=1, 2, \ldots$, 
and $\mu_{eh}$ is the effective electron-hole reduced mass, given by
\begin{equation}
\mu^*_{eh} = \left(\frac{1}{m^*_e}+\frac{1}{m^*_h}\right)^{-1}\,,
\end{equation}
where $m_e^*$ ($m_h^*$) is the effective electron (hole) mass.  
In this approximation, the electron-hole pair is assumed to be subject to a screened Coulomb potential characterized by the 
dielectric constant $\varepsilon$.  This is a good approximation only if the exciton radius, $a_n$, is much larger than the lattice constant ({\it Wannier} exciton).  
The exciton radius is given by
\begin{equation}
a_n = \frac{\varepsilon m_e n^2}{\mu_{eh}} \, a_0\,,
\end{equation}
where $a_0$ is the (hydrogen) Bohr radius.  
The relevant values for the materials we considered in the letter are given in Table~\ref{tab:exciton-info}, which also lists the binding energy and size of the various 1s exciton states (i.e.~with $n=1$).  

The 1s-exciton radii listed in Table~\ref{tab:exciton-info} for GaAs, Ge, and Si are much larger than the lattice constants given in 
Table~\ref{tab:qedark-parameters}, so that the approximation of the binding energies with Eq.~(\ref{eq:exciton-binding}) is 
expected to be reasonable.  
For NaI and CsI, the approximation is expected to be worse, but not dramatically so.  
We can thus use this simple estimate of the binding energies to reach at least qualitative conclusions for how the 
inclusion of exciton effects might affect the DM-mass threshold and the DM-electron scattering-rate calculation.  

First, we see from Table~\ref{tab:exciton-info} that the 1s-exciton binding energies for the low-band-gap materials, GaAs, Ge, and Si, 
are very small, $\sim 10$~meV, but even for the insulators, NaI and CsI, the binding energy only reaches about $\sim 100$~meV.  
This lowers the mass threshold by $\sim 1-30$~keV, depending on the material, an effect that is smaller than the numerical uncertainty 
of the rate calculation without excitons. 

Second, recall that the electron's recoil energy after a DM scattering event is typically several eV.  
The typical recoil energy is thus larger than the band gap energy for semiconductors like GaAs, Ge, and Si.   
A moderate increase in the density of states from the inclusion of exciton effects $10$~meV below the band gap, as well as just above it, is thus not expected to be important in the rate calculation. 
For the insulators NaI and CsI with band gaps around 6 eV, an increase in the density of states below and above the conduction band's bottom could be somewhat important, since the electron will largely prefer to scatter to those states rather than higher-energy ones. 

The calculation of exciton effects in the DM-electron scattering requires a dedicated effort. 
One reason for this is that existing numerical codes usually calculate exciton effects for photon absorption or emission.  
However, a photon being absorbed by an electron does not significantly change the momentum of the electron, so that 
the transition from valence to conduction band occurs at roughly the same $k$-point.  
Instead, DM scattering off an electron does transfer a sizeable momentum, comparable with the crystal momentum. 

The above discussion shows that it would be desirable to include exciton effects for NaI and CsI in the future. Neglecting the exciton effects, as we have done in our calculations, gives an overall conservative estimate for the DM-electron scattering rates. 

\vskip -4mm
\section{RECOIL SPECTRA FOR GALLIUM ARSENIDE, SODIUM IODINE, AND CESIUM IODINE} \vskip -3mm

Fig.~\ref{fig:spectra} shows the electron recoil spectra from DM-electron scattering for GaAs, NaI, and CsI.  
as a function of total deposited energy $E_e$, for two DM masses and two choices for the DM form factor.  
We include also spectra for Ge and Si for comparison (see also~\cite{Essig:2015cda}). 
As expected, the spectra extend to higher recoil energies for higher DM masses, and $F_{\rm DM}\propto 1/q^2$ spectra decrease faster 
than those for $F_{\rm DM}=1$, since lower momentum transfers are preferred.   
Bump-like features in the spectra are explained by comparing the energy at which they occur with the energies of the 
available valence bands. 

\vskip -4mm
\section{DENSITY OF STATES AND  BAND STRUCTURES} \vskip -3mm

Fig.~\ref{fig:DOS} shows our calculated band structure and density of states (DoS) for GaAs, NaI, and CsI.  
For completeness, we include slightly modified plots from~\cite{Essig:2015cda} for Ge and Si. 
We show all valence electron levels included in our DM-electron-scattering-rate calculation as well as the bottom of the conduction band.

\end{document}